\documentclass{aastex61}
\usepackage{amsmath}
\received{\today}
\revised{}
\accepted{}
\submitjournal{ApJ}

\shorttitle{CH$_3$NH$_2$ chemistry}
\shortauthors{Suzuki et al.}
\begin{document}

\title{Survey observation of CH$_3$NH$_2$ and its Formation Process}

\author[0000-0003-3278-2513]{Taiki Suzuki}
\affil{Astrobiology Center, Osawa 2-21-1, Mitaka, Tokyo 181-8588, Japan}
\affil{National Astronomical Observatory of Japan, Osawa 2-21-1, Mitaka, Tokyo 181-8588, Japan}
\email{taiki.suzuki@nao.ac.jp}

\author[0000-0001-7031-8039]{Liton Majumdar}
\affiliation{School of Earth and Planetary Sciences, National Institute of Science Education and Research, HBNI, Jatni 752050, Odisha, India}
%\affiliation{Jet Propulsion Laboratory, California Institute of Technology, 4800 Oak Grove Drive, Pasadena, CA 91109, USA}
\email{liton@niser.ac.in}
\email{liton.majumdar@jpl.nasa.gov}
 
\author{Kazuki Tokuda}
\affiliation{Department of Physical Science, Graduate School of Science, Osaka Prefecture University, 1-1 Gakuen-cho, Naka-ku, Sakai, Osaka 599-8531, Japan}

\author{Harumi Minamoto}
\affiliation{Department of Earth and Planetary Sciences, Tokyo Institute of Technology, 2-12-1 Ookayama, Meguro-ku, Tokyo, 152-8551, Japan}

\author[0000-0003-2775-7487]{Masatoshi Ohishi}
\affil{National Astronomical Observatory of Japan, Osawa 2-21-1, Mitaka, Tokyo 181-8588, Japan}

\author{Masao Saito}
\affil{National Astronomical Observatory of Japan, Osawa 2-21-1, Mitaka, Tokyo 181-8588, Japan}

\author{Tomoya Hirota}
\affil{National Astronomical Observatory of Japan, Osawa 2-21-1, Mitaka, Tokyo 181-8588, Japan}

\author{Hideko Nomura}
\affiliation{Department of Earth and Planetary Sciences, Tokyo Institute of Technology, 2-12-1 Ookayama, Meguro-ku, Tokyo, 152-8551, Japan}
\affil{National Astronomical Observatory of Japan, Osawa 2-21-1, Mitaka, Tokyo 181-8588, Japan}

\author{Yoko Oya}
\affil{Department of Physics, The University of Tokyo, 7-3-1, Hongo, Bunkyo-ku, Tokyo 113-0033, Japan}
%% Note that the \and command from previous versions of AASTeX is now
%% depreciated in this version as it is no longer necessary. AASTeX 
%% automatically takes care of all commas and "and"s between authors names.

%% AASTeX 6.1 has the new \collaboration and \nocollaboration commands to
%% provide the collaboration status of a group of authors. These commands 
%% can be used either before or after the list of corresponding authors. The
%% argument for \collaboration is the collaboration identifier. Authors are
%% encouraged to surround collaboration identifiers with ()s. The 
%% \nocollaboration command takes no argument and exists to indicate that
%% the nearby authors are not part of surrounding collaborations.

%% Mark off the abstract in the ``abstract'' environment. 
\begin{abstract}
We present the observational result of a glycine precursor, methylamine (CH$_3$NH$_2$), together with methanol (CH$_3$OH) and methanimine (CH$_2$NH) towards high-mass star-forming regions, NGC6334I, G10.47+0.03, G31.41+0.3, and W51~e1/e2 using ALMA.
The molecular abundances toward these sources were derived using the rotational diagram method and compared with our state-of-the-art chemical model.
We found that the observed ratio of ``CH$_3$NH$_2$/CH$_3$OH'' is in between 0.11 and 2.2.
We also found that the observed ``CH$_3$NH$_2$/CH$_3$OH'' ratio agrees well with our chemical model by considering the formation of CH$_3$NH$_2$ on the grain surface via hydrogenation process to HCN.
This result clearly shows the importance of hydrogenation processes to form CH$_3$NH$_2$.
NGC63343I MM3, where CH$_3$NH$_2$ was not detected in this study and showed ``CH$_3$NH$_2$/CH$_3$OH'' ratio of less than 0.02, is clearly distinguished from the other cores.
%
%The physical evolution may be different from the ckemical model for this source.
\end{abstract}

%% Keywords should appear after the \end{abstract} command. 
%% See the online documentation for the full list of available subject
%% keywords and the rules for their use.

%% From the front matter, we move on to the body of the paper.
%% Sections are demarcated by \section and \subsection, respectively.
%% Observe the use of the LaTeX \label
%% command after the \subsection to give a symbolic KEY to the
%% subsection for cross-referencing in a \ref command.
%% You can use LaTeX's \ref and \label commands to keep track of
%% cross-references to sections, equations, tables, and figures.
%% That way, if you change the order of any elements, LaTeX will
%% automatically renumber them.

%% We recommend that authors also use the natbib \citep
%% and \citet commands to identify citations.  The citations are
%% tied to the reference list via symbolic KEYs. The KEY corresponds
%% to the KEY in the \bibitem in the reference list below. 

\section{Introduction}
It is believed that the first chemical evolution toward the origin of life would have started in molecular clouds and continued within the protoplanetary disk, followed by their delivery to the early Earth by comets and asteroids.
However, the synthesis and evolution of organic molecules, which form the building blocks of more complex biotic molecules, is not well understood.
Over the last several years, there have been significant advances in this field thanks to dedicated searches for molecules of biological importance in the ISM and in the atmospheres of comets.
Among them, observations in the direction of the Galactic center towards Sgr B2(N) with the Green Bank Telescope (GBT) have led to the detections of interstellar aldehydes, namely propenal (CH$_2$CHCHO) and propanal (CH$_3$CH$_2$CHO) \citep{Hollis04a} and simple aldehyde sugars like glycoaldehyde (CH$_2$OHCHO) \citep{Hollis04b}, the first keto ring molecule to be found in the ISM.
Recently, propylene oxide (CH$_3$CHCH$_2$O) \citep{McGuire16} has also been detected in the same source.
This is the first molecule detected in interstellar space that has the property of chirality, making it a leap forward in our understanding of how prebiotic molecules are made in the Universe.
In the era of the Atacama Large Millimeter Array (ALMA), the detection of the branched alkyl molecule iso-propyl cyanide (i-C$_3$H$_7$CN) in Sgr B2(N) also gave us the clues to the presence of amino acids in the ISM due to its key side-chain structure \citep{Belloche14}.
The amino acids are the building blocks of life and this is why the search for amino acids and their complex organic precursors at different stages of star and planet formation is one of the exciting topics in modern astronomy.
Since glycine is the simplest amino acid and the only non-chiral member out of 20 standard amino acids, it has gathered attention of researchers.
Recently, volatile glycine (NH$_2$CH$_2$COOH) was detected in the coma of comet 67P/Churyumov-Gerasimenko by the ROSINA (Rosetta Orbiter Spectrometer for Ion and Neutral Analysis) mass spectrometer \citep{Altwegg16}, supporting glycine's interstellar origin.

Revealing the formation pathways to glycine is important topic for astrochemistry and astrobiology, since most of the prebiotic molecules contain N atoms and the general chemical evolution of N-bearing molecules in the ISM is poorly known.
The high-mass star-forming regions are the best sources to study the chemical evolution of COMs (Complex Organic Molecules) with their high gas density and the warm environment that trigger the molecular evolution through thermal hopping of molecules on grains.
In this context, formation processes to glycine in the high-mass star-forming regions have been studied by many authors:
[I]  It is suggested that protonated hydroxylamine (NH$_2$OH$^+$) will react with acetic acid (CH$_3$COOH) in the gas phase \cite{Blagojevic03} to form glycine.
[II] The importance of grain surface chemistry is emphasized as well.
CH$_3$NH$_2$ will react with CO$_2$ to form glycine under the radiation of UV photons or cosmic rays \cite{Holtom05,Lee09}.
[III] \cite{Singh13} suggested that glycine can be formed via successive gas phase radical-radical, radical-molecule reactions of simple species, such as CH$_2$, NH$_2$, CH, and CO.
Of the precursors of glycine, recent works have improved the understanding of CH$_3$NH$_2$ chemistry.
Experimentally, \cite{Kim11} reported the formation of CH$_3$NH$_2$ after electron and photon irradiation on interstellar ice analogues consisting of CH$_4$ and NH$_3$.
This path would involve the recombination of radical species, CH$_3$ and NH$_2$ (CH$_3$ + NH$_2$ $\longrightarrow$ CH$_3$NH$_2$), which are the products of decomposition of CH$_4$ and NH$_3$.
Another candidate pathway would be a successive hydrogenation processes to HCN (HCN + 2H $\longrightarrow$ CH$_2$NH, and CH$_2$NH + 2H $\longrightarrow$ CH$_3$NH$_2$) \citep{Woon02}.
\cite{Theule11} experimentally demonstrated this formation process using the interstellar ice analogues containing HCN.

The study of the chemical model, where the evolution of molecular abundances are numerically solved with thousands of reactions along with the physical evolution of the star, is an essential tool to test the importance of formation paths.
The formation processes of glycine's precursors have been discussed with the chemical model studies.
\cite{Garrod13} has compared several possible processes to form glycine, including both gas phase and grain surface reactions and suggested that with a model for high-mass star-forming region (fast warm-up model), glycine is most efficiently formed via the reaction between CH$_2$NH$_2$ and COOH radicals, where COOH radical is formed from the destruction of HCOOH.
\cite{Suzuki18} extended this work and suggested the acceleration of glycine formation by the photochemical reactions of CH$_3$NH$_2$ and CO$_2$, which was supported by experiments \citep{Lee09}.
In either cases, CH$_3$NH$_2$ may play an important role in the formation of glycine.
The detections of CH$_3$NH$_2$ in Murchison meteorite and coma of comet 67P/Churyumov-Gerasimenko suggest the formation of CH$_3$NH$_2$ in the extra terrestrial condition \citep{Pizzarello94,Altwegg16}.
\cite{Altwegg16} has pointed out this chemical link considering the co-existence of glycine and CH$_3$NH$_2$ in the coma of comet 67~P as well.
With the chemical model, \cite{Suzuki16} showed that CH$_3$NH$_2$ is formed on grains via the successive hydrogenation processes to HCN and CH$_2$NH, which agrees with previous terrestrial studies \citep{Theule11}.
On the other hand, we got an unexpected result that the grain surface chemistry to CH$_2$NH is less important to explain the gas phase CH$_2$NH due to its rapid conversion to CH$_3$NH$_2$ on grains.
Instead, CH$_2$NH is efficiently formed via a gas phase reaction of ``CH$_3$ + NH $\longrightarrow$ CH$_2$NH + H".

Such theoretically predicted formation processes must be confirmed by the observation toward actual star-forming regions.
The number of studies for glycine precursors is very small for the low-mass star-forming regions.
CH$_2$NH is also detected toward a Solar-Like protostar, IRAS16293-2422B, as a part of ALMA-PILS project \citep{Ligterink18}, and toward the molecular cloud L183 and the translucent cloud CB17 as well \citep{Turner99}.
However, they could not detect CH$_3$NH$_2$ and currently the detection of CH$_3$NH$_2$ is not achieved toward low-mass star-forming regions.

On the other hand, the detections of glycine's precursors toward high-mass star-forming regions have been achieved by many authors.
So far, many authors claimed the detection of CH$_2$NH toward Sgr B2 region.
This molecule is detected toward Sgr B2(OH) \citep{Godfrey73,Turner89}, Sgr B2(N) \citep{Halfen13}, and Sgr B2(M) \citep{Sutton91}.
\cite{Nummelin98} also successfully detect it in Sgr B2(N), (M), and (NW).
\cite{Jones08, Jones11} found the distribution of CH$_2$NH from Sgr B2(N) to (S) at 3 mm and 7 mm with the MOPRA telescope.
CH$_2$NH is reported toward other high-mass star-forming regions, W51 e1/e2, Orion KL, and G34.3+0.15 \citep{Dickens97,White03}, and G19.61-0.23 \citep{Qin10}.
We extended CH$_2$NH survey and reported new detections of CH$_2$NH toward G10.47+0.03, G31.41+0.3, NGC6334F, and DR21(OH) \citep{Suzuki16}.
NGC6334F is also referred as NGC6334I and hereafter we use the term of NGC6334I to avoid confusion with the works by other authors.
\cite{Widicus17} added new CH$_2$NH sources of GCM+0.693-0.027, the shocked region located in the Sgr B2 complex, a massive hot core G12.91-0.26, and G24.33+00.11 MM1.
CH$_3$NH$_2$ was reported early toward Sgr B2 \citep{Kaifu74,Fourikis74}, and \cite{Belloche13} presented the detailed analysis toward Sgr B2 (N) and (M) with IRAM 30~m telescope.
\cite{Halfen13} showed that CH$_2$NH and CH$_3$NH$_2$ toward Sgr B2 (N) have the different excitation temperatures, 44$\pm$13 and 159 $\pm$30~K, respectively, suggesting that they exist in the different environment.
Though the survey observation toward other high-mass star-forming regions by \cite{Ligterink15} could not confirm CH$_3$NH$2$, \cite{Ohishi19} succeeded the detection of CH$_3$NH$_2$ toward G10.47+0.03 with NRO~45m telescope.
Since the rotation temperature of CH$_3$NH$_2$ was less than 45~K, the observed CH$_3$NH$_2$ would be in the cold envelope surrounding the hot core.
After that, \cite{Bogelund19} confirmed the detection of CH$_3$NH$_2$ in NGC6334I, resolving three clumps called MM1, MM2, and MM3.
Furthermore, tentative detection of CH$_3$NH$_2$ were reported toward Orion Hot core \citep{Pagani17}.
Since glycine's precursors are well known in the high-mass star-forming regions, it is possible to discuss their formation paths with chemical model using the typical abundance of these species.

Since CH$_3$OH is the most abundant complex organic molecules in the star-forming regions, it will be helpful to understand the chemical evolution of CH$_3$NH$_2$.
CH$_3$OH is known to be formed via the hydrogenation process to CO (CO + 2H $\rightarrow$ H$_2$CO, and H$_2$CO + 2H $\rightarrow$ CH$_3$OH), which is very similar to the predicted formation path of CH$_3$NH$_2$ (HCN + 2H $\rightarrow$ CH$_2$NH, and CH$_2$NH + 2H $\rightarrow$ CH$_3$NH$_2$).
Therefore the comparison of the column densities between CH$_3$NH$_2$ and CH$_3$OH would be useful to reveal the formation process of CH$_3$NH$_2$.
Since the molecular surveys of CH$_3$NH$_2$, CH$_2$NH, and CH$_3$OH have been mainly performed with single dish telescope, the column density ratio utilizing such single dish data will suffer from the unknown source sizes that lead to the different degree of the beam dilution.
Therefore, the interferometric observations of these species are strongly desired.
%
%The ratio of ``CH$_3$OH/CH$_3$NH$_2$'' obtained in \cite{Bogelund19} should be revised with the accurate observation of CH$_3$OH since CH$_3$OH was not their prime target.
%
In this paper, we will report the observations of CH$_3$NH$_2$, CH$_2$NH, and CH$_3$OH toward G10.47+0.03, NGC6334I, G31.41+0.3, and W51~e1/e2 region by ALMA telescope.
The detail of our observation is described in Section~2.
The observational result is presented in Section~3.
Our result is compared with the chemical model in Section~4.
We summarize our work in Section~5.

\section{Observation and Analysis}
\subsection{Source Selection}
We selected high-mass star-forming regions where we have already detected CH$_2$NH with NRO~45m telescope \citep{Suzuki16}.
In this work, we performed a survey observation of the glycine precursor, CH$_2$NH, toward CH$_3$OH-rich high-mass star-forming regions.
As a result, we detected CH$_2$NH toward eight sources.
One finding of our work was that CH$_2$NH showed wide range of fractional abundances among our sources.
Since CH$_2$NH-rich sources showed the different strength of  a recombination line, H54$\beta$, the different abundance of CH$_2$NH may be due to the different evolutionary phase.
Therefore, we selected four sources, NGC6334I, G10.47+0.03, G31.41+0.3, and W51~e1/e2, from \cite{Suzuki16}, so that we can cover the different evolutionary phase of the star-formation.
Their coordinates, the source velocities, the distances, and previously reported CH$_2$NH fractional abundances are summarized in Table~\ref{tbl:source_list}.

\subsection{Spectroscopy}
The molecular line parameters of CH$_3$OH, CH$_3$NH$_2$, and CH$_2$NH are provided by the database of SLAIM from the on-line database splatalogue\footnote{\url{http://www.cv.nrao.edu/php/splat/}}.
The quantum number of CH$_3$OH transitions are described by  $J$, $\pm$ $K_{\rm a}$, and $\Gamma$.
For transitions with $\Gamma = A$, the $\Gamma$ is associated with a '+' or '-' sign to represent the parity.
On the other hand, $\Gamma = E_1$ and $E_2$ are presented by the identical alphabet of $\Gamma = E$, but denoted by the positive and negative sign of K, respectively, due to the absence of a parity entry.
CH$_2$NH is a simple asymmetric top with all atoms being on the simple plane, and the transitions are described using labels J, K$_{\rm a}$, and K$_{\rm c}$.
The transitions of CH$_3$NH$_2$ are labeled by the torsion-inversion rotation irreducible representations $\Gamma$, with its states of  labeled with the $A_1$, $A_2$, $B_1$, $B_2$, $E_{1\pm1}$, and $E_{2\pm1}$ \citep{Ilyushin07}.
The selection rules are $\Gamma$ = $A_1 \leftrightarrow A_2$, $B_1 \leftrightarrow B_2$, $E_{1\pm1} \leftrightarrow E_{1\pm1}$, and $E_{2\pm1} \leftrightarrow E_{2\pm1}$.
The symmetry levels have nuclear spin-statistical weights of 1 for the $A_1$, $A_2$, and $E_2$ states and 3 for the $B_1$, $B_2$, and $E_1$ components.
The hyper fine structures of CH$_2$NH and CH$_3$NH$_2$ are not considered in this analysis.

\subsection{Observation}
Our observations were carried out with ALMA cycle~5 in May, 2018, using the ALMA Band 5 and 6 receivers.
This interferometric observations have the advantage to obtain the spatial distributions of molecules.
Although we could not get the source size by our previous single dish observation, the information of the source size are essential to discuss the accurate column densities.
The observing parameters, such as the phase centers, the spectral resolutions, the maximum recoverable scales, and the angular resolutions, are summarized in Tables~\ref{tbl:observation_parameter1} and \ref{tbl:observation_parameter2}.
The observed frequency ranges are the same for all sources, as the center frequencies, the band widths and number of channels are summarized in Table~\ref{tbl:observation_frequency_range}.
These receiver setups were designed to cover the number of molecular transitions of CH$_3$OH, CH$_3$NH$_2$, and CH$_2$NH that can cover the wide range of the upper state energy level (Table~\ref{tbl:observed_transitions}).
With the on-line database splatalogue, we excluded the possibility of the contamination by other strong molecular lines for these transitions.

\subsection{Analysis}
\begin{figure}
 \begin{tabular}{ll}
\includegraphics[scale=0.5]{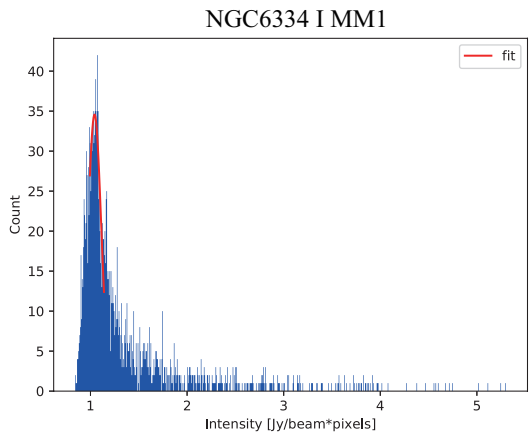}\\
 \end{tabular}
\caption{
An example of the intensity histogram obtained in NGC6334I MM1 region.
The Gaussian fitting is performed to obtain the mean and the standard deviation, which are used to subtract the continuum emission with casa.
\label{fig:continuum_subtraction}
}
\end{figure}

Our data were calibrated by Common Astronomy Software Applications (CASA) V.5.1.1, with the ALMA Cycle 5 pipeline.
For the subtraction of the continuum emission in the $(u,v)$ domain (the raw visibilities measured by the interferometer), we determine the continuum levels statistically using the method described in \cite{Sanchez-Monge18}.
First, the spectra before the continuum subtraction is obtained.
Second, the histograms of the intensity are created using the all channels in the spectra.
Finally, we perform the Gaussian fitting to the lowest peak in the histogram to obtain the peak, the mean and the standard deviation of the Gaussian.
As an example, we shown the histogram of the intensity and the result of the Gaussian fitting in Figure~\ref{fig:continuum_subtraction}.
The continuum level is determined as the mean value of the Gaussian.
We regarded the channels with the intensity less than three-sigma level from the continuum level, and the CASA task {\sc uvcontsub} is used to subtract the continuum emission.

\begin{figure}
 \begin{tabular}{ll}
\includegraphics[scale=0.5]{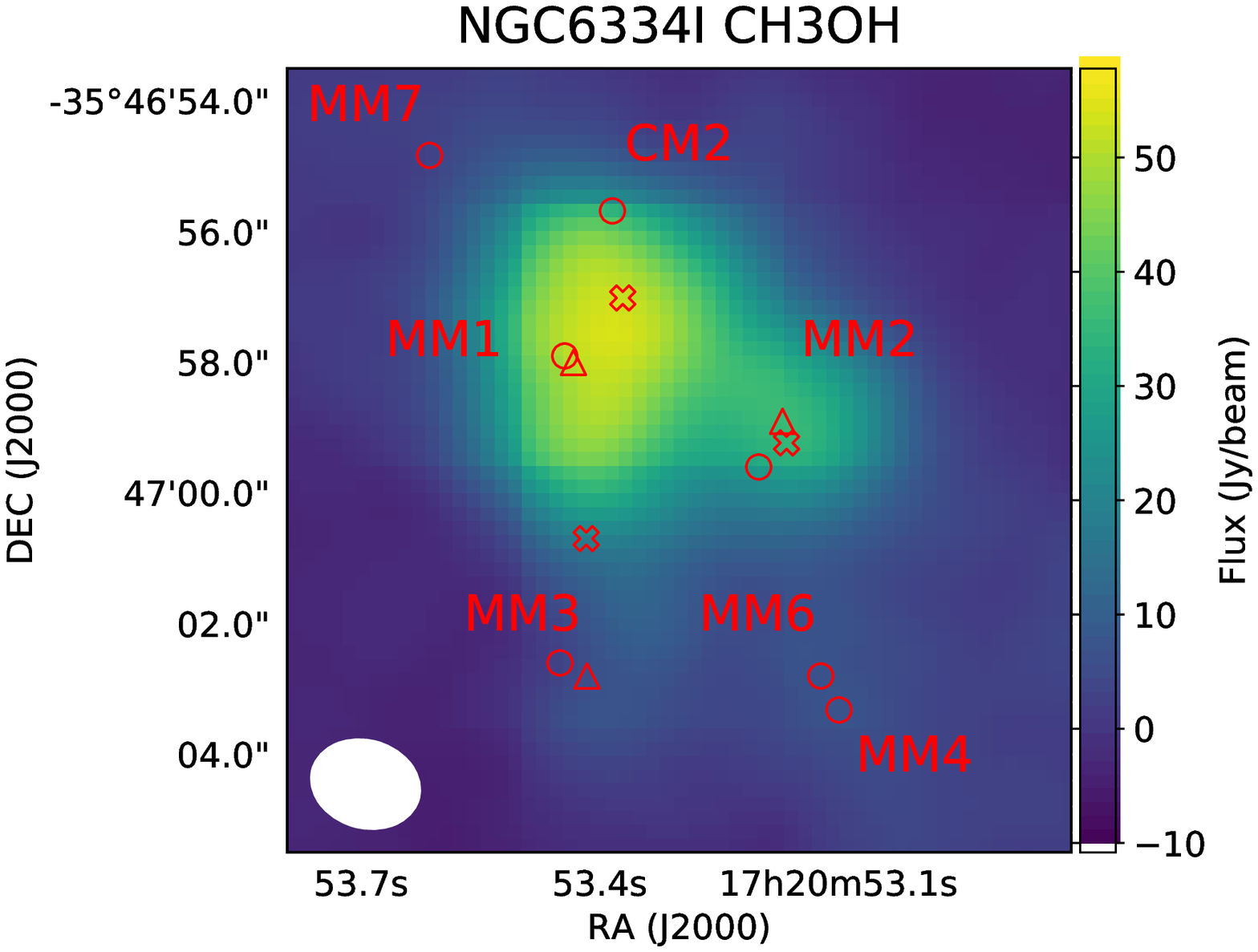}&
\includegraphics[scale=0.5]{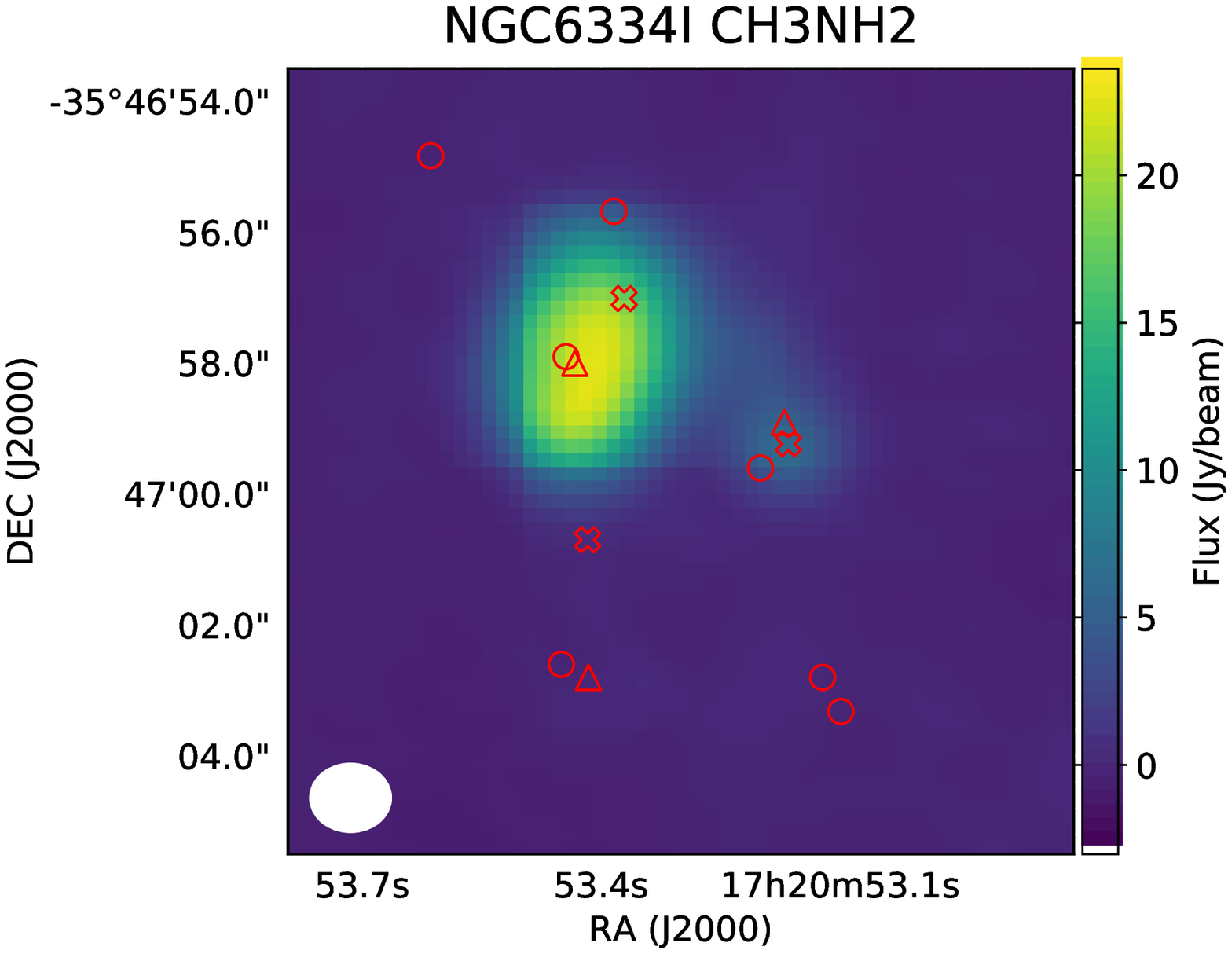}\\
\includegraphics[scale=0.5]{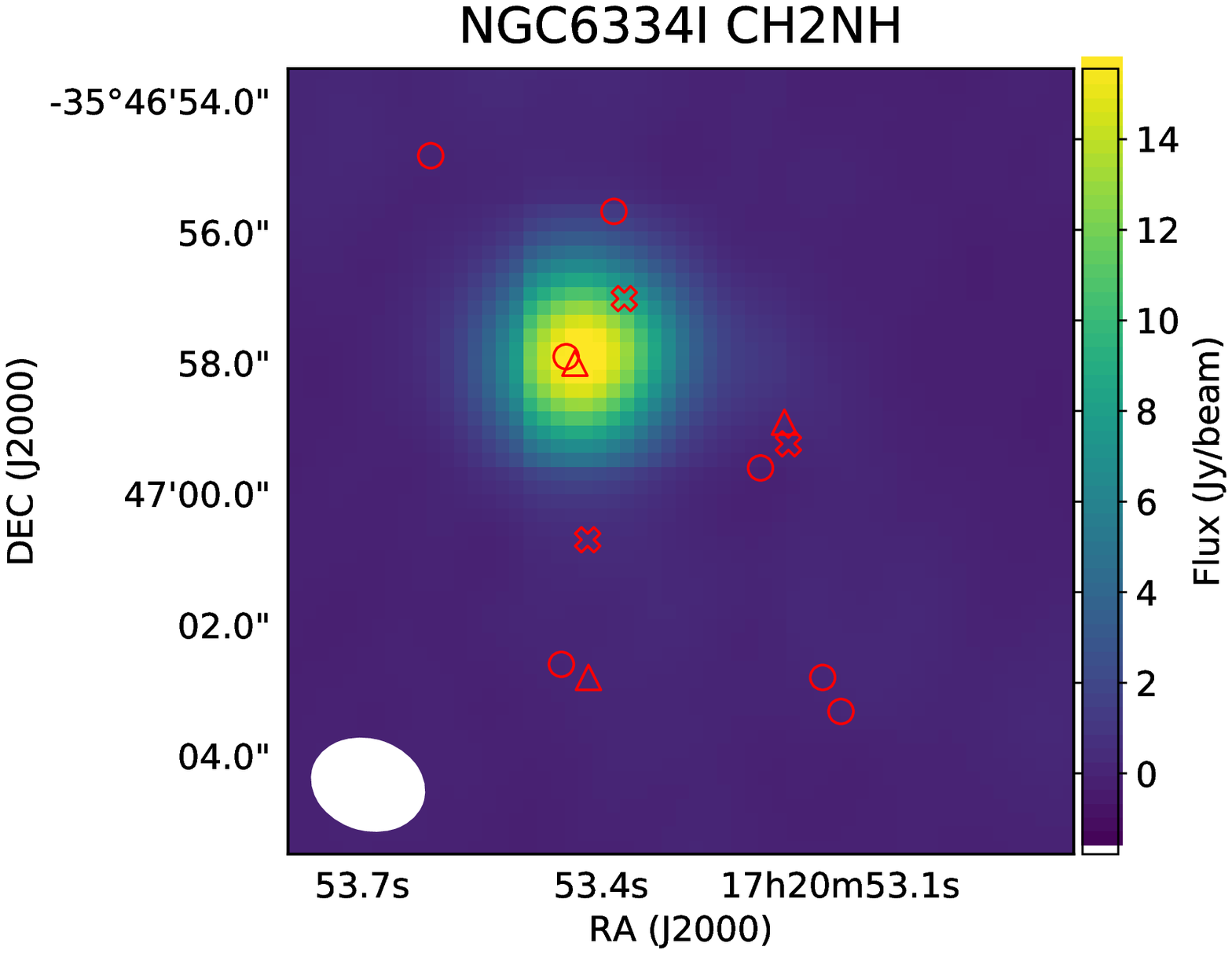}&\\
 \end{tabular}
\vspace{35mm}
\caption{
The integrated intensity maps of CH$_3$OH 4, 1, $E$ $\rightarrow$ 3 ,1, $E$, CH$_3$NH$_2$ 4, 1, $E_{1+1}$ $\rightarrow$ 3, 0, $E_{1+1}$, and CH$_2$NH 3, 2 ,2 $\rightarrow$ 2, 2, 1 transition toward NGC6334I.
The positions of previously known continuum sources \citep{Hunter06,Brogan16} are shown by circles on the map.
The triangles shows the positions of hot cores, MM1, MM2, and MM3, where the spectra was extracted.
The positions observed by \cite{Bogelund19} are marked by X position.
The velocity ranges are from -14.6 to -0.9, from -14.2 to -5.9, and from -13.9 to -0.2~km s$^{-1}$ for CH$_3$OH, CH$_3$NH$_2$, and CH$_2$NH, respectively.
\label{fig:map_N63}
}
\end{figure}

\begin{figure}
 \begin{tabular}{ll}
\includegraphics[scale=0.5]{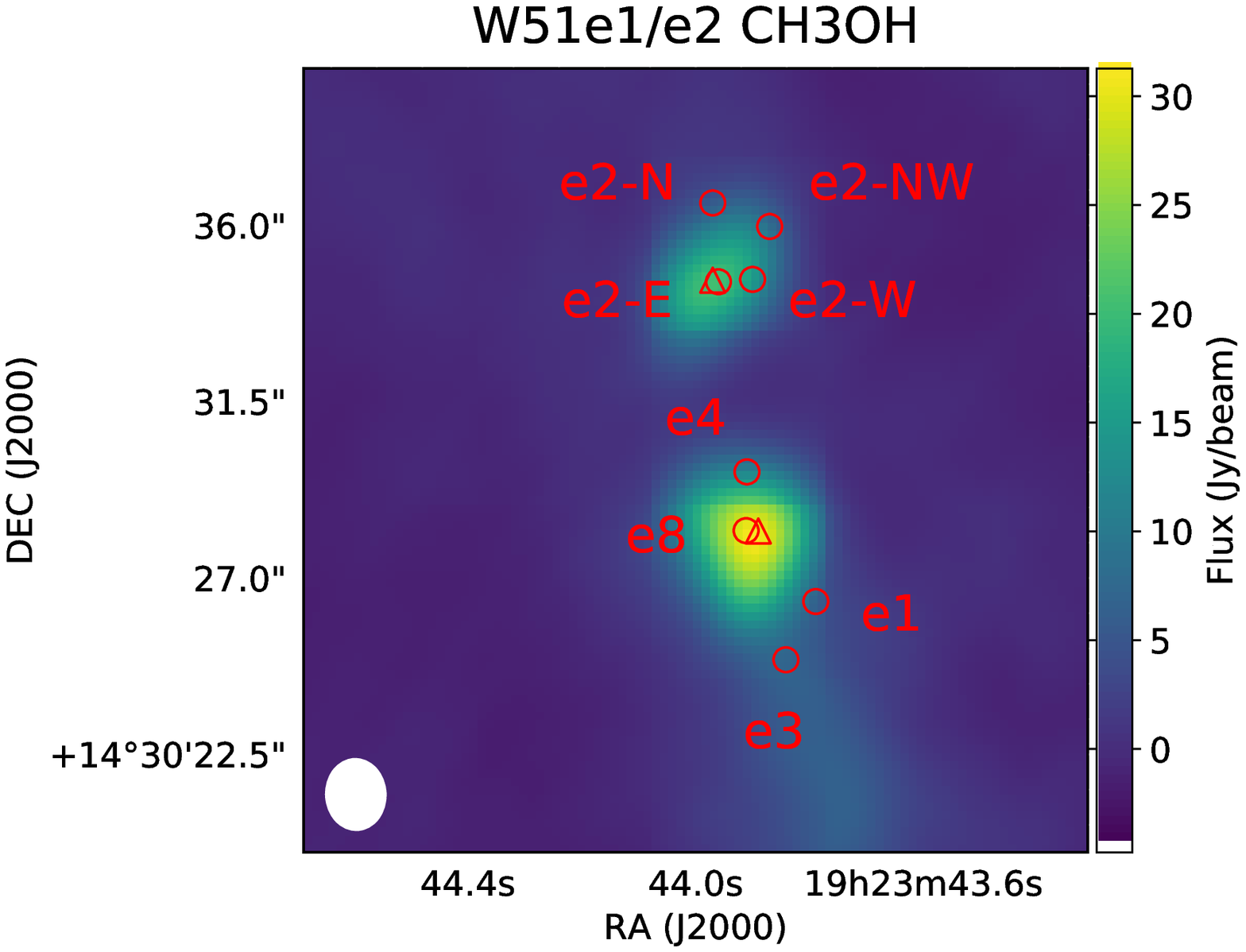}&
\includegraphics[scale=0.5]{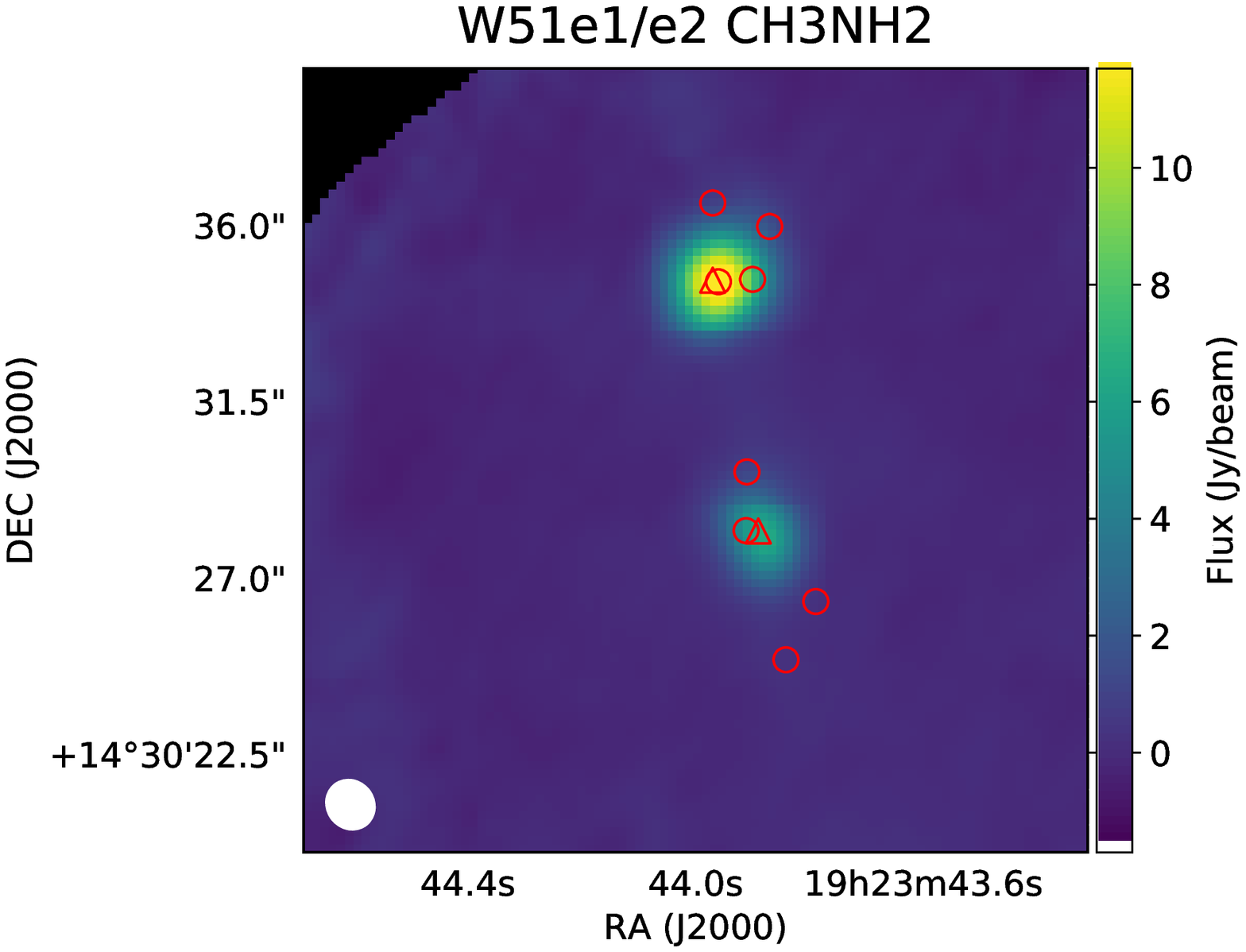}\\
\includegraphics[scale=0.5]{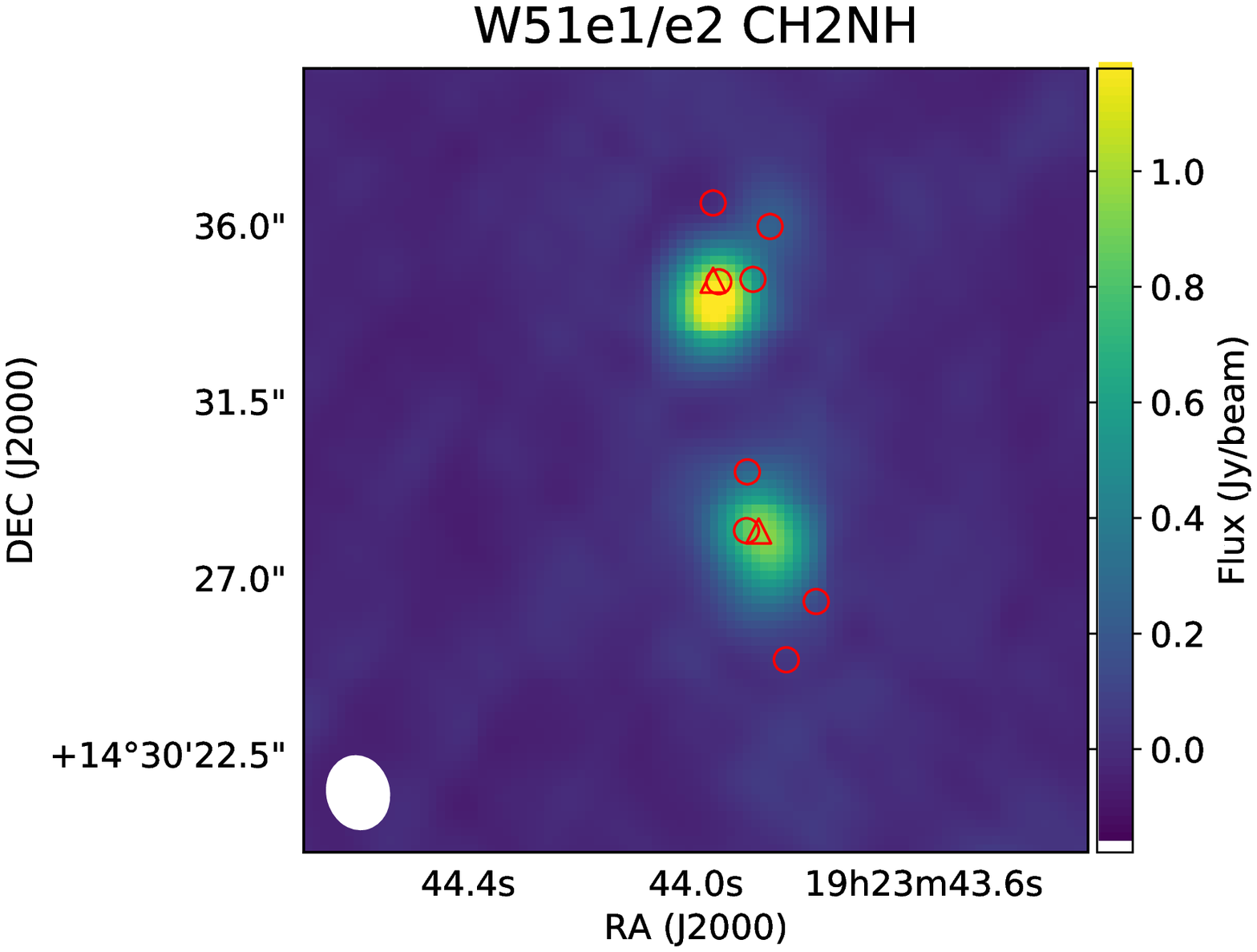}&\\
 \end{tabular}
\vspace{35mm}
\caption{
The integrated intensity maps of CH$_3$OH 4, 1, $E$ $\rightarrow$ 3 ,1, $E$, CH$_3$NH$_2$ 4, 1, $E_{1+1}$ $\rightarrow$ 3, 0, $E_{1+1}$, and CH$_2$NH 3, 2 ,2 $\rightarrow$ 2, 2, 1 transition toward W51~e1/e2. 
The positions of previously known continuum sources \citep{Gaume93,Shi10} are shown on the map by circles.
The triangles shows the positions of hot cores, where the spectra was extracted.
The velocity ranges are from 44.9 to 58.5, from 49.8 to 58.1, and from 47.0 to 60.7~km s$^{-1}$ for CH$_3$OH, CH$_3$NH$_2$, and CH$_2$NH, respectively.
\label{fig:map_W51}
}
\end{figure}

%Then, the data have been averaged within primary beam.
%
Then, the deconvolved image was obtained by the CASA task {\sc tclean} by applying natural weighting.
Prior to the detailed analysis of molecular abundance, we created the moment 0 map of the intensity through the CASA task {\sc immoment}.
The distribution of the integrated intensities for ``4, 1, $E$ $\rightarrow$ 3 ,1, $E$'' transition of CH$_3$OH , ``4, 1, $E_{1+1}$ $\rightarrow$ 3, 0, $E_{1+1}$'' transition of CH$_3$NH$_2$, and ``3, 2 ,2 $\rightarrow$ 2, 2, 1'' transition of CH$_2$NH, were as shown through Figures~\ref{fig:map_N63} to \ref{fig:map_W51}.
These lines are less likely to be contaminated by other transitions and detected toward all sources.
%
%Although it is quite difficult to select the completely contamination-free transitions from our molecular-rich sources, above transitions would enable us to get the spatial distributions of these molecules.
%
The upper energy levels and the products of the intrinsic intensity and the square of the permanent dipole moments of these transitions are, 44.3, 44.9, and 25.9~K, and 3.1, 2.9, and 3.7~D$^2$.
If we use the analysis result described in Section~3, CH$_3$OH ``4, 1, $E$ $\rightarrow$ 3 ,1, $E$'' transition shows the relatively high optical thickness of 1.1 in NGC 6334I MM1, but less than 0.6 for the other sources.
The optical depths of ``4, 1, $E_{1+1}$ $\rightarrow$ 3, 0, $E_{1+1}$'' transition of CH$_3$NH$_2$, and ``3, 2 ,2 $\rightarrow$ 2, 2, 1'' transitions are less than 0.3 for all sources.
Hence they would trace the warm and somewhat dense regions.

\section{Result}
\subsection{Spatial Distribution}
Figure~\ref{fig:map_N63} presents the integrated intensity maps of CH$_3$OH 4, 1, $E$ $\rightarrow$ 3 ,1, $E$, CH$_3$NH$_2$ 4, 1, $E_{1+1}$ $\rightarrow$ 3, 0, $E_{1+1}$, and CH$_2$NH 3, 2 ,2 $\rightarrow$ 2, 2, 1 transitions toward NGC6334I.
\cite{Hunter06} performed the mapping observation of 1.3~mm continuum emission with Submillimeter Array (SMA), and reported the structures named SMA1, SMA2, SMA3, and SMA4.
Later, with Karl G. Jansky Very Large Array (VLA) and ALMA, \cite{Brogan16} carried out the the high angular resolution mapping as high as 0.$\arcsec$17 (220~AU) from 5~cm to 1.3~mm.
They found the  MM1, MM2, MM3, MM4, MM6, MM7, MM9, and CM2, where MM1, MM2, MM3, and MM4 are respectively corresponding to previously known SMA1, SMA2, SMA3, and SMA4.
Of these components, MM3 is known as an UCHI\hspace*{-1pt}I region associated with free-free emission \citep{Hunter06}.
Many organic molecules have been reported toward MM1 and MM2 \citep[e.g.,][]{Walsh10,Zernickel12}.
The molecular distribution overlap with the positions of MM1, MM2, and MM3.
In our observation, CH$_3$OH clearly showed extended distribution than CH$_3$NH$_2$ and CH$_2$NH.
The CH$_3$OH distribution is complex with its size of $\sim$5 and $\sim$3~$\arcsec$, respectively for MM1 and MM2, but $\sim$10~$\arcsec$ if we consider the whole emission including the southern weak emission.
This whole distribution is close to the maximum recoverable scale.
CH$_3$NH$_2$ and CH$_2$NH emissions are dominant from MM1, with their source sizes of $\sim$3 and $\sim$2$\arcsec$, respectively.
Though CH$_3$NH$_2$ and CH$_2$NH emissions in MM2 are weaker than MM1, there are compact ($\sim$1$\arcsec$) sources. 
Though the previous continuum mapping observation by \cite{Brogan16} suggested the existence of at least seven and two unresolved cores in MM1 and MM2, respectively, our spatial distribution is not sufficient to resolve such structures. 
Future high-resolution mapping observations of CH$_3$OH would resolve the distribution of CH$_3$OH and probably CH$_3$NH$_2$ and CH$_2$NH inside MM1 and MM2.
The distributions of CH$_3$NH$_2$ and CH$_2$NH have already been reported by \cite{Bogelund19} toward MM1 and MM2 with the different frequency, which agree with our results.
In NG6334I region, we extracted the spectra from the peak positions of the CH$_2$NH intensity in MM1 region, since the distribution of CH$_2$NH is compact.
The coordinate of NGC6334I MM1 is offsetted from the observed position of \cite{Bogelund19} by about 2$\arcsec$.
The MM2 position is selected as the peak intensity position of CH$_3$NH$_2$, which is very close to the MM2 region analyzed in \cite{Bogelund19} and the peak intensity of the continuum.
The MM3 position is selected as the peak intensity position of CH$_3$OH since both CH$_2$NH and CH$_3$NH$_2$ intensities are so weak.
Our MM3 coordinate is offseted toward north by about 3$\arcsec$ than \cite{Bogelund19}.

\begin{figure}
 \begin{tabular}{ll}
\includegraphics[scale=0.5]{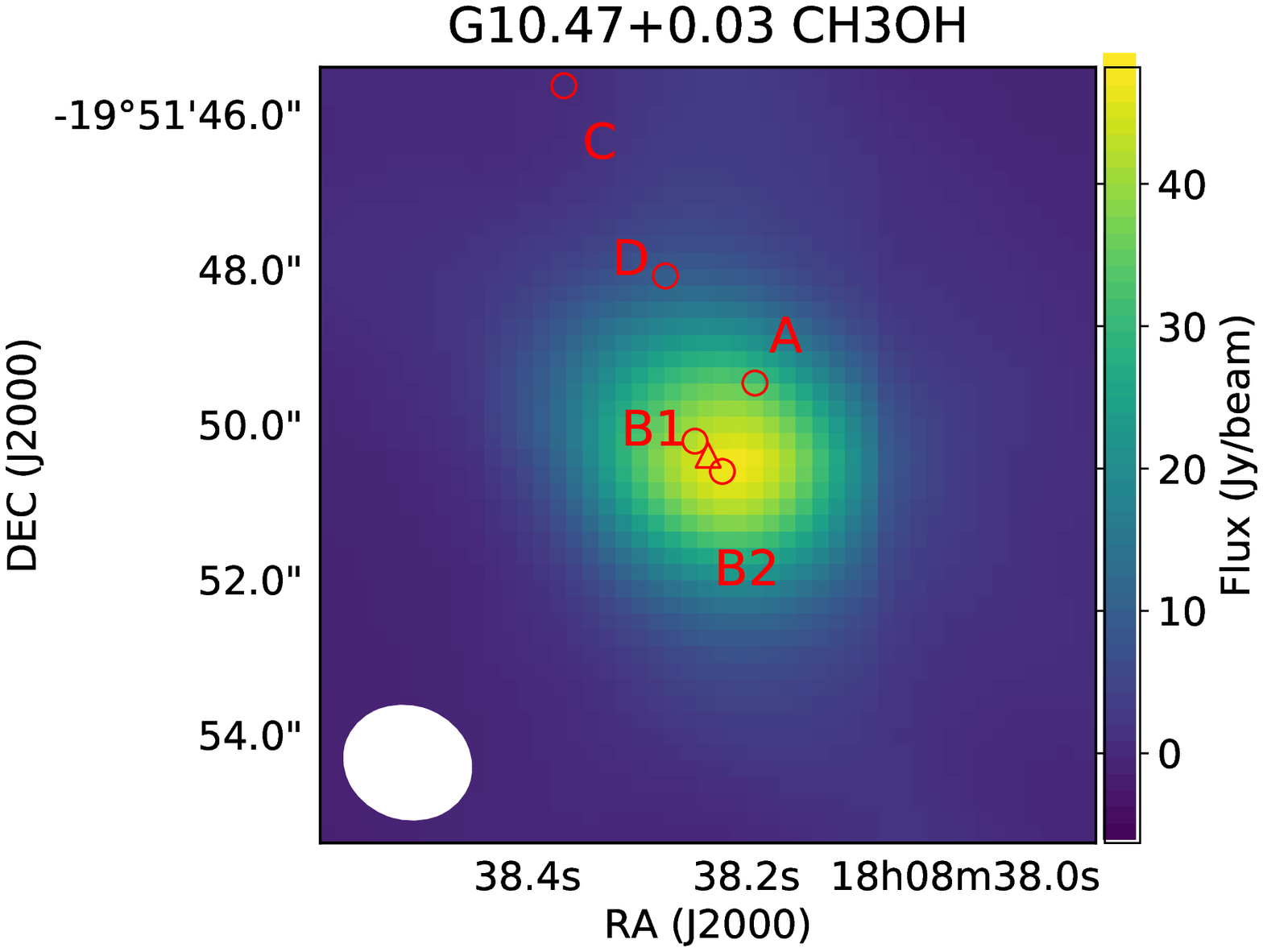}&
\includegraphics[scale=0.5]{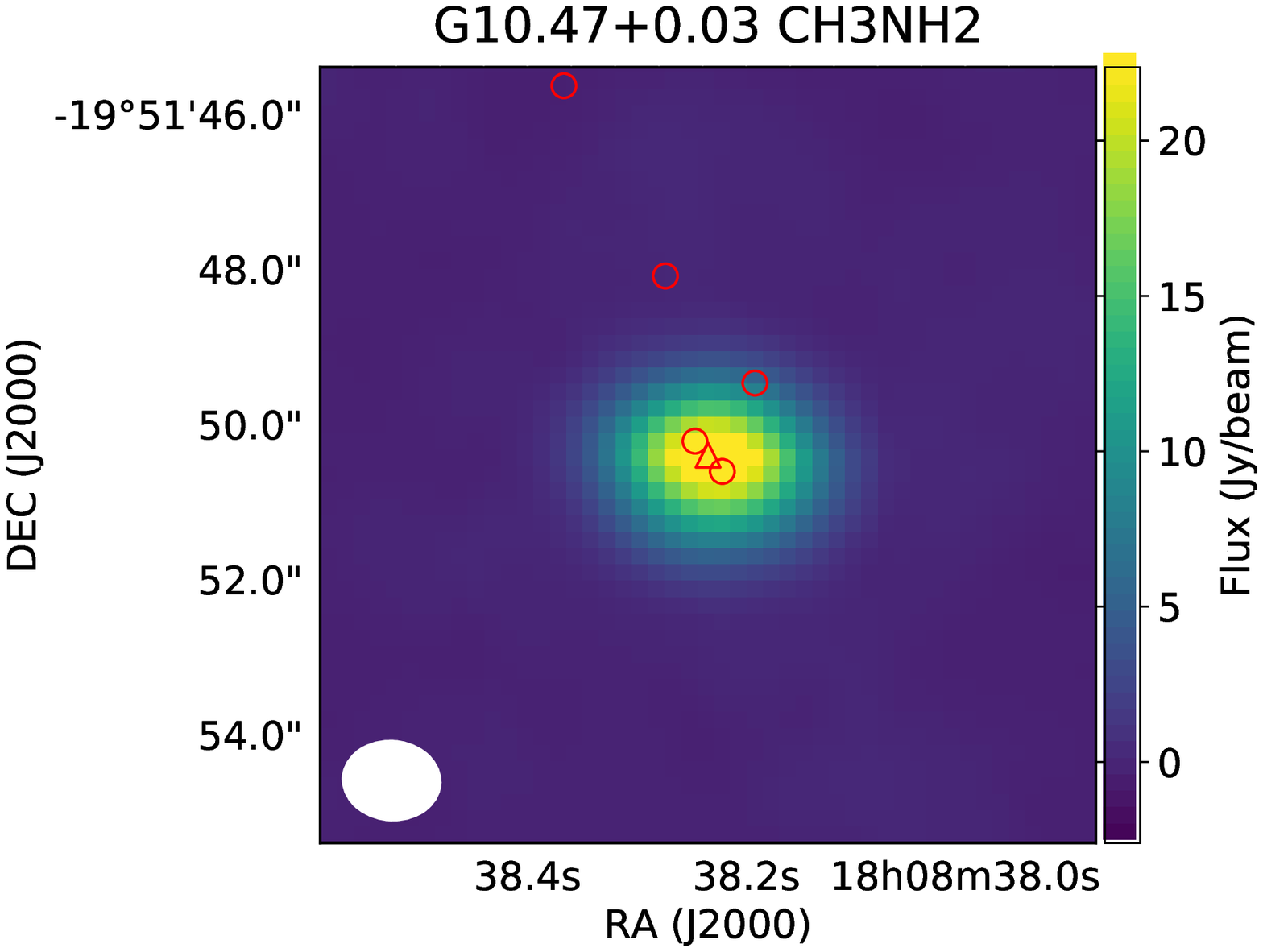}\\
\includegraphics[scale=0.5]{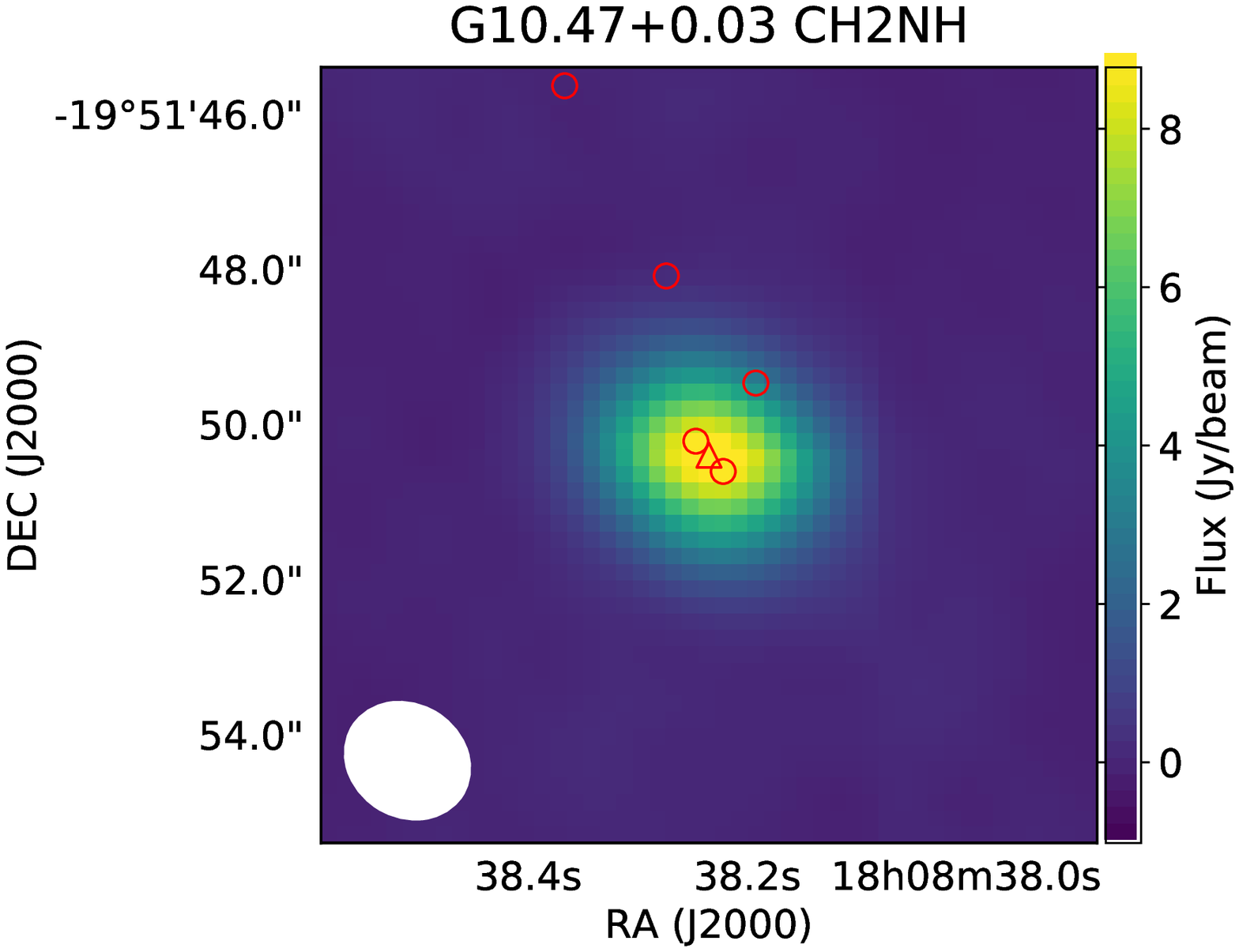}&\\
 \end{tabular}
\vspace{35mm}
\caption{
The integrated intensity maps of CH$_3$OH 4, 1, $E$ $\rightarrow$ 3 ,1, $E$, CH$_3$NH$_2$ 4, 1, $E_{1+1}$ $\rightarrow$ 3, 0, $E_{1+1}$, and CH$_2$NH 3, 2 ,2 $\rightarrow$ 2, 2, 1 transition toward G10.41+0.03. 
The positions of previously known continuum sources \citep{Wood89,Cesaroni98,Cesaroni10} are shown by circles on the map.
The triangle shows the position of hot core, where the spectra was extracted.
The velocity ranges are from 57.6 to 71.3, from 58.8 to 67.1, and from 62.9 to 76.6~km s$^{-1}$ for CH$_3$OH, CH$_3$NH$_2$, and CH$_2$NH, respectively.
\label{fig:map_G10}
}
\end{figure}

G10.47+0.03 is a well known massive star-forming region associated with UCHI\hspace*{-1pt}I and two HCHI\hspace*{-1pt}I regions.
In our mapping region, UCHI\hspace*{-1pt}I regions, A, B, and C, are knwon through the 2 and 6~cm mapping by VLA \citep{Wood89}.
The high resolution mapping observation at 1.3~cm with VLA further resolved the component B into HCHI\hspace*{-1pt}I regions B1 and B2, and interpreted this source as face-on rotating disks \citep{Cesaroni98}.
They showed the absorption of the NH$_3$ (4,4) transition toward these B1 and B2 cores, probably due to the molecular outflow originating from HCHI\hspace*{-1pt}I regions.
The component D was detected with 3.6~cm continuum emission \citep{Cesaroni10}, while other components A, B1, B2 were resolved with 6, 2, and 1.3 ~cm continuum as well.
\cite{Cesaroni10} also provided the evidence of the infalling gas toward the the embedded cores associated with the outflow.
While these components are not resolved in Figure~\ref{fig:map_G10} with our low spatial resolution, the extension of CH$_3$OH transition toward north indicate that component~D only show CH$_3$OH not being associated with CH$_3$NH$_2$ and CH$_2$NH.
Other spatial structures are not resolved even with the other transitions of CH$_3$OH, CH$_3$NH$_2$, and CH$_2$NH.
Since the southern component is not reported by \cite{Cesaroni10}, this feature may be due to very young source where continuum emission is so weak.
These components are depicted on our integrated intensity maps in Figure~\ref{fig:map_G10}.
The source size is $\sim$3$\arcsec$ for CH$_3$OH, while they are $\sim$2$\arcsec$ for CH$_3$NH$_2$ and CH$_2$NH.
The distribution of CH$_3$OH is slightly elongated toward north than CH$_3$NH$_2$ and CH$_2$NH, suggesting that CH$_3$OH is also abundant in cores A and D.
On the other hand, the distribution of CH$_3$NH$_2$ and CH$_2$NH would be limited to B1 and B2, though our data cannot distinguish these cores.
The spectra in G10.47+0.03 was extracted from the peak intensity position of CH$_3$NH$_2$, which is corresponding to the B1 and B2 positions.

\begin{figure}
 \begin{tabular}{ll}
\includegraphics[scale=0.5]{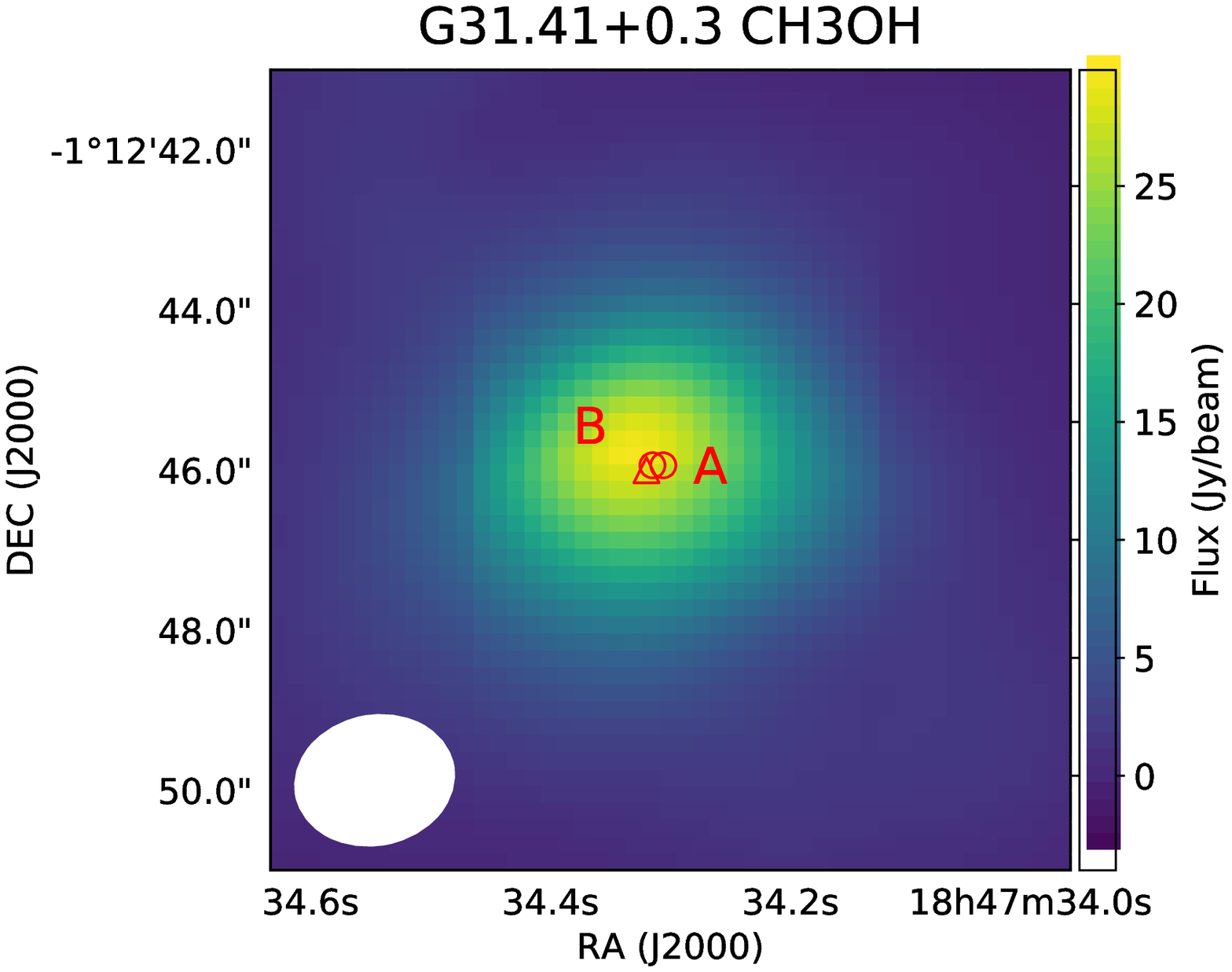}&
\includegraphics[scale=0.5]{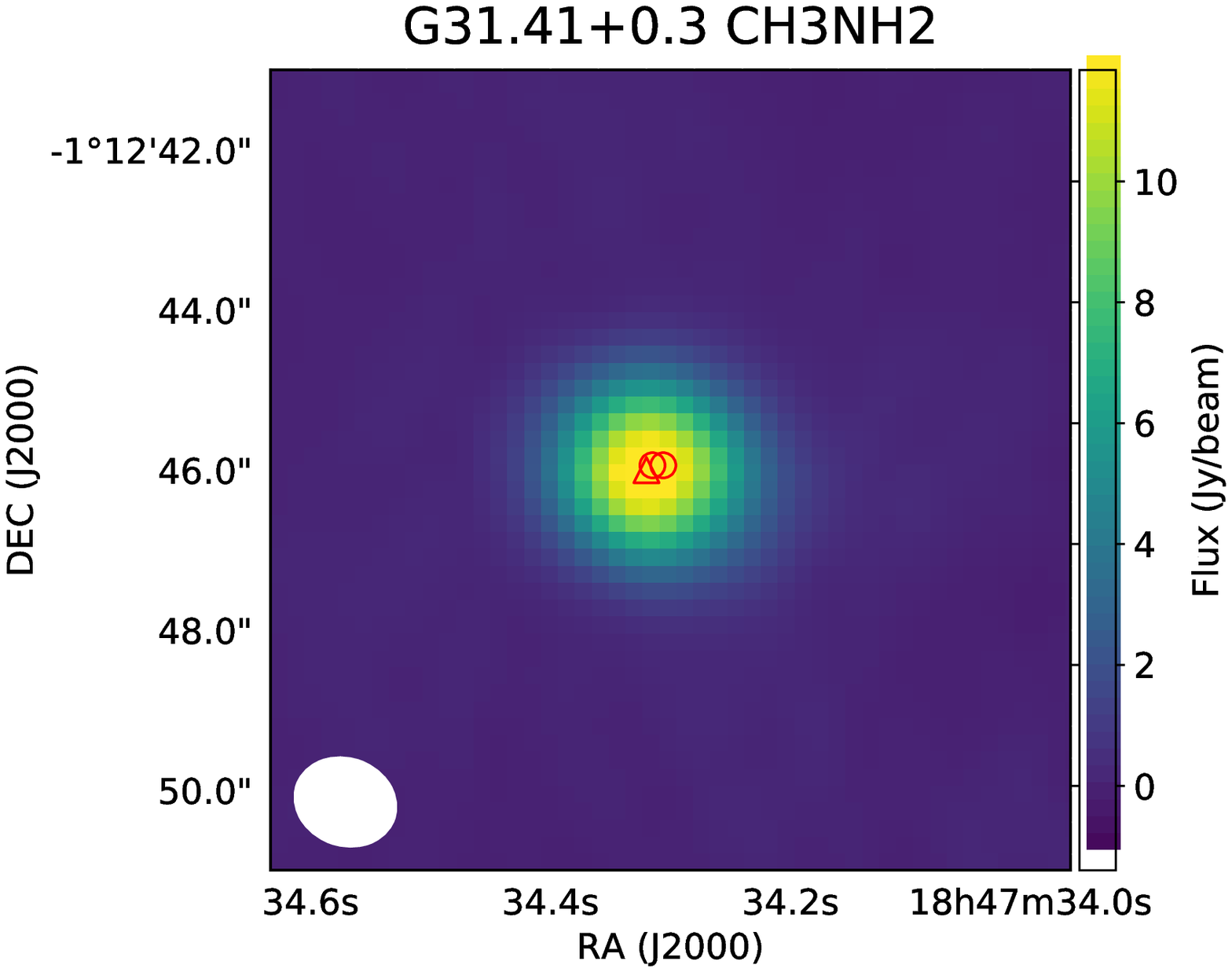}\\
\includegraphics[scale=0.5]{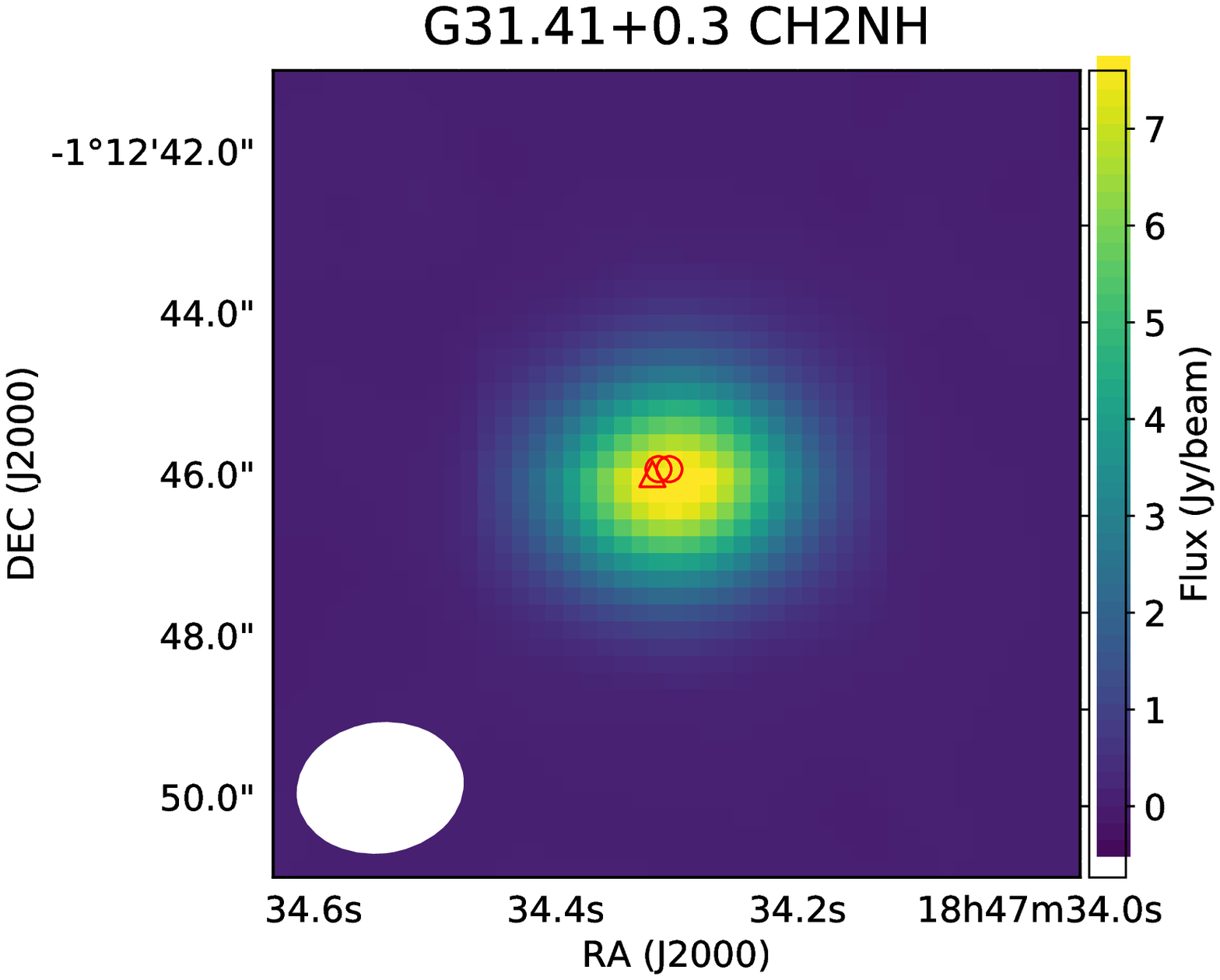}&\\
 \end{tabular}
\vspace{35mm}
\caption{
The integrated intensity maps of CH$_3$OH 4, 1, $E$ $\rightarrow$ 3 ,1, $E$, CH$_3$NH$_2$ 4, 1, $E_{1+1}$ $\rightarrow$ 3, 0, $E_{1+1}$, and CH$_2$NH 3, 2 ,2 $\rightarrow$ 2, 2, 1 transition toward G31.41+0.3. 
The positions of previously known continuum sources \citep{Cesaroni10} are shown by circles on the map.
The triangle shows the position of hot core, where the spectra was extracted.
The velocity ranges are from 85.6 to 99.2, from 89.8 to 98.1, and from 90.8 to 104.5~km s$^{-1}$ for CH$_3$OH, CH$_3$NH$_2$, and CH$_2$NH, respectively.
\label{fig:map_G31}
}
\end{figure}

G31.41+0.3 is a massive star-forming region similar to G10.47+0.03.
Although this source is known to have UCHI\hspace*{-1pt}I region, the position of hot molecular core is away from the UCHI\hspace*{-1pt}I region by about 5$\arcsec$ \citep{Cesaroni94}, which is the outside of Figure~\ref{fig:map_G31}.
\cite{Cesaroni10} detected weak continuum sources A and B inside the hot molecular core of G31.41+0.3.
These continuum emissions are thought to be originated from the thermal jet rather than UCHI\hspace*{-1pt}I regions.
The non-detection of UCHI\hspace*{-1pt}I region inside of G31.41+0.3 implies that this source is less evolved than G10.47+0.03 \citep{Cesaroni10}.
Our peak positions agree with the position of previously known hot molecular core.
The source size is $\sim$4$\arcsec$ for CH$_3$OH, while they are $\sim$2$\arcsec$ for CH$_3$NH$_2$ and CH$_2$NH.
The positions of sources A and B are shown on our integrated intensity maps in Figure~\ref{fig:map_G31}.
These sources are too close to be distinguished with each core with our spatial resolution.
The detailed spatial structure of this source is not resolved even with the other transitions of CH$_3$OH, CH$_3$NH$_2$, and CH$_2$NH.
We extracted the spectra from the peak intensity position of CH$_3$NH$_2$, which is almost the identical to sources A and B.

W51~e1/e2 region is the well known protoclustar region, which possess UCHI\hspace*{-1pt}I regions e1, e2, e3, e4, and e8 in our mapping regions \citep{Gaume93,Zhang97}.
These structures are shown on our integrated intensity maps (Figure~\ref{fig:map_W51}).
The emission peaks coincide with e2 and e8 components.
Through the mapping observation of 870~$\micron$ dust continuum emission with SMA, \cite{Tang09} detected the extension of dust ridge toward northwest of e2 with an overall length of $\sim$2$\arcsec$, and southwest of e8 with an overall length of $\sim$3$\arcsec$, probably being controlled by magnetic field.
The southwest distribution of CH$_3$OH from e8 are clearly seen in our map, overlapping with the above dust ridge.
The characteristic northwest distribution of CH$_2$NH from e2 traces another dust ridge.
W51~e2 is the strongest HI\hspace*{-1pt}I region which is thought to be powered by 08-type star \citep{Shi10}.
\cite{Shi10} performed continuum mapping observation toward W51~e2 at 0.85 and 1.3~mm with VLA, and 7 and 13~mm with ALMA.
They identified sub cores named e2-N, e2-W, e2-E, and e2-NW.
Only the emission from e2-N are thought to be free-free emission from the HI\hspace*{-1pt}I region, while the other continuum emission would be from dust.
Of four sources, e2-E is the only source being associated with the the hydrogen recombination lines.
\cite{Goddi16} found that e2-E and e2-NW are associated with NH$_3$ and CH$_3$OH emissions, while e2-W is traced by the absorption.
These detailed structure in e2 is not resolved in our maps..
The transitions of CH$_3$OH 4, 1, $E$ $\rightarrow$ 3 ,1, $E$, CH$_3$NH$_2$ 4, 1, $E_{1+1}$ $\rightarrow$ 3, 0, $E_{1+1}$, and CH$_2$NH 3, 2 ,2 $\rightarrow$ 2, 2, 1 distribute toward both e2 and e8.
The source size is $\sim$3$\arcsec$ for CH$_3$OH, while they are $\sim$2$\arcsec$ for CH$_3$NH$_2$ and CH$_2$NH for e2, while they are $\sim$3, $\sim$1, and $\sim$2$\arcsec$, respectively, for CH$_3$OH, CH$_3$NH$_2$, and CH$_2$NH, in e8.
These distributions are consistent with the other transitions of CH$_3$OH, CH$_3$NH$_2$, and CH$_2$NH.
We extracted the spectra from the peak positions of CH$_3$NH$_2$ intensity, which are corresponding to W51~e2-E and e8, and hereafter we call W51~ e2-E as simply W51 e2.

In total, molecular abundances of CH$_3$OH, CH$_3$NH$_2$, CH$_2$NH were examined toward seven cores as summarized in Table~\ref{tbl:detected_source}.

\subsection{Spectra and Analysis}
\begin{figure}
 \begin{tabular}{l}
\includegraphics[scale=0.7]{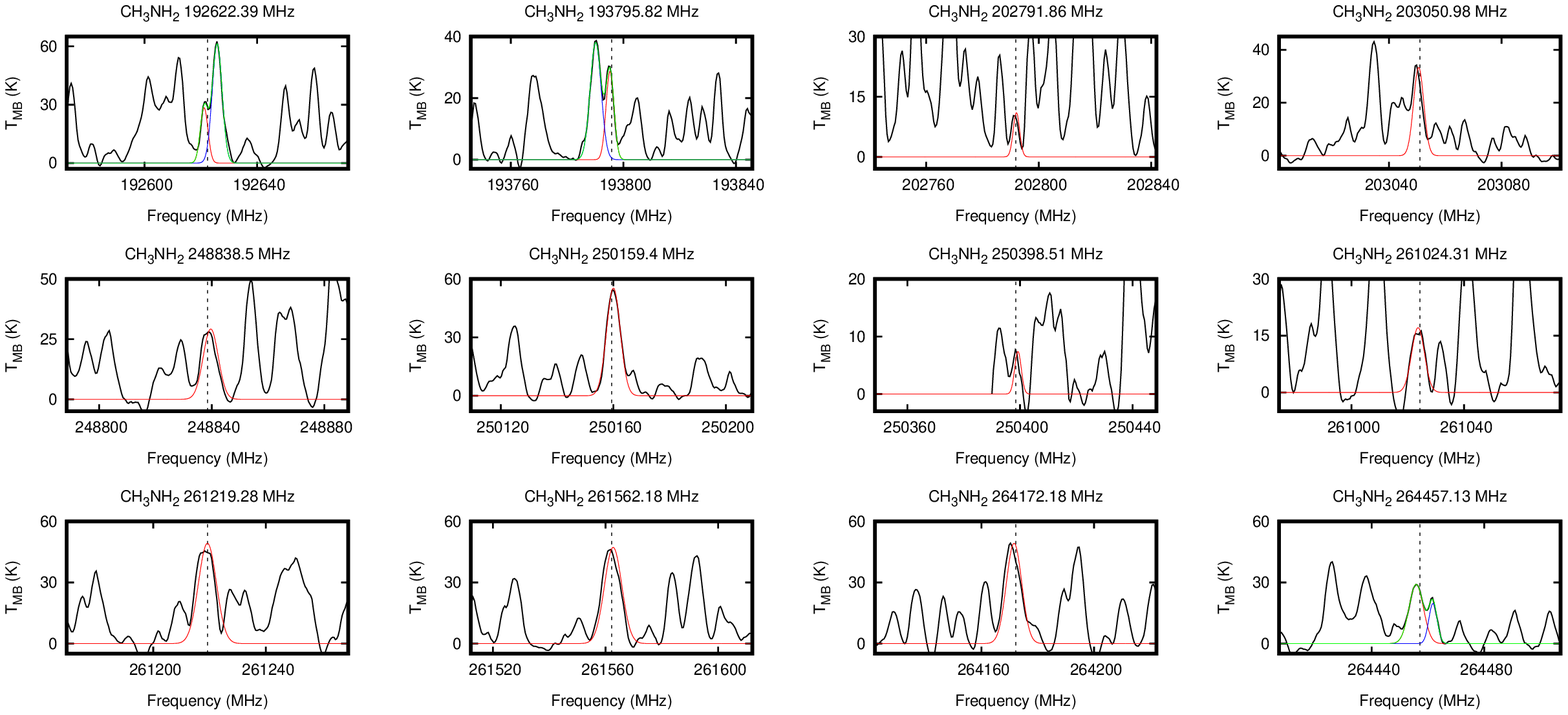}\\
\includegraphics[scale=0.7]{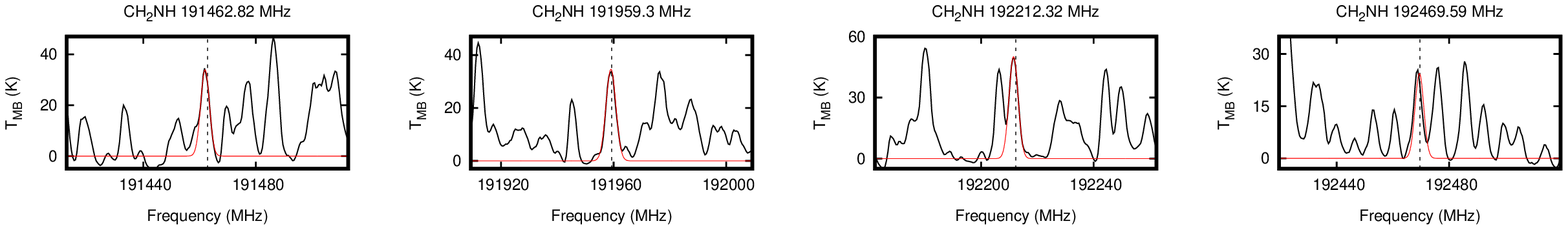}\\
 \end{tabular}
\vspace{15mm}
\caption{
The observed transitions of CH$_3$NH$_2$ and CH$_2$NH toward NGC6334I MM1 at the position in \label{tbl:detected_source}.
T$_{rm MB}$ is the brightness temperature.
The vertical dotted lines represent the rest frequency calculated from the velocity of source.
The results of Gaussian fitting is overlapped on the spectra with red line.
For some cases, we performed least-squares fitting assuming more than two components of Gaussian, and other Gaussian components and the shape of the sum of all Gaussians are shown by blue lines.
\label{fig:spectle_N63MM1}
}
\end{figure}

All the molecular lines used in our analysis are shown through Figures~\ref{fig:spectle_N63MM1} to \ref{fig:spectle_W51e2}.
We note that the rest frequencies of CH$_2$NH 7, 1, 6 $\rightarrow$ 7, 0, 7 transition at 250.16168~GHz and CH$_3$NH$_2$ 2, 2, B2 $\rightarrow$ 2, 1, B1 transition at 250.1594~GHz are blended with each other.
Since the rest frequencies of these transitions are separated by 2.2~MHz, we assigned the transitions using the peak frequency obtained from the Gaussian fitting.

Some line shapes observed toward NGC6334I MM1 are not well fitted by Gaussian.
This would be due to the subcomponents in this source which are not resolved with our spatial resolution.
For other sources, molecular lines are relatively well fitted by Gaussian, though some transitions are suffering from the contamination by other molecular lines.
In this case, least-squares fitting is performed assuming two components of Gaussian.
The rest frequencies of transitions calculated from the source velocities are presented by the dotted lines.
%
%The absorption feature of CH$_3$OH ``4, 1, $E$ $\rightarrow$ 3, 1, $E$'' toward W51~e2 would be due to the core W51~e2-W \citep{Goddi16}, which is not resolved in our map.

The FWHM line widths and the peak values are obtained through the Gaussian fitting.
The detected lines are shown through Tables~\ref{tbl:NGC6334FMM1_line} to \ref{tbl:W51e8} with their line parameters.
While CH$_3$OH and CH$_2$NH are detected toward all sources, CH$_3$NH$_2$ is not confirmed toward NGC6334I MM3.
We calculated column densities using the rotation-diagram method described in \cite{Turner91}, and the following equation is employed:
\begin{equation}
\log \frac{3kW}{8\pi^3 \nu S\mu^2 g_{\rm I}g_{\rm K}} = \log \frac{N}{Q_{\rm rot}} - \frac{E_{\rm u}}{k} \frac{\log e}{T_{\rm rot}}
\end{equation}
where $W$ is the integrated intensity, $S$ is the intrinsic line strength, $\mu$ is the permanent electric dipole moment, $g_{\rm I}$ and $g_{\rm K}$ are the nuclear spin degeneracy and the $K$-level degeneracy, respectively, $N$ is the column density, $Q_{\rm rot}$ is the rotational partition function, $E_{\rm u}$ is the upper level energy, and $T_{\rm rot}$ is the rotation temperature.
The rotation diagrams for all sources are shown in Figure~\ref{fig:rotation}.
The column density and the excitation temperature are derived by utilizing the least-squares fitting.
The column density is derived from the interception of a diagram, and its slope gives us the excitation temperature.

In addition, the column density of the molecular hydrogen is estimated from the continuum emission from the dust with the same way as \cite{Hernandez14}, where the gas mass is calculated as
\begin{equation}
M_{\rm gas} = \frac{S_\nu D^2 R_{\rm d}}{B_\nu(T_{\rm d})\kappa_\nu}.
\end{equation}
where $S_{\nu}$, $D$, $R_{\rm d}$, $\kappa_\nu$, and $B_\nu (T_{\rm d})$ are the flux density, distance
to the core, gas-to-dust ratio, the dust opacity per unit dust mass, and the Planck function at the dust temperature ($T_{\rm d}$), respectively.
As described in \cite{Hernandez14}, by assuming $R_{\rm d}$ of 100, and $\kappa_\nu$ of 0.74 cm$^2$ g$^{-1}$, and using the Rayleigh-Jeans approximation, the equation to obtain the hydrogen column density  at 1.3~mm is obtained as follow:
\begin{equation}
\left[\frac{N_{\rm H_{2}}}{{\rm cm}^2}\right] = \frac{2.35\times10^{16}}{\theta^2}\left[\frac{S_\nu}{\rm Jy}\right]\left[\frac{T_{\rm d}}{\rm K}\right]^{-1},
\end{equation}
where $\theta$ is the source size in radians.
In our work, we use the angular resolution of our observations as $\theta$.
$S_\nu$ is obtained by the procedure that we employ to subtract the continuum emission.
We use the spectral window whose center frequency is 205.849 in Set~1, since the largest number of channels in this window would be useful to obtain the dust continuum level.
We approximately use the excitation temperature of CH$_3$OH as $T_{\rm d}$.

\subsection{Abundances}
The obtained column densities and the excitation temperatures of H$_2$, CH$_3$OH, CH$_3$NH$_2$, and CH$_2$NH are summarized in Table~\ref{tbl:observation_result}.
CH$_3$NH$_2$ column densities are high in NGC6334I MM1 and G10.47+0.03, with their column densities of 1.0 $\times$ 10$^{18}$ and 9.4 $\times$ 10$^{17}$~cm$^{-2}$, respectively.
In addition, CH$_3$OH column density of 1.3$\times$ 10$^{18}$~cm$^{-2}$ in G10.47+0.03 is the highest in our sources.
The high abundance of N-bearing species in NGC6334I is contrary to our previous study in \cite{Suzuki18}, where we found that N-bearing species is rich in G10.47+0.03 while they are poor in NGC6334I region.
Since the spatial distribution of CH$_3$OH is much extended than those of CH$_2$NH and CH$_3$NH$_2$ in NGC 6334I region, it is possible that the abundances of N-bearing species are underestimated than O-bearing species in \cite{Suzuki18} due to the difference of source size.
The obtained excitation temperature of G10.47+0.03 is 1466~K, which is much higher than the report of \cite{Ohishi19}, where the excitation temperature of G10.47+0.03 was less than 45~K.
Probably the single dish observation by \cite{Ohishi19} detected extended structure surrounding this core, which is resolved out with our interferometric observation.

Since we could not detect CH$_3$NH$_2$ transition toward NGC6334I MM3 region, the upper limit of CH$_3$NH$_2$ column density is obtained by assuming the excitation temperature of 170~K, corresponding to that of CH$_3$OH.
With this excitation temperature, 16, 2, B2 $\rightarrow$ 16, 1, B1 transition gives us the upper limit of the column density.
With the r.m.s level of 0.5~K for this transition and line width of 6.0~km~s$^{-1}$, we get the upper limit of column density of 1.0$\times$10$^{16}$~cm$^{-2}$ with 3 sigma noise level.
In addition, CH$_2$NH abundances are not determined from the rotation diagram method toward NGC6334I MM1 and MM3 due to the limited range of detected transitions.
Therefore the excitation temperatures are fixed to be 120 and 170~K, respectively, for MM1 and MM3 to derive the column densities.
Then the range of CH$_2$NH column densities for MM1 and MM3 are, respectively, from 0.3 to 1.3~$\times$10$^{17}$ and from 0.4 to 1.5~$\times$10$^{15}$~cm$^{-2}$.
Three cores in NGC6334I regions, MM1, MM2, and MM3, are the chemically interesting regions.
As it has been reported by \cite{Bogelund19}, the abundances of CH$_3$NH$_2$ and CH$_2$NH differ by orders of magnitude in NGC6334I MM1, MM2, and MM3.
Our estimated column densities of CH$_2$NH and CH$_3$NH$_2$ agree well with \cite{Bogelund19} except for CH$_3$NH$_2$ in NGC6334I MM1, where our obtained abundance is higher than their work by about a factor of three.
Since our observed position in NGC6334I is offsetted by about 2$\arcsec$ compared to \cite{Bogelund19}, this difference would not be surprising.
In addition, our receiver set enables us to obtain the column density of CH$_3$OH toward MM1, MM2, and MM3 simultaneously.
The similar column densities of CH$_3$OH in NGC6334I MM1, MM2, and MM3 clearly show the depletion of only N bearing species toward MM2 and MM3.

For G31.41+0.03, W51 e2, and W51 e8, we report the abundance of CH$_3$NH$_2$ for the first time.
Their excitation temperatures and column densities are 79~K and 3.2 $\times$ 10$^{17}$~cm$^{-2}$ for G31.41+0.03, 95~K and 1.9 $\times$ 10$^{17}$~cm$^{-2}$ for W51 e2, and 131~K and 2.5 $\times$ 10$^{17}$~cm$^{-2}$ for W51 e8. 
The observed abundance ratios of ``CH$_3$NH$_2$/CH$_2$NH'' and ``CH$_3$NH$_2$/CH$_3$OH'' are shown in Table~\ref{tbl:ratios}.

\section{Comparison with Chemical Modeling}
In this section, we evaluate our observational results using our chemical model \citep{Suzuki18}.
Since the formation process of CH$_3$OH is well known to be via the grain surface hydrogenation processes to CO, the molecular ratio of ``CH$_3$NH$_2$/CH$_3$OH'' is the essential to test if CH$_3$NH$_2$ is formed via the successive hydrogenation processes to HCN.
%
%On the other hand, CH$_2$NH are thought to be a product of the gas phase chemistry, since CH$_2$NH is quickly converted CH$_3$NH$_2$ on grains before the sublimation \citep{Suzuki16}.

%
%The gas phase abundance of CH$_2$NH increases gradually after the warm-up phase, while the abundance of CH$_3$NH$_2$ decreases by the destruction by reactive ions, radicals, and potochemical processes.
%
%Since CH$_3$NH$_2$ does not increase after the evaporation, the ratio of ``CH$_3$NH$_2$/CH$_2$NH'' will decrease after the warm-up phase by the production of CH$_2$NH.
%
%To compare the observed molecular ratios of ``CH$_3$NH$_2$/CH$_3$OH'' and ``CH$_3$NH$_2$/CH$_2$NH'' with chemical model,  we summarize the observed molecular ratios in Table~\ref{tbl:ratios}.
%
%Only the upper limit of the column density is determined toward NGC6334I MM3.
%
%The ratio of ``CH$_3$NH$_2$/CH$_2$NH'' is more than unity for all sources, while the ratios of ``CH$_3$NH$_2$/CH$_2$NH'' is between 0.11 and 2.2 except for NGC633I MM3, where only the upper limit of CH$_3$NH$_2$ column density is determined.

We use the chemical model described in \cite{Suzuki18} for this comparison.
In this work, we used the gas-phase chemical network kida.uva.2014\footnote{\url{http://kida.obs.u-bordeaux1.fr/}} \citep{Wakelam15}, and the grain surface reactions in \cite{Garrod13}.
With these reaction sets, we updated the formation process of CH$_3$NH$_2$.
We added the hydrogenation processes of HCN on the grain surface (HCN + 2H $\rightarrow$ H$_2$CN, and H$_2$CN + 2H $\rightarrow$ CH$_3$NH$_2$) as described in \cite{Suzuki16}, resulting in the increase of CH$_3$NH$_2$ abundance.
In total, 489 species composed of 13 elements (H, He, C, N, O, Si, S, Fe, Na, Mg, Cl, P, F) were included.
To simulate the physical conditions in hot cores, we used the two stage physical model, where the free-fall collapse is followed by a dynamically-static warm up by following \cite{Garrod13}.
In our standard model, the cold collapse phase started from n$_{H_{2}}$ =$ 3\times10^3$ cm$^{-3}$ to final post collapse density of n$_{H_{2}}$ =$ 2\times10^7$ cm$^{-3}$.
The increase in visual extinction during collapse leads to the minimum dust-grain temperature of 8 K followed by a warm-up from 8 to 400 K; during this phase, the gas and dust temperatures are assumed to be well coupled and the gas density is fixed. 
We employed the fast warm-up models described in \cite{Garrod13}, whose timescale for the warm-up phases is 7.12$\times$10$^{4}$~years, but continue calculation fixing the temperature.
Since the weak hydrogen recombination lines of our sources suggest the young ages of the cores, we stopped the simulations at 1.0$\times$10$^{6}$~years.
The gravitational constant B is the parameter to slow down the collapse assuming the turbulence or the magnetic field \citep{Nejad90}.
The standard model used B to be unity, which is the case of free fall.
The values of the gravitational constant B, the peak temperature, the peak density, and the timescale of warm-up phase are free parameters.

We show our simulation results in Figure~\ref{fig:simulation} with our standard model.
The simulated fractional abundances of CH$_3$OH, CH$_3$NH$_2$, and CH$_2$NH, compared to the total proton density in the gas phase and on grains (sum of molecular abundances in the grain mantle and on the grain surface) are shown by dotted and solid lines, respectively.
The time of zero year is corresponding to the beginning of the warm-up phase.
As we discussed in the previous studies, CH$_2$NH is expected to be efficiently formed in the gas phase reaction of ``CH$_3$ + NH $\rightarrow$ CH$_2$NH + H'', while CH$_3$NH$_2$ is built on grain surface via successive hydrogenation processes to HCN \citep{Suzuki16,Suzuki18}, similar to CH$_3$OH.
When the grain surface temperature gets high enough, these species sublimate from grains, leading to the sudden increase of the gas phase molecular abundance.
The sublimation of CH$_3$NH$_2$ happens at $\sim$6.2$\times$10$^{4}$~years, when the dust temperature is $\sim$130~K, while the sublimation of CH$_3$OH happens when the dust temperature is $\sim$110~K due to the smaller binding energy.
Since CH$_3$NH$_2$ is frozen before 6.2$\times$10$^{4}$~years, the gas phase abundance ratio before this age is not meaningful.
We compare the fractional abundances of CH$_3$OH, CH$_3$NH$_2$, and CH$_2$NH with our observation.
Considering the errors and assumptions both in the simulations and the observations, we set a criteria to be factor of 10 to see if our model roughly reproduce the observed fractional abundances.
The observed fractional abundances of CH$_3$OH, CH$_3$NH$_2$, and CH$_2$NH compared to the number of atomic hydrogen are, respectively, ranging from 3.6$\times$10$^{-9}$ - 8.5$\times$10$^{-8}$, 5.0$\times$10$^{-9}$ - 3.8$\times$10$^{-8}$, and 1.3$\times$10$^{-10}$ - 2.3$\times$10$^{-9}$.
Hence our criteia to constrain the simulated fractional abundances of CH$_3$OH, CH$_3$NH$_2$, and CH$_2$NH are, respectively, 3.6$\times$10$^{-10}$ - 8.5$\times$10$^{-7}$, 5.0$\times$10$^{-10}$ - 3.8$\times$10$^{-7}$, and 1.3$\times$10$^{-11}$ - 2.3$\times$10$^{-8}$.
In Figure~\ref{fig:simulation}, the simulated fractional abundances of CH$_3$NH$_2$ and CH$_2$OH agree with the observation at the later phase of the simulation.
However, the fractional abundance of CH$_2$NH is always higher than the observed value.
As discussed in \cite{Suzuki16}, the formation rate of CH$_2$NH from radicals are well investigated (see also KIDA database) and therefore this overproduction would not be due to the wrong formation rate.
We confirmed this overproduction of CH$_2$NH under the different gas temperatures of 200 and 300~K, and hence the temperature would not be a key to solve this discrepancy.
Other possibility would be the lack of the destruction processes of CH$_2$NH in the model, or the overproduction of NH and CH$_3$ radicals, which are the precursors of CH$_2$NH, by the cosmic rays or the secondary UV photons in the gas phase.

The molecular ratios of ``CH$_3$NH$_2$/CH$_3$OH'' after the sublimation of CH$_3$NH$_2$ are presented in Figure~\ref{fig:simulation}.
Both CH$_3$OH and CH$_3$NH$_2$ are formed on grains during the warm-up phase and destroyed simultaneously in the gas phase after the sublimation by reactive radicals, ions, and cosmic rays, and hence the range of ``CH$_3$NH$_2$/CH$_3$OH'' is very limited.
The ratio of ``CH$_3$NH$_2$/CH$_3$OH'' is 0.1 at first, and it increases slightly to 0.4 at the end of the simulation, since CH$_3$OH is efficiently destroyed by C and H$_3$O$^+$, while the destruction of CH$_3$NH$_2$ by these species are not included in kida.uva.2014.
Therefore the slight increase of ``CH$_3$NH$_2$/CH$_3$OH'' may be due to the lack of destruction processes of CH$_3$NH$_2$.
The ratio of ``CH$_3$NH$_2$/CH$_3$OH'' is the ideal tool to discuss the CH$_3$NH$_2$ chemistry.
First, the ``CH$_3$NH$_2$/CH$_3$OH'' ratio is free from the uncertainty of the hydrogen column density.
Second, the similarity in CH$_3$NH$_2$ and CH$_3$OH chemistry can reduce the uncertainty of the physical parameters, enabling us to focus on the formation process of CH$_3$NH$_2$.
%

%
%On the other hand, the ratio of ``CH$_3$NH$_2$/CH$_2$NH'' has obvious time dependency.
%
%Since CH$_2$NH is produced rather than desteroyed in the gas phase, the ``CH$_3$NH$_2$/CH$_2$NH'' ratio decreases from 11 to 0.02 during the simulation.
%
%This time dependence would be useful to constran the age of the actual star-forming regions.
%
%As shown in Table~\ref{tbl:ratios}, the observed ``CH$_3$NH$_2$/CH$_2$NH'' ratios are between 3.9 and 43 for almost all hot cores, but it is less than 11 in NGC6334I MM3.
%

%Under the nine different parameter sets in the middle of Table~\ref{tbl:ratios}, we simulate the chemical evolutions.
%
To assume the different physical evolution of hot cores, we perform nine simulations under the different gravitational constant B, the temperature, the density, and the warm-up timescale.
The physical parameters are basically same as the standard model but only one parameter is changed to develop other models.
Here, the gravitational constant B is set to be 0.7, 0.2, and 0.1.
The temperature is changed from 400 to 200~K.
The gas density is put to be 1$\times$10$^6$ and 1$\times$10$^8$~cm$^{-3}$.
The two different warm-up timescales, 7.12 $\times$ 10$^3$ and 7.12 $\times$ 10$^5$~years, are prepared and named as long and short warm-up time scale model.
For the comparison with the observational result, we set the criteria that the fractional abundances of CH$_3$OH and CH$_3$NH$_2$ are, respectively, in the range of 3.6$\times$10$^{-10}$ - 8.5$\times$10$^{-7}$ and 5.0$\times$10$^{-10}$ - 3.8$\times$10$^{-7}$.
%we set the criteria that the rtio of ``CH$_3$NH$_2$/CH$_2$NH'' is less than 0.1 to reproduce the observed high ``CH$_3$NH$_2$/CH$_2$NH'' ratio with roughly a factor of 10.
%
%For all models, the ``CH$_3$NH$_2$/CH$_2$NH'' ratio gets lower than 0.1 at $\sim$ 3 $\times$10$^5$~years after the beginning of the warm-up phase.
%
%Then, the typical timescale to satisfy``CH$_3$NH$_2$/CH$_2$NH''$>$0.1 after the sublimation process of molecules is 2 $\times$ 10$^{5}$~years independent of medels.
%
%In other words, the chemical model results suggest that the timescale of hot chemistry to form CH$_2$NH in our hot cores would be less than 2 $\times$ 10$^{5}$~years.
%
%The range of ``CH$_3$NH$_2$/CH$_2$NH'' ratio is also shown in Table~\ref{tbl:ratios}.
%
Then, we get the ``CH$_3$NH$_2$/CH$_3$OH'' ratio of between 0.09 and 0.39, except for the cases of different value of B, as is summarized in Table~\ref{tbl:ratios}.
%a
It is interesting that only the constant B strongly changes the ``CH$_3$NH$_2$/CH$_3$OH'' ratio among our free parameters.
The parameter B controls the timescale of collapsing phase, it leads to the different HCN and CO abundances on grains at the collapsing phase prior to the warm-up phase.
With its low binding energy, the abundance of CO decreases by the gas phase chemistry if the timescale of the collapsing phase is long due to high value of B.
Since CH$_3$OH is the products of hydrogenation processes to CO, it changes the abundance of CH$_3$OH and ``CH$_3$NH$_2$/CH$_3$OH'' ratio as well.

The abundance ratios reported by the model of \cite{Garrod13} are also summarized in this table to discuss the CH$_3$NH$_2$ chemistry.
Though both of our model and the model of \cite{Garrod13} include the radical-radical reactions to form CH$_3$NH$_2$ (e.g., CH$_3$ + NH$_2$) on grain surface, only our model includes the hydrogenation process to HCN and therefore our model shows much higher abundance of CH$_3$NH$_2$.
If we exclude the hydrogenation process to HCN on grains from our model, we get the close CH$_3$NH$_2$ abundance to \cite{Garrod13}.
Thus, the result of \cite{Garrod13} showed the ``CH$_3$NH$_2$/CH$_3$OH'' ratio of less than 0.007, which is almost 50 times smaller than our model with the same physical evolution.
According to Table~\ref{tbl:ratios}, the range of observed ``CH$_3$NH$_2$/CH$_3$OH'' ratio is between 0.11 and 2.2.
These high values are not explained by the model of \cite{Garrod13}.
Since our simulations show the ``CH$_3$NH$_2$/CH$_3$OH'' ratio of more than 0.2 for all cases, they show good agreement with the observed ratios.
These results clearly indicate the importance of hydrogenation process to form CH$_3$NH$_2$.

Our conclusion that CH$_3$NH$_2$ is built through the successive hydrogenation to HCN is different from previous study of \cite{Bogelund19}.
In \cite{Bogelund19}, the authors observed NGC6334I and derived the CH$_3$NH$_2$ abundances to be 2.7$\times$10$^{17}$, 6.2 $\times$10$^{16}$, and 3.0$\times$10$^{15}$~cm$^{-2}$ for MM1, MM2, and MM3, respectively.
With these ratios, they obtained the ``CH$_3$NH$_2$/CH$_3$OH'' ratio ranging from $\sim5\times10^{-3}$ to $\sim5\times10^{-4}$, which agrees with the peak abundance ratio by the fast warm-up model in \cite{Garrod13} within a factor of five.
Therefore they suggested that CH$_3$NH$_2$ is built via recombination process of radical species (CH$_3$ + NH$_2$).
Their low value of ``CH$_3$NH$_2$/CH$_3$OH'' ratio is contrary to our result.
Especially there is a critical disagreement in ``CH$_3$NH$_2$/CH$_3$OH'' ratio toward NGC6334 region, although our CH$_3$NH$_2$ column density is not so different with their work.
This disagreement comes from the CH$_3$OH column density to obtain ``CH$_3$NH$_2$/CH$_3$OH'' ratio.

We believe our ``CH$_3$NH$_2$/CH$_3$OH'' ratios are more reliable for few reasons.
First, \cite{Bogelund19} observed only one transition of the isotope of CH$_3$OH, which would lead the uncertainty of the CH$_3$OH abundance.
Second, if we apply their observation results and assume the CH$_3$NH$_2$ column density of 2.7$\times$10$^{17}$~cm$^{-2}$ and ``CH$_3$NH$_2$/CH$_3$OH'' ratio of 1$\times$10$^{-3}$, then the CH$_3$OH column density should be 2.7$\times$10$^{20}$~cm$^{-2}$.
This surprisingly high CH$_3$OH column density is not consistent with the previous studies.
The column density of CH$_3$OH can be roughly estimated from the previous observation of single dish telescope by assuming the source size.
\cite{Ikeda01} found that the CH$_3$OH column density toward NGC6334I was 3.4$\times$10$^{16}$~cm$^{-2}$ under the assumption of the source size of 20$\arcsec$.
Our mapping result suggests that the source size of CH$_3$OH is $\sim$5$\arcsec$ in NGC6334I region.
Then, from \cite{Ikeda01}, the CH$_3$OH column density can be estimated to be $3.4\times10^{16}\times(20\arcsec/5\arcsec)^2 = 5.4\times10^{17}$~cm$^{-2}$.
Therefore the column density of CH$_3$OH used in \cite{Bogelund19} would be strongly overestimated.
The column density of $5.4\times10^{17}$~cm$^{-2}$ rather agrees with our CH$_3$OH column densities toward NGC6334I region.

\subsection{Comparison with Other Sources}
The CH$_3$NH$_2$ and CH$_3$OH abundances were investigated toward Sgr B2 (M) and (N) by \cite{Belloche13} and \cite{Neill14}.
According to \cite{Belloche13} and \cite{Neill14}, the ratio of ``CH$_3$NH$_2$/CH$_3$OH'' in Sgr B2 (N) are 0.03 and 0.1, respectively.
The different ratios claimed by different authors would come from the smaller source size of CH$_3$NH$_2$, since the spatial resolution by \cite{Neill14} is higher than \cite{Belloche13}.
The ``CH$_3$NH$_2$/CH$_3$OH'' ratio of 0.1 is close to our values in NGC6334I MM2, G10.47+0.03, G31.41+0.3, and W51~e8.
The ``CH$_3$NH$_2$/CH$_3$OH'' ratio in NGC6334I MM1 and W51 e2 are higher than Sgr B2.
On the other hand, the small value of ``CH$_3$NH$_2$/CH$_3$OH'' in Sgr B2(M) is close to NGC6334I MM3.

We also show the analysis of CH$_3$NH$_2$ abundance toward Orion~KL Hot core using archival data of ALMA in the Appendix of this paper.
The derived column density of CH$_3$NH$_2$ toward Orion KL is 1.6$\times$10$^{16}$~cm$^{-2}$.
The CH$_3$OH column densities toward Orion KL-W region is found in \cite{Peng12}.
Their CH$_3$OH contour map suggests that the CH$_3$OH abundance at the position of the Hot core would be close to the position of KL-W region.
In this case, CH$_3$OH column density should be 9.2$\times$10$^{17}$~cm$^{-2}$.
Therefore we get the ``CH$_3$NH$_2$/CH$_3$OH'' ratio of 0.017 for Orion~KL Hot Core.

The small ``CH$_3$NH$_2$/CH$_3$OH'' ratio in Sgr B2 (M) and Orion~KL Hot core is similar to NGC6334I.
This small ratio is not explained by our chemical model under the possible range of the physical parameters.
The disagreement may be due to the different physical evolution from the other sources.

Finally, we note that this survey was performed toward CH$_2$NH-rich sources reported in \cite{Suzuki16}.
Therefore our source samples may be biased toward N-bearing species rich sources.
Further expansion of CH$_3$NH$_2$ survey towards other hot cores would be useful to reveal more general picture of hot core chemistry.
The survey observation of low-mass protostar is also an interesting topic.
IRAS16293-2422 is the only hot corino where ``CH$_3$NH$_2$/CH$_3$OH'' ratio was investigated.
The reported ``CH$_3$NH$_2$/CH$_3$OH'' ratio of less than 5.3$\times$10$^{-5}$ is extraordinary lower than our result.
It would be worth to explore the chemical difference between the high-mass and the low-mass protostars by future surveys.

\section{Conclusion}
The main results of this paper are summarized as follows:

\begin{enumerate}
\item We performed a survey observation of CH$_3$NH$_2$, CH$_2$NH, and CH$_3$OH toward NGC6334I, G10.47+0.03, G31.41+0.3, and W51~e1/e2 region with ALMA.
NGC6334I were resolved into MM1, 2, and 3, and W51~e1/e2 are resolved into e2 and e8.
In total, we analyzed the molecular abundances at seven hot cores.
CH$_3$NH$_2$, CH$_2$NH, and CH$_3$OH are detected for all sources except for NGC6334I MM3, where CH$_3$NH$_2$ was not detected.

\item The excitation temperature and column density of CH$_3$NH$_2$, CH$_2$NH, and CH$_3$OH were obtained by the rotation diagram method.
The CH$_3$NH$_2$ abundances are obtained for G31.41+0.03, W51 e2, and W51 e8 for the first time.
NGC6334I and G10.47+0.03 are especially CH$_3$NH$_2$-rich sources with their column densities of $\sim$10$^{18}$~cm$^{-2}$.
For other sources, the column densities of CH$_3$NH$_2$ are typically $\sim$10$^{17}$~cm$^{-2}$, but less than 3$\times$10$^{15}$~cm$^{-2}$ for NGC6334I MM3.

\item The observed fractional abundances for CH$_3$OH, CH$_3$NH$_2$, and CH$_2$NH are, respectively, $\sim$10$^{-8}$, $\sim$10$^{-8}$, and  $\sim$10$^{-9}$.
The observed fractional abundance of CH$_2$NH is not explained by our modeling within a factor of 10, while those of CH$_3$OH and CH$_3$NH$_2$ show the agreement.
This discrepancy would be due to the lack of the destruction process of CH$_2$NH, or the overproduction of radicals to form CH$_2$NH.

\item We obtained the observed ``CH$_3$NH$_2$/CH$_3$OH'' ratio of between 0.11 and 2.2. 
This high ``CH$_3$NH$_2$/CH$_3$OH'' ratio is not explained by the recombination of radicals as suggested by \cite{Garrod13} and the grain surface hydrogenation process discussed in \cite{Suzuki16} is required.
Our conclusion is different from the previous observation toward NGC6334I region by \cite{Bogelund19}, though the obtained CH$_3$NH$_2$ column density agrees well.
The difference is due to the method to measure CH$_3$OH column density, and our work would be more accurate than \cite{Bogelund19}. 

\item NGC6334I MM3 shows extremely low ``CH$_3$NH$_2$/CH$_2$NH'' ratio of $<$0.02, which is close to Orion KL Hot core and Sgr B2 (M).
Such small molecular ratio is not explained by our current chemical model by changing the physical parameters.
The detailed discussion of these sources are remained to the future works.
In addition, it is possible that our source selection is biased to N-bearing molecule rich sources by selecting the target from our previous CH$_2$NH survey.
The extension of this survey toward various sources would be important to further expand our knowledge.
\end{enumerate}

\acknowledgments
This paper makes use of the following ALMA data: ADS/JAO ALMA$\#$2017.1.01248.S. ALMA is a partnership of ESO (representing its member states), NSF (USA) and NINS (Japan), together with NRC (Canada), MOST and ASIAA (Taiwan), and KASI (Republic of Korea), in cooperation with the Republic of Chile. The Joint ALMA Observatory is operated by ESO, AUI/NRAO and NAOJ. The National Radio Astronomy Observatory is a facility of the National Science Foundation operated under cooperative agreement by Associated Universities, Inc.
This study was supported by the Astrobiology Program of National Institutes of Natural Sciences (NINS) and by the JSPS Kakenhi Grant  Numbers 19K03936, 19H05069, and 19K14753.

%% To help institutions obtain information on the effectiveness of their
%% telescopes, the AAS Journals has created a group of keywords for telescope
%% facilities. A common set of keywords will make these types of searches
%% significantly easier and more accurate. In addition, they will also be
%% useful in linking papers together which utilize the same telescopes
%% within the framework of the National Virtual Observatory.
%% See the AASTeX Web site at http://www.journals.uchicago.edu/AAS/AASTeX
%% for information on obtaining the facility keywords.

%% After the acknowledgments section, use the following syntax and the
%% \facility{} macro to list the keywords of facilities used in the research
%% for the paper.  Each keyword will be checked against the master list during
%% copy editing.  Individual instruments or configurations can be provided 
%% in parentheses, after the keyword, but they will not be verified.

%{\it Facilities:} \facility{Nickel}, \facility{HST (STIS)}, \facility{CXO (ASIS)}.

%% Appendix material should be preceded with a single \appendix ,nd.
%% There should be a \section ,nd for each appendix. Mark appendix
%% subsections with the same markup you use in the main body of the paper.

%% Each Appendix (indicated with \section) will be lettered A, B, C, etc.
%% The equation counter will reset when it encounters the \appendix
%% ,nd and will number appendix equations (A1), (A2), etc.

\appendix
The Orion-KL is the nearest massive star-forming region, which is located approximately 418 $\pm$ 6 pc away from the sun \citep{Kim08}. 
Its proximity and rich molecular composition make this region well suited for astrochemical study.
Since this region was discovered in 1967 \citep*{Kleinmann67}, numerous studies including line survey have been conducted so far \citep[e.g.,][]{Pagani17, Feng15, Gong15, Turner91}.

In Orion-KL, several remarkable sources called Hot core and Compact ridge are known as regions where many organic molecules exist \citep{Blake87}.
Hot core is known as a high-density (10$^6$ cm$^{-3}$) region with a warm ($\sim$ 150 K), compact ($<$ 0.05 pc) clump \citep{Zapata11}, and Compact ridge is also known to be a warm and dense region. 
Many previous studies have revealed that the chemical composition is different between these regions.
While many molecules containing nitrogen (e.g., NH$_3$, CH$_3$CN, etc.) are observed 
in Hot core, 
O-bearing species (e.g.,CH$_3$OH, CH$_3$OCH$_3$, etc.) are observed in Compact ridge \citep{Favre11}.

The abundance of CH$_3$NH$_2$ is analyzed using archive data of Orion~KL Hot core (ADS/JAO ALMA$\#$2013.1.00553.S. and 2011.0.00009.SV).
The description of this observation is found in \cite{Pagani17}.
We used CASA software V.5.0.0 during the procedure to analyze observational data.
Cycle 2 data cube was already calibrated by \cite{Pagani17}, and the reduced data are available on the web site of CDS (Centre de Donnees astronomiques de Strasbourg)\footnote{http://cdsarc.u-strasbg.fr/viz-bin/qcat?J/A+A/604/A32}.
The continuum emission in SV data was subtracted using the line free channels in the $(u,v)$ domain by CASA task {\sc uvcontsub} with channels.
After that, CASA task {\sc tclean} was used to deconvolve the images by applying natural weighting.

\begin{figure}
\begin{tabular}{ll}
\includegraphics[scale=0.4]{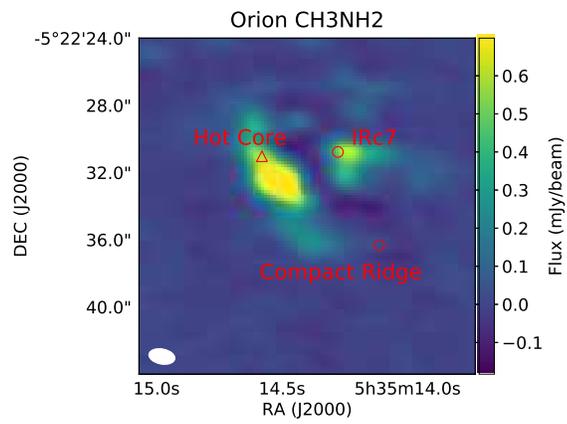}&\\
\end{tabular}
\vspace{35mm}
\caption{
Integrated intensity map of CH$_{3}$NH$_{2}$ 5, 2, $B_{1}$ $\rightarrow$ 5, 1, $B_{2}$ transition toward Orion KL.
The triangle shows the positions of the Orion Hot core, where the spectra was extracted.
The circles denote the positions of IRc7, and Compact ridge.
The velocity range is from 3.0 to 12.0~km s$^{-1}$.
\label{fig:map_Orion}
}
\end{figure}

%HERE
Figure~\ref{fig:map_Orion} shows the integrated intensity maps of the 6 transitions in Table~\ref{tbl:Ori}, created by the CASA task {\sc immoment}.
CH$_3$NH$_2$ emission appears mainly at Hot core and partially at IRc7.
According to previous work \citep[e.g.,][]{Feng15, Gong15}, N-bearing species tend to have the similar peak at or near Hot core.
The distribution of CH$_3$NH$_2$ also shows the same trend.

The spectrum were extracted from the region of $1\arcsec.0$ in diameter around Hot core (RA$_{J2000}: 05^{\rm{h}}35^{\rm{m}}14^{\rm{s}}.580$, Dec$_{J2000}:-05^{\circ}22'31\arcsec.029$).
The Gaussian fitting was performed to obtain FWHM line widths $\Delta$v and the mean local standard of rest velocity $V_{\rm {LSR}}$ of the emission lines.
Table~\ref{tbl:Ori} shows the all the detected CH$_3$NH$_2$ transitions with their line parameters.
A transition at 242.262~GHz is reported for the first time in this work, while other lines are already detected in \cite{Pagani17}
The average LSR velocity and FWHM line width are estimated to be 4.84~km~s$^{-1}$ and 4.16~km~s$^{-1}$, respectively.
$V_{\rm {LSR}}$ are consistent with those reported by \citet{Feng15} for N-bearing COMs observed toward Hot core (e.g, 4.9~km~s$^{-1}$ for CH$_2$CHCN, 5.1~km~s$^{-1}$ for CH$_3$CH$_2$CN).
On the other hand, the line width of CH$_3$NH$_2$ is narrower than those of other molecules in Hot core \citep[typically 5--15~km~s$^{-1}$,][]{Pagani17}.
With the rotation diagram method, the column density and the excitation temperature of CH$_3$NH$_2$ is obtained to be 1.6$\pm$0.6$\times$10$^{16}$cm$^{-2}$ and 77$\pm$17~K.
Since \cite{Pagani17} obtained the CH$_3$NH$_2$ column density of 1$\times$10$^{16}$~cm$^{-2}$ through LTE fitting on spectra, our result agrees well with their result.

\clearpage

\begin{deluxetable}{lllrrrr}
\tabletypesize{\scriptsize}
\tablecaption{List of Observed Sources}
\tablewidth{0pt}
\tablehead{
\colhead{Source}
& \colhead{\shortstack {RA$_{J2000}$ \\  $^{\rm h}$ $^{\rm m}$ $^{\rm s}$}  }
& \colhead{\shortstack {Dec$_{J2000}$  \\  $\degr$  '  "} } 
& \colhead{\shortstack {V$_{\rm LSR}$ \\ (km~s~$^{-1}$)}} 
& \colhead{\shortstack {distance \\ (kpc)}}
& \colhead{\shortstack {X[CH$_2$NH] \\ (x10$^{-8}$)}}
& \colhead{reference} 
}
\startdata
%NAME&RA&Dec&Vlsr&distance
NGC6334I&17 20 53.4&-35 47 1.0&-7&1.3&0.24&1, 3\\
G10.47+0.03&18 08 38.13&-19 51 49.4&67&8.5&3.1&1, 3\\
G31.41+0.3&18 47 34.6&-01 12 43.0&97&7.9&0.88&1, 4\\
W51~e1/e2&19 23 43.77&+14 30 25.9&57&5.4&0.28&1, 3\\
\enddata
\tablecomments{
The coordinate of the phase center, radial velocity, and the distance to sources are summarized. 
The fractional abundance of CH$_2$NH compared to hydrogen reported in \cite{Suzuki16} is also shown.
 References. (1) \cite{Ikeda01} (2)\cite{Menten07} (3)\cite{Reid14} (4)\cite{Churchwell90}
}
\label{tbl:source_list}
\end{deluxetable}
\clearpage

\begin{deluxetable}{lllll}
\tabletypesize{\scriptsize}
\tablecaption{Observation Parameters}
\tablewidth{0pt}
\tablehead{
\colhead{}
& \colhead{NGC6334I Set1}
& \colhead{NGC6334I Set2}
& \colhead{G10.47+0.03 Set1}
& \colhead{G10.47+0.03 Set2} 
}
\startdata
Observation date&2018 May 13&2018 May 14&2018 May 13&2018 May 14\\
Configuration&C43-2&C43-1&C43-2&C43-1\\
Phase Center ($\degr$  '  ")&07:20:53.4, -35:47:1.0&07:20:53.4, -35:47:1.0&08:08:38.1, -19:51:49.4&08:08:38.1, -19:51:49.4\\
Band&Band 5&Band 6&Band 5&Band 6\\
Time on Source (minutes)&21&6&40&11\\
Number of antennas&43&46&43&43\\
Spectral Resolution (MHz)&487.5&487.5&487.5&487.5\\
Maximum Recoverable Scale (")&11.8&8.8&11.8&8.8\\
Bandpass Calibrator&J1617-5848&J1924-2914&J1924-2914&J1924-2914\\
Phase Calibrator&J1733-3722&J1733-3722&J1832-2039&J1832-2039\\
Flux Calibrator&J1617-5848&J1924-2914&J1924-2914&J1924-2914\\
Angular Resolution ( " $\times$  ")&1.50 $\times$ 1.18&1.04 $\times$ 1.56&1.38 $\times$ 1.25&1.20 $\times$ 0.94\\
\enddata
\tablecomments{
}
\label{tbl:observation_parameter1}
\end{deluxetable}

\begin{deluxetable}{lllll}
\tabletypesize{\scriptsize}
\tablecaption{Observation Parameters (continued)}
\tablewidth{0pt}
\tablehead{
\colhead{}
& \colhead{G31.41+0.3 Set1}
& \colhead{G31.41+0.3 Set2}
& \colhead{W51 e1/e2 Set1}
& \colhead{W51 e1/e2 Set2} 
}
\startdata
Observation date&2018 May 13&2018 May 15&2018 May 13&2018 May 14\\
Configuration&C43-2&C43-2&C43-2&C43-2\\
Phase Center&18:47:34.6, -01:12:43.0&18:47:34.6, -01:12:43.0&19:23:43.8, 14:30:25.9&19:23:43.8, 14:30:25.9\\
Band&Band 5&Band 6&Band 5&Band 6\\
Time on Source (minutes)&42&12&33&13\\
Number of antennas&43&46&43&46\\
Spectral Resolution (MHz)&487.5&487.5&487.5&487.5\\
Maximum Recoverable Scale (")&11.6&8.8&11.8&8.8\\
Bandpass Calibrator&J2000-1748&J2000-1748&J2000-1748&J1751+0939\\
Phase Calibrator&J1851+0035&J1851+0035&J1922+1530&J1922+1530\\
Flux Calibrator&J2000-1748&J1751+0939&J2000-1748&J1751+0939\\
Angular Resolution (  "$\times$  ")&1.25 $\times$ 0.79&1.13 $\times$ 0.98&1.32 $\times$ 1.19&1.15 $\times$ 1.05\\
\enddata
\tablecomments{
}
\label{tbl:observation_parameter2}
\end{deluxetable}
\clearpage

\begin{deluxetable}{rrrrrr}
\tabletypesize{\scriptsize}
\tablecaption{Observed Frequency Range}
\tablewidth{0pt}
\tablehead{
\multicolumn{3}{c}{Set~1} &  \multicolumn{3}{c}{Set~2}\\
\cline{1-3} \cline{4-6} \\
\colhead{  \shortstack {Center Frequency \\(rest, GHz)}}
& \colhead{  \shortstack {Band Width \\(GHz)}}
& \colhead{Number of Channels}
& \colhead{  \shortstack {Center Frequency \\(rest, GHz)}}
& \colhead{  \shortstack {Band Width \\(GHz)}}
& \colhead{Number of Channels}
}
\startdata
192.212&0.234&480&247.611&0.234&480\\
191.959&0.234&480&247.967&0.234&480\\
191.732&0.234&480&248.838&0.234&480\\
191.462&0.234&480&249.192&0.234&480\\
193.795&0.234&480&249.443&0.234&480\\
193.415&0.234&480&250.161&0.234&480\\
192.622&0.234&480&250.219&0.234&480\\
192.469&0.234&480&250.507&0.234&480\\
203.050&0.938&1920&261.024&0.234&480\\
205.849&0.938&1920&261.219&0.234&480\\
&&&261.562&0.234&480\\
&&&261.805&0.234&480\\
&&&263.986&0.234&480\\
&&&264.172&0.234&480\\
&&&264.457&0.234&480\\
&&&264.752&0.234&480\\
\enddata
\tablecomments{
}
\label{tbl:observation_frequency_range}
\end{deluxetable}
\clearpage

\begin{deluxetable}{lllllllllrr}
\tabletypesize{\scriptsize}
\tablecaption{The Observed Transitions}
\tablewidth{0pt}
\tablehead{
\colhead{Species}
& \colhead{ \shortstack {Frequency \\ (MHz)}}
& \colhead{ transition}
& \colhead{ transition}
& \colhead{ transition}
& \colhead{ $\rightarrow$}
& \colhead{ transition}
& \colhead{ transition}
& \colhead{ transition}
& \colhead{ \shortstack {E$_{\rm u}$ \\ (K)}}
& \colhead{ \shortstack {S$\mu^2$ \\ (D$^2$)}}
}
\startdata
\hline 
&&$J'$&$K_{\rm a}'$&$\Gamma'$&$\rightarrow$&$J''$&$K_{\rm a}''$&$\Gamma''$&& \\
\hline
CH$_3$OH&191732.99&6&3&$E$&$\rightarrow$&7&2&$E$&96.5&1.4\\
&191810.50&4&1&$A^+$&$\rightarrow$&3&1&$A^+$&37.6&3.0\\
&193415.33&4&0&$E$&$\rightarrow$&3&0&$E$&36.3&3.2\\
&193441.61&4&-1&$E$&$\rightarrow$&3&-1&$E$&28.8&3.0\\
&193454.37&4&0&$A^+$&$\rightarrow$&3&0&$A^+$&23.2&3.2\\
&193471.43&4&3&$A^+$&$\rightarrow$&3&3&$A^+$&73.0&1.4\\
&193471.54&4&3&$A^-$&$\rightarrow$&3&3&$A^-$&73.0&1.4\\
&193474.42&4&3&$E$&$\rightarrow$&3&3&$E$&70.9&1.4\\
&193488.05&4&2&$A^-$&$\rightarrow$&3&2&$A^-$&60.9&2.4\\
&193488.96&4&-3&$E$&$\rightarrow$&3&-3&$E$&85.9&1.4\\
&193506.57&4&1&$E$&$\rightarrow$&3&1&$E$&44.3&3.1\\
&193510.75&4&2&$A^+$&$\rightarrow$&3&2&$A^+$&60.9&2.4\\
&193511.23&4&-2&$E$&$\rightarrow$&3&-2&$E$&49.1&2.4\\
&193511.34&4&2&$E$&$\rightarrow$&3&2&$E$&45.5&2.4\\
&205791.23&1&1&$A^+$&$\rightarrow$&2&0&$A^+$&16.8&1.0\\
&247611.04&18&3&$A^-$&$\rightarrow$&18&2&$A^+$&446.6&17.4\\
&247967.93&23&1&$E$&$\rightarrow$&23&0&$E$&661.4&11.7\\
&248885.48&16&3&$A^-$&$\rightarrow$&16&2&$A^+$&365.4&15.3\\
&249192.86&16&-3&$E$&$\rightarrow$&15&-4&$E$&378.3&4.7\\
&249419.90&15&3&$A^-$&$\rightarrow$&15&2&$A^+$&328.3&14.3\\
&249443.34&7&4&$A^-$&$\rightarrow$&8&3&$A^-$&145.3&1.2\\
&249451.89&7&4&$A^+$&$\rightarrow$&8&3&$A^+$&145.3&1.2\\
&250291.13&13&3&$A^-$&$\rightarrow$&13&2&$A^+$&261.0&12.3\\
&250507.02&11&0&$A^+$&$\rightarrow$&10&1&$A^+$&153.1&10.6\\
&261805.74&2&1&$E$&$\rightarrow$&1&0&$E$&28.0&1.3\\
\hline 
&&$J'$&$K_{\rm a}'$&$K_{\rm b}'$&$\rightarrow$&$J''$&$K_{\rm a}''$&$K_{\rm b}''$&& \\
\hline
CH$_2$NH&191462.82&3&0&3&$\rightarrow$&2&0&2&18.4&5.3\\
&191959.30&3&2&2&$\rightarrow$&2&2&1&49.9&2.9\\
&192212.32&4&1&3&$\rightarrow$&4&0&4&39.8&9.8\\
&192469.59&3&2&1&$\rightarrow$&2&2&0&50.0&2.9\\
&250161.68&7&1&6&$\rightarrow$&7&0&7&97.2&13.8\\
&263986.06&3&2&1&$\rightarrow$&4&1&4&50.0&0.8\\
\hline 
&&$J'$&$K_{\rm a}'$&$\Gamma'$&$\rightarrow$&$J''$&$K_{\rm a}''$&$\Gamma''$&& \\
\hline
CH$_3$NH$_2$&192622.39&14&2&$E_{1-1}$&$\rightarrow$&14&1&$E_{1+1}$&239.0&11.9\\
&193795.82&16&2&$B_2$&$\rightarrow$&16&1&$B_1$&305.3&15.6\\
&202791.86&3&1&$E_{2-1}$&$\rightarrow$&2&0&$E_{2+1}$&16.7&0.9\\
&203050.98&12&2&$E_{1-1}$&$\rightarrow$&12&1&$E_{1+1}$&181.6&8.6\\
&203096.32&16&2&$E_{1+1}$&$\rightarrow$&16&1&$E_{1+1}$&305.9&0.6\\
&248838.50&3&2&$B_1$&$\rightarrow$&3&1&$B_2$&28.9&2.3\\
&249527.26&10&5&$E_{2+1}$&$\rightarrow$&11&4&$E_{2+1}$&214.8&1.5\\
&250110.22&4&2&$A_1$&$\rightarrow$&4&1&$A_2$&37.4&3.1\\
&250159.40&2&2&$B_2$&$\rightarrow$&2&1&$B_1$&22.6&1.3\\
&250255.53&5&2&$E_{1+1}$&$\rightarrow$&5&1&$E_{1-1}$&47.9&1.3\\
&250398.51&12&1&$E_{1-1}$&$\rightarrow$&11&2&$E_{1+1}$&168.6&0.7\\
&260963.40&11&1&$B_2$&$\rightarrow$&10&2&$B_1$&145.9&3.9\\
&261024.31&4&1&$E_{2+1}$&$\rightarrow$&3&0&$E_{2+1}$&25.9&3.1\\
&261219.28&4&1&$B_1$&$\rightarrow$&3&0&$B_2$&25.6&3.9\\
&261252.83&6&2&$B_1$&$\rightarrow$&6&1&$B_2$&60.9&4.5\\
&261562.18&8&0&$B_1$&$\rightarrow$&7&1&$B_2$&76.8&4.9\\
&264172.18&4&1&$E_{1+1}$&$\rightarrow$&3&0&$E_{1+1}$&25.9&3.7\\
&264457.13&9&2&$E_{1+1}$&$\rightarrow$&9&1&$E_{1-1}$&111.8&3.4\\
&264752.35&7&2&$B_2$&$\rightarrow$&7&1&$B_1$&75.8&5.1\\
\enddata
\tablecomments{
The observed transitions of CH$_3$NH$_2$, CH$_2$NH, and CH$_3$OH are shown.
E$_{\rm u}$ represents the energy of the upper state level, while S$\mu^2$ shows the product of the intrinsic line strength and the square of the permanent dipole moment.
}
\label{tbl:observed_transitions}
\end{deluxetable}
\clearpage

\begin{deluxetable}{lll}
\tabletypesize{\scriptsize}
\tablecaption{Detected Sources}
\tablewidth{0pt}
\tablehead{
\colhead{Source}
& \colhead{\shortstack {$\alpha$(J2000) \\  $^{\rm h}$ $^{\rm m}$ $^{\rm s}$}}
& \colhead{\shortstack {$\delta$(J2000)  \\  $\degr$  '  "}} 
}
\startdata
NGC6334I MM1&17 20 53.433&-35 46 58.0\\
NGC6334I MM2&17 20 53.170&-35 46 58.6\\
NGC6334I MM3&17 20 53.416&-35 47 02.8\\
G10.47+0.03&18 08 38.258&-19 51 50.2\\
G31.41+0.3&18 47 34.320&-01 12 46.0\\
W51 e2&19 23 43.986&+14 30 34.6\\
W51 e8&19 23 43.890&+14 30 28.2\\
Orion KL Hot core&05 35 14.580&-05 22 31.029\\
\enddata
\tablecomments{
The coordinate, radial velocity, and the distance to sources are summarized. 
The fractional abundance of CH$_2$NH compared to hydrogen reported in \cite{Suzuki16} is also shown.
 References. (1) \cite{Ikeda01} (2)\cite{Menten07} (3)\cite{Reid14} (4)\cite{Churchwell90}
}
\label{tbl:detected_source}
\end{deluxetable}
\clearpage

\startlongtable
\begin{deluxetable}{llllllllrrrrrr}
\tabletypesize{\scriptsize}
\tablewidth{5pt}
\tablecaption{Observed Lines towards NGC6334I MM1}
\tablehead{
\colhead{Species}
& \colhead{}
& \colhead{}
& \colhead{}
& \colhead{$\rightarrow$}
& \colhead{}
& \colhead{}
& \colhead{}
&\colhead{\shortstack {Rest. \\ freq \\ (MHz)}}  
& \colhead{\shortstack {Obs \\ freq \\ (MHz)}}
& \colhead{\shortstack {T$_{\rm MB}$ \\ (K)} }
& \colhead{\shortstack {$\Delta$v \\ (km~s$^{-1}$)} }
& \colhead{\shortstack {V$_{\rm LSR}$ \\ (km~s$^{-1}$) }}
& \colhead{\shortstack {rms \\ (K) }}
}
\startdata
\hline 
&$J'$&$K_{\rm a}'$&$\Gamma'$&$\rightarrow$&$J''$&$K_{\rm a}''$&$\Gamma''$&&&&&& \\
\hline
CH$_3$OH&6&3&$E$&$\rightarrow$&7&2&$E$&191732.99&191732.55&80.0&7.7&-6.3&1.32\\
&4&1&$A^+$&$\rightarrow$&3&1&$A^+$&191810.50&191810.23&67.9&8.2&-6.6&2.23\\
&4&0&$E$&$\rightarrow$&3&0&$E$&193415.33&193414.74&73.6&8.9&-6.1&2.63\\
&4&-1&$E$&$\rightarrow$&3&-1&$E$&193441.61&193440.61&73.9&8.6&-5.5&2.43\\
&4&0&$A^+$&$\rightarrow$&3&0&$A^+$&193454.37&193453.49&65.9&10.8&-5.6&0.81\\
&4&1&$E$&$\rightarrow$&3&1&$E$&193506.57&193504.87&82.0&5.2&-4.4&1.93\\
&1&1&$A^+$&$\rightarrow$&2&0&$A^+$&205791.23&205790.68&79.0&7.7&-6.2&4.86\\
&18&3&$A^-$&$\rightarrow$&18&2&$A^+$&247611.04&247608.40&46.0&4.3&-3.8&1.86\\
&23&1&$E$&$\rightarrow$&23&0&$E$&247967.93&247966.57&63.8&6.0&-5.4&2.71\\
&16&3&$A^-$&$\rightarrow$&16&2&$A^+$&248885.48&248882.36&54.3&4.9&-3.2&1.89\\
&16&-3&$E$&$\rightarrow$&15&-4&$E$&249192.86&249191.94&57.8&7.6&-5.9&0.44\\
&15&3&$A^-$&$\rightarrow$&15&2&$A^+$&249419.90&249417.29&61.5&6.3&-3.9&2.80\\
&7&4&$A^-$&$\rightarrow$&8&3&$A^-$&249443.34&249441.38&66.1&4.9&-4.6&1.86\\
&7&4&$A^+$&$\rightarrow$&8&3&$A^+$&249451.89&249450.71&56.7&6.4&-5.6&1.80\\
&13&3&$A^-$&$\rightarrow$&13&2&$A^+$&250291.13&250288.95&59.6&4.5&-4.4&2.65\\
&11&0&$A^+$&$\rightarrow$&10&1&$A^+$&250507.02&250504.36&57.5&5.0&-3.8&2.15\\
&2&1&$E$&$\rightarrow$&1&0&$E$&261805.74&261804.19&54.2&6.2&-5.2&1.24\\
\hline 
&$J'$&$K_{\rm a}'$&$K_{\rm b}'$&$\rightarrow$&$J''$&$K_{\rm a}''$&$K_{\rm b}''$&&&&&& \\
\hline
CH$_2$NH&3&0&3&$\rightarrow$&2&0&2&191462.82&191461.94&34.0&5.7&-5.6&0.26\\
&3&2&2&$\rightarrow$&2&2&1&191959.30&191959.04&34.6&6.5&-6.6&0.15\\
&4&1&3&$\rightarrow$&4&0&4&192212.32&192211.53&50.1&5.9&-5.8&0.26\\
&3&2&1&$\rightarrow$&2&2&0&192469.59&192469.42&24.4&5.7&-6.7&0.16\\
\hline 
&$J'$&$K_{\rm a}'$&$\Gamma'$&$\rightarrow$&$J''$&$K_{\rm a}''$&$\Gamma''$&&&&&& \\
\hline
CH$_3$NH$_2$&14&2&$E_{1-1}$&$\rightarrow$&14&1&$E_{1+1}$&192622.39&192621.18&28.4&5.0&-5.1&0.23\\
&16&2&$B_2$&$\rightarrow$&16&1&$B_1$&193795.82&193795.24&28.7&5.2&-6.1&0.19\\
&3&1&$E_{2-1}$&$\rightarrow$&2&0&$E_{2+1}$&202791.86&202792.09&11.0&4.0&-7.3&0.31\\
&12&2&$E_{1-1}$&$\rightarrow$&12&1&$E_{1+1}$&203050.98&203050.54&33.3&6.3&-6.4&0.16\\
&3&2&$B_1$&$\rightarrow$&3&1&$B_2$&248838.50&248839.64&29.2&7.6&-8.4&0.41\\
&2&2&$B_2$&$\rightarrow$&2&1&$B_1$&250159.40&250160.03&55.3&6.9&-7.8&0.38\\
&12&1&$E_{1-1}$&$\rightarrow$&11&2&$E_{1+1}$&250398.51&250399.29&7.4&3.3&-7.9&0.15\\
&4&1&$E_{2+1}$&$\rightarrow$&3&0&$E_{2+1}$&261024.31&261024.77&15.6&4.1&-7.5&0.40\\
&4&1&$B_1$&$\rightarrow$&3&0&$B_2$&261219.28&261219.25&49.2&8.6&-7.0&0.97\\
&8&0&$B_1$&$\rightarrow$&7&1&$B_2$&261562.18&261562.72&47.2&8.3&-7.6&0.49\\
&4&1&$E_{1+1}$&$\rightarrow$&3&0&$E_{1+1}$&264172.18&264171.58&49.2&7.3&-6.3&0.43\\
&9&2&$E_{1+1}$&$\rightarrow$&9&1&$E_{1-1}$&264457.13&264455.82&29.1&6.8&-5.5&0.79\\
\enddata
\tablecomments{
The observed molecular transitions toward NGC6334I are summarized.
T$_{\rm MB}$, $\Delta$v, and V$_{\rm LSR}$ are, respectively, the brightness temperature, the FWHM line width, and the local standard of rest velocity.
We showed the quantumn number of CH$_3$OH, CH$_3$NH$_2$, and CH$_2$NH, respecively, the rest frequency, the observed frequency, and r.m.s noise level.
The line parameters were obtained through the least-squares fitting assuming the shape of gaussian.
}
\label{tbl:NGC6334FMM1_line}
\end{deluxetable}
\clearpage

\startlongtable
\begin{deluxetable}{llllllllrrrrrr}
\tabletypesize{\scriptsize}
\tablewidth{5pt}
\tablecaption{Observed Lines towards NGC6334I MM2}
\tablehead{
\colhead{Species}
& \colhead{}
& \colhead{}
& \colhead{}
& \colhead{$\rightarrow$}
& \colhead{}
& \colhead{}
& \colhead{}
& \colhead{\shortstack {Rest. \\ freq \\ (MHz)}}  
& \colhead{\shortstack {Obs \\ freq \\ (MHz)}}
& \colhead{\shortstack {T$_{\rm MB}$ \\ (K)} }
& \colhead{\shortstack {$\Delta$v \\ (km~s$^{-1}$)} }
& \colhead{\shortstack {V$_{\rm LSR}$ \\ (km~s$^{-1}$) }}
& \colhead{\shortstack {rms \\ (K) }}
}
\startdata
\hline 
&$J'$&$K_{\rm a}'$&$\Gamma'$&$\rightarrow$&$J''$&$K_{\rm a}''$&$\Gamma''$&&&&&& \\
\hline
&6&3&$E$&$\rightarrow$&7&2&$E$&191732.99&191733.87&77.0&6.2&-8.4&1.32\\
&4&1&$A^+$&$\rightarrow$&3&1&$A^+$&191810.50&191811.46&75.5&6.5&-8.5&2.23\\
&4&0&$E$&$\rightarrow$&3&0&$E$&193415.33&193416.28&74.3&7.1&-8.5&2.63\\
&4&-1&$E$&$\rightarrow$&3&-1&$E$&193441.61&193442.54&66.3&7.7&-8.4&2.43\\
&4&0&$A^+$&$\rightarrow$&3&0&$A^+$&193454.37&193455.50&68.8&7.4&-8.8&0.81\\
&4&1&$E$&$\rightarrow$&3&1&$E$&193506.57&193507.85&77.1&7.4&-9.0&1.93\\
&1&1&$A^+$&$\rightarrow$&2&0&$A^+$&205791.23&205792.37&80.7&6.2&-8.7&4.86\\
&18&3&$A^-$&$\rightarrow$&18&2&$A^+$&247611.04&247611.93&86.7&5.7&-8.1&1.86\\
&23&1&$E$&$\rightarrow$&23&0&$E$&247967.93&247969.04&55.1&4.3&-8.3&2.71\\
&16&3&$A^-$&$\rightarrow$&16&2&$A^+$&248885.48&248886.37&82.1&5.8&-8.1&1.89\\
&16&-3&$E$&$\rightarrow$&15&-4&$E$&249192.86&249193.89&60.6&5.2&-8.2&0.44\\
&15&3&$A^-$&$\rightarrow$&15&2&$A^+$&249419.90&249420.86&82.0&6.3&-8.2&2.80\\
&7&4&$A^-$&$\rightarrow$&8&3&$A^-$&249443.34&249444.49&71.2&5.8&-8.4&1.86\\
&7&4&$A^+$&$\rightarrow$&8&3&$A^+$&249451.89&249453.00&71.2&6.3&-8.3&1.80\\
&13&3&$A^-$&$\rightarrow$&13&2&$A^+$&250291.13&250292.01&80.6&7.3&-8.1&2.65\\
&11&0&$A^+$&$\rightarrow$&10&1&$A^+$&250507.02&250508.32&61.2&7.8&-8.6&2.15\\
&2&1&$E$&$\rightarrow$&1&0&$E$&261805.74&261807.05&71.8&6.9&-8.5&1.24\\
\hline 
&$J'$&$K_{\rm a}'$&$K_{\rm b}'$&$\rightarrow$&$J''$&$K_{\rm a}''$&$K_{\rm b}''$&&&&&& \\
\hline
CH$_2$NH&3&0&3&$\rightarrow$&2&0&2&191462.82&191463.43&5.8&4.1&-8.0&0.39\\
&3&2&2&$\rightarrow$&2&2&1&191959.30&191959.52&2.6&3.7&-7.3&0.22\\
&4&1&3&$\rightarrow$&4&0&4&192212.32&192213.20&12.3&3.9&-8.4&0.39\\
&3&2&1&$\rightarrow$&2&2&0&192469.59&192470.21&2.9&3.0&-8.0&0.23\\
&7&1&6&$\rightarrow$&7&0&7&250161.68&250162.08&10.5&4.6&-7.5&0.28\\
\hline
&$J'$&$K_{\rm a}'$&$\Gamma'$&$\rightarrow$&$J''$&$K_{\rm a}''$&$\Gamma''$&&&&&& \\
\hline
CH$_3$NH$_2$&14&2&$E_{1-1}$&$\rightarrow$&14&1&$E_{1+1}$&192622.39&192622.81&5.1&3.1&-7.7&0.34\\
&16&2&$B_2$&$\rightarrow$&16&1&$B_1$&193795.82&193796.54&2.5&2.9&-8.1&0.28\\
&12&2&$E_{1-1}$&$\rightarrow$&12&1&$E_{1+1}$&203050.98&203051.46&1.3&3.4&-7.7&0.24\\
&10&5&$E_{2+1}$&$\rightarrow$&11&4&$E_{2+1}$&250110.22&250111.65&2.6&2.4&-8.7&0.31\\
&4&1&$E_{2+1}$&$\rightarrow$&3&0&$E_{2+1}$&261219.28&261221.37&9.6&6.3&-9.4&0.94\\
&4&1&$E_{1+1}$&$\rightarrow$&3&0&$E_{1+1}$&264172.18&264173.54&12.9&4.6&-8.5&2.65\\
\enddata
\tablecomments{
Same as Table~\ref{tbl:NGC6334FMM2_line} but toward NGC6334I MM2.
}
\label{tbl:NGC6334FMM2_line}
\end{deluxetable}
\clearpage

\startlongtable
\begin{deluxetable}{llllllllrrrrrr}
\tabletypesize{\scriptsize}
\tablewidth{5pt}
\tablecaption{Observed Lines towards NGC6334I MM3}
\tablehead{
\colhead{Species}
& \colhead{}
& \colhead{}
& \colhead{}
& \colhead{$\rightarrow$}
& \colhead{}
& \colhead{}
& \colhead{}
& \colhead{\shortstack {Rest. \\ freq \\ (MHz)}}  
& \colhead{\shortstack {Obs \\ freq \\ (MHz)}}
& \colhead{\shortstack {T$_{\rm MB}$ \\ (K)} }
& \colhead{\shortstack {$\Delta$v \\ (km~s$^{-1}$)} }
& \colhead{\shortstack {V$_{\rm LSR}$ \\ (km~s$^{-1}$) }}
& \colhead{\shortstack {rms \\ (K) }}
}
\startdata
\hline 
&$J'$&$K_{\rm a}'$&$\Gamma'$&$\rightarrow$&$J''$&$K_{\rm a}''$&$\Gamma''$&&&&&& \\
\hline
&6&3&$E$&$\rightarrow$&7&2&$E$&191732.99&191733.87&77.0&6.2&-8.4&2.43\\
&4&1&$A^+$&$\rightarrow$&3&1&$A^+$&191810.50&191811.46&75.5&6.5&-8.5&4.11\\
&4&0&$E$&$\rightarrow$&3&0&$E$&193415.33&193416.28&74.3&7.1&-8.5&4.70\\
&4&-1&$E$&$\rightarrow$&3&-1&$E$&193441.61&193442.54&66.3&7.7&-8.4&4.34\\
&4&0&$A^+$&$\rightarrow$&3&0&$A^+$&193454.37&193455.50&68.8&7.4&-8.8&1.45\\
&4&1&$E$&$\rightarrow$&3&1&$E$&193506.57&193507.85&77.1&7.4&-9.0&3.43\\
&1&1&$A^+$&$\rightarrow$&2&0&$A^+$&205791.23&205792.37&80.7&6.2&-8.7&3.53\\
&18&3&$A^-$&$\rightarrow$&18&2&$A^+$&247611.04&247611.93&86.7&5.7&-8.1&0.67\\
&16&3&$A^-$&$\rightarrow$&16&2&$A^+$&248885.48&248886.37&82.1&5.8&-8.1&0.40\\
&16&-3&$E$&$\rightarrow$&15&-4&$E$&249192.86&249193.89&60.6&5.2&-8.2&0.29\\
&15&3&$A^-$&$\rightarrow$&15&2&$A^+$&249419.90&249420.86&82.0&6.3&-8.2&0.69\\
&7&4&$A^-$&$\rightarrow$&8&3&$A^-$&249443.34&249444.49&71.2&5.8&-8.4&0.57\\
&7&4&$A^+$&$\rightarrow$&8&3&$A^+$&249451.89&249453.00&71.2&6.3&-8.3&0.65\\
&13&3&$A^-$&$\rightarrow$&13&2&$A^+$&250291.13&250292.01&80.6&7.3&-8.1&0.77\\
&2&1&$E$&$\rightarrow$&1&0&$E$&261805.74&261807.05&71.8&6.9&-8.5&0.66\\
\hline 
&$J'$&$K_{\rm a}'$&$K_{\rm b}'$&$\rightarrow$&$J''$&$K_{\rm a}''$&$K_{\rm b}''$&&&&&& \\
\hline
CH$_2$NH&3&0&3&$\rightarrow$&2&0&2&191462.82&191464.20&0.6&2.6&-9.2&0.72\\
&3&2&2&$\rightarrow$&2&2&1&191959.30&191961.06&0.5&3.6&-9.7&0.41\\
&4&1&3&$\rightarrow$&4&0&4&192212.32&192213.72&0.7&3.1&-9.2&0.70\\
&3&2&1&$\rightarrow$&2&2&0&192469.59&192470.84&0.5&3.0&-8.9&0.42\\
\enddata
\label{tbl:NGC6334FMM3_line}
\tablecomments{
Same as Table~\ref{tbl:NGC6334FMM1_line} but toward NGC6334I MM3.
}
\end{deluxetable}
\clearpage

\startlongtable
\begin{deluxetable}{llllllllrrrrrr}
\tabletypesize{\scriptsize}
\tablewidth{5pt}
\tablecaption{Observed Lines towards G10.47+0.03}
\tablehead{
\colhead{Species}
& \colhead{}
& \colhead{}
& \colhead{}
& \colhead{$\rightarrow$}
& \colhead{}
& \colhead{}
& \colhead{}
& \colhead{\shortstack {Rest. \\ freq \\ (MHz)}}  
& \colhead{\shortstack {Obs \\ freq \\ (MHz)}}
& \colhead{\shortstack {T$_{\rm MB}$ \\ (K)} }
& \colhead{\shortstack {$\Delta$v \\ (km~s$^{-1}$)} }
& \colhead{\shortstack {V$_{\rm LSR}$ \\ (km~s$^{-1}$) }}
& \colhead{\shortstack {rms \\ (K) }}
}
\startdata
\hline 
&$J'$&$K_{\rm a}'$&$\Gamma'$&$\rightarrow$&$J''$&$K_{\rm a}''$&$\Gamma''$&&&&&& \\
\hline
CH$_3$OH&6&3&$E$&$\rightarrow$&7&2&$E$&191732.99&191734.01&69.3&11.1&65.4&3.43\\
&4&1&$A^+$&$\rightarrow$&3&1&$A^+$&191810.50&191811.63&69.0&12.9&65.2&1.14\\
&4&0&$E$&$\rightarrow$&3&0&$E$&193415.33&193417.12&68.5&12.2&64.2&0.73\\
&4&-1&$E$&$\rightarrow$&3&-1&$E$&193441.61&193443.24&61.8&9.8&64.5&1.35\\
&4&0&$A^+$&$\rightarrow$&3&0&$A^+$&193454.37&193456.95&61.9&11.4&63.0&1.35\\
&4&1&$E$&$\rightarrow$&3&1&$E$&193506.57&193506.82&58.7&10.9&66.6&1.46\\
&1&1&$A^+$&$\rightarrow$&2&0&$A^+$&205791.23&205792.69&55.5&12.0&64.9&0.21\\
&18&3&$A^-$&$\rightarrow$&18&2&$A^+$&247611.04&247612.26&69.4&14.2&65.5&2.07\\
&23&1&$E$&$\rightarrow$&23&0&$E$&247967.93&247969.58&53.6&9.4&65.0&1.19\\
&16&3&$A^-$&$\rightarrow$&16&2&$A^+$&248885.48&248886.58&71.2&10.0&65.7&2.07\\
&16&-3&$E$&$\rightarrow$&15&-4&$E$&249192.86&249194.14&53.8&11.3&65.5&0.59\\
&15&3&$A^-$&$\rightarrow$&15&2&$A^+$&249419.90&249420.85&80.1&10.4&65.9&2.67\\
&7&4&$A^-$&$\rightarrow$&8&3&$A^-$&249443.34&249443.69&56.2&9.7&66.6&2.07\\
&7&4&$A^+$&$\rightarrow$&8&3&$A^+$&249451.89&249453.19&56.8&10.4&65.4&2.37\\
&13&3&$A^-$&$\rightarrow$&13&2&$A^+$&250291.13&250292.50&80.1&11.2&65.4&2.67\\
&11&0&$A^+$&$\rightarrow$&10&1&$A^+$&250507.02&250508.50&56.5&8.9&65.2&2.67\\
&2&1&$E$&$\rightarrow$&1&0&$E$&261805.74&261807.91&55.7&9.5&64.5&3.85\\
\hline 
&$J'$&$K_{\rm a}'$&$K_{\rm b}'$&$\rightarrow$&$J''$&$K_{\rm a}''$&$K_{\rm b}''$&&&&&& \\
\hline
CH$_2$NH&3&0&3&$\rightarrow$&2&0&2&191462.82&191462.67&25.3&11.1&67.2&0.10\\
&3&2&2&$\rightarrow$&2&2&1&191959.30&191959.61&14.8&8.0&66.5&0.10\\
&4&1&3&$\rightarrow$&4&0&4&192212.32&192212.82&29.8&9.0&66.2&0.31\\
&3&2&1&$\rightarrow$&2&2&0&192469.59&192469.59&11.9&9.1&67.0&0.21\\
&7&1&6&$\rightarrow$&7&0&7&250161.68&250162.00&39.3&9.2&66.6&1.19\\
\hline 
&$J'$&$K_{\rm a}'$&$K_{\rm b}'$&$\rightarrow$&$J''$&$K_{\rm a}''$&$K_{\rm b}''$&&&&&& \\
\hline
CH$_3$NH$_2$&11&1&$B_2$&$\rightarrow$&10&2&$B_1$&260963.40&260963.40&15.4&7.3&67.0&0.89\\
&4&1&$E_{2+1}$&$\rightarrow$&3&0&$E_{2+1}$&261024.31&261023.82&11.5&5.4&67.6&0.59\\
&4&1&$B_1$&$\rightarrow$&3&0&$B_2$&261219.28&261220.78&18.0&8.4&65.3&0.89\\
&8&0&$B_1$&$\rightarrow$&7&1&$B_2$&261562.18&261564.24&33.8&10.9&64.6&0.69\\
&4&1&$E_{1+1}$&$\rightarrow$&3&0&$E_{1+1}$&264172.18&264172.89&40.8&10.7&66.2&0.69\\
\enddata
\label{tbl:G10.47+0.03_line}
\tablecomments{
Same as Table~\ref{tbl:NGC6334FMM1_line} but toward G10.47+0.03.
}
\end{deluxetable}
\clearpage

\startlongtable
\begin{deluxetable}{llllllllrrrrrr}
\tabletypesize{\scriptsize}
\tablewidth{5pt}
\tablecaption{Observed Lines towards G31.41+0.3}
\tablehead{
\colhead{Species}
& \colhead{}
& \colhead{}
& \colhead{}
& \colhead{$\rightarrow$}
& \colhead{}
& \colhead{}
& \colhead{}
& \colhead{\shortstack {Rest. \\ freq \\ (MHz)}}  
& \colhead{\shortstack {Obs \\ freq \\ (MHz)}}
& \colhead{\shortstack {T$_{\rm MB}$ \\ (K)} }
& \colhead{\shortstack {$\Delta$v \\ (km~s$^{-1}$)} }
& \colhead{\shortstack {V$_{\rm LSR}$ \\ (km~s$^{-1}$) }}
& \colhead{\shortstack {rms \\ (K) }}
}
\startdata
\hline 
&$J'$&$K_{\rm a}'$&$\Gamma'$&$\rightarrow$&$J''$&$K_{\rm a}''$&$\Gamma''$&&&&&& \\
\hline
CH$_3$OH&6&3&$E$&$\rightarrow$&7&2&$E$&191732.99&191733.95&47.1&10.9&95.5&0.45\\
&18&3&$A^-$&$\rightarrow$&18&2&$A^+$&247611.04&247610.99&49.9&10.5&97.1&0.60\\
&23&1&$E$&$\rightarrow$&23&0&$E$&247967.93&247967.96&33.9&7.7&97.0&0.39\\
&16&3&$A^-$&$\rightarrow$&16&2&$A^+$&248885.48&248885.37&53.0&10.2&97.1&0.60\\
&16&-3&$E$&$\rightarrow$&15&-4&$E$&249192.86&249193.20&37.9&9.8&96.6&0.42\\
&15&3&$A^-$&$\rightarrow$&15&2&$A^+$&249419.90&249420.04&53.6&10.1&96.8&0.19\\
&7&4&$A^-$&$\rightarrow$&8&3&$A^-$&249443.34&249443.38&47.7&8.6&97.0&0.42\\
&7&4&$A^+$&$\rightarrow$&8&3&$A^+$&249451.89&249452.04&44.4&7.0&96.8&0.78\\
&13&3&$A^-$&$\rightarrow$&13&2&$A^+$&250291.13&250291.06&56.5&9.8&97.1&0.93\\
&11&0&$A^+$&$\rightarrow$&10&1&$A^+$&250507.02&250507.24&50.5&9.3&96.7&1.69\\
\hline 
&$J'$&$K_{\rm a}'$&$K_{\rm b}'$&$\rightarrow$&$J''$&$K_{\rm a}''$&$K_{\rm b}''$&&&&&& \\
\hline
CH$_2$NH&3&0&3&$\rightarrow$&2&0&2&191462.82&191463.20&14.7&7.6&96.4&0.27\\
&3&2&2&$\rightarrow$&2&2&1&191959.30&191959.39&12.7&6.5&96.9&0.06\\
&4&1&3&$\rightarrow$&4&0&4&192212.32&192212.40&24.0&7.4&96.9&0.25\\
&3&2&1&$\rightarrow$&2&2&0&192469.59&192469.67&13.0&6.7&96.9&0.15\\
&7&1&6&$\rightarrow$&7&0&7&250161.68&250161.13&31.6&8.0&97.7&0.42\\
\hline 
&$J'$&$K_{\rm a}'$&$\Gamma'$&$\rightarrow$&$J''$&$K_{\rm a}''$&$\Gamma''$&&&&&& \\
\hline
CH$_3$NH$_2$&12&2&$E_{1-1}$&$\rightarrow$&12&1&$E_{1+1}$&203050.98&203050.53&5.0&8.6&97.7&0.18\\
&3&2&$B_1$&$\rightarrow$&3&1&$B_2$&248838.50&248840.06&15.6&7.3&95.1&0.39\\
&4&2&$A_1$&$\rightarrow$&4&1&$A_2$&250110.22&250110.06&7.6&5.8&97.2&0.20\\
&12&1&$E_{1-1}$&$\rightarrow$&11&2&$E_{1+1}$&250398.51&250399.89&9.7&8.6&95.3&0.26\\
&4&1&$E_{2+1}$&$\rightarrow$&3&0&$E_{2+1}$&261024.31&261023.38&10.7&6.8&98.1&0.26\\
&4&1&$B_1$&$\rightarrow$&3&0&$B_2$&261219.28&261219.99&19.1&7.7&96.2&0.45\\
&8&0&$B_1$&$\rightarrow$&7&1&$B_2$&261562.18&261563.50&37.8&9.9&95.5&0.75\\
&4&1&$E_{1+1}$&$\rightarrow$&3&0&$E_{1+1}$&264172.18&264173.14&22.1&10.3&95.9&0.75\\
\enddata
\label{tbl:G31.41+0.3_line}
\tablecomments{
Same as Table~\ref{tbl:NGC6334FMM2_line} but toward G31.41+0.3.
}
\end{deluxetable}
\clearpage

\startlongtable
\begin{deluxetable}{llllllllrrrrrr}
\tabletypesize{\scriptsize}
\tablewidth{5pt}
\tablecaption{Observed Lines towards W51 e2}
\tablehead{
\colhead{Species}
& \colhead{}
& \colhead{}
& \colhead{}
& \colhead{$\rightarrow$}
& \colhead{}
& \colhead{}
& \colhead{}
& \colhead{\shortstack {Rest. \\ freq \\ (MHz)}}  
& \colhead{\shortstack {Obs \\ freq \\ (MHz)}}
& \colhead{\shortstack {T$_{\rm MB}$ \\ (K)} }
& \colhead{\shortstack {$\Delta$v \\ (km~s$^{-1}$)} }
& \colhead{\shortstack {V$_{\rm LSR}$ \\ (km~s$^{-1}$) }}
& \colhead{\shortstack {rms \\ (K) }}
}
\startdata
\hline 
&$J'$&$K_{\rm a}'$&$\Gamma'$&$\rightarrow$&$J''$&$K_{\rm a}''$&$\Gamma''$&&&&&& \\
\hline
CH$_3$OH&6&3&$E$&$\rightarrow$&7&2&$E$&191732.99&191731.90&35.9&5.4&58.7&0.74\\
&4&1&$A^+$&$\rightarrow$&3&1&$A^+$&191810.50&191809.38&37.2&4.7&58.8&2.40\\
&4&0&$E$&$\rightarrow$&3&0&$E$&193415.33&193413.94&43.9&5.5&59.2&0.96\\
&4&-1&$E$&$\rightarrow$&3&-1&$E$&193441.61&193440.30&42.2&5.4&59.0&2.97\\
&4&0&$A^+$&$\rightarrow$&3&0&$A^+$&193454.37&193452.95&41.2&5.5&59.2&3.08\\
&4&1&$E$&$\rightarrow$&3&1&$E$&193506.57&193505.20&46.6&5.7&59.1&1.94\\
&1&1&$A^+$&$\rightarrow$&2&0&$A^+$&205791.23&205790.03&37.6&5.0&58.7&1.71\\
&13&3&$A^-$&$\rightarrow$&13&2&$A^+$&250291.13&250289.38&28.8&5.1&59.1&3.59\\
&11&0&$A^+$&$\rightarrow$&10&1&$A^+$&250507.02&250504.85&28.6&5.1&59.6&4.70\\
&2&1&$E$&$\rightarrow$&1&0&$E$&261805.74&261804.15&39.5&4.7&58.8&3.87\\
\hline 
&$J'$&$K_{\rm a}'$&$K_{\rm b}'$&$\rightarrow$&$J''$&$K_{\rm a}''$&$K_{\rm b}''$&&&&&& \\
\hline
CH$_2$NH&3&0&3&$\rightarrow$&2&0&2&191462.82&191463.12&5.0&7.0&56.5&0.42\\
&3&2&2&$\rightarrow$&2&2&1&191959.30&191960.24&4.6&6.0&55.5&0.14\\
&4&1&3&$\rightarrow$&4&0&4&192212.32&192213.30&11.2&6.3&55.5&0.37\\
&3&2&1&$\rightarrow$&2&2&0&192469.59&192470.51&3.8&4.8&55.6&0.15\\
&7&1&6&$\rightarrow$&7&0&7&250161.68&250162.38&13.6&5.7&56.2&0.94\\
\hline 
&$J'$&$K_{\rm a}'$&$\Gamma'$&$\rightarrow$&$J''$&$K_{\rm a}''$&$\Gamma''$&&&&&& \\
\hline
CH$_3$NH$_2$&16&2&$B_2$&$\rightarrow$&16&1&$B_1$&193795.82&193796.61&3.7&4.8&55.8&0.05\\
&12&2&$E_{1-1}$&$\rightarrow$&12&1&$E_{1+1}$&203050.98&203051.55&4.0&7.5&56.2&0.13\\
&4&2&$A_1$&$\rightarrow$&4&1&$A_2$&250110.22&250111.74&4.6&4.9&55.2&0.21\\
&4&1&$E_{2+1}$&$\rightarrow$&3&0&$E_{2+1}$&261024.31&261023.80&5.1&2.9&57.6&0.72\\
&8&0&$B_1$&$\rightarrow$&7&1&$B_2$&261562.18&261562.24&21.0&6.1&56.9&0.85\\
&4&1&$E_{1+1}$&$\rightarrow$&3&0&$E_{1+1}$&264172.18&264173.94&22.2&7.5&55.0&0.97\\
\enddata
\label{tbl:W51e2}
\tablecomments{
Same as Table~\ref{tbl:NGC6334FMM2_line} but toward W51 e2.
}
\end{deluxetable}
\clearpage

\startlongtable
\begin{deluxetable}{llllllllrrrrrr}
\tabletypesize{\scriptsize}
\tablewidth{5pt}
\tablecaption{Observed Lines towards W51 e8}
\tablehead{
\colhead{Species}
& \colhead{}
& \colhead{}
& \colhead{}
& \colhead{$\rightarrow$}
& \colhead{}
& \colhead{}
& \colhead{}
& \colhead{\shortstack {Rest. \\ freq \\ (MHz)}}  
& \colhead{\shortstack {Obs \\ freq \\ (MHz)}}
& \colhead{\shortstack {T$_{\rm MB}$ \\ (K)} }
& \colhead{\shortstack {$\Delta$v \\ (km~s$^{-1}$)} }
& \colhead{\shortstack {V$_{\rm LSR}$ \\ (km~s$^{-1}$) }}
& \colhead{\shortstack {rms \\ (K) }}
}
\startdata
\hline 
&$J'$&$K_{\rm a}'$&$\Gamma'$&$\rightarrow$&$J''$&$K_{\rm a}''$&$\Gamma''$&&&&&& \\
\hline
CH$_3$OH&6&3&$E$&$\rightarrow$&7&2&$E$&191732.99&191732.82&36.6&9.5&57.3&0.59\\
&4&1&$A^+$&$\rightarrow$&3&1&$A^+$&191810.50&191810.43&50.3&10.3&57.1&0.77\\
&4&0&$E$&$\rightarrow$&3&0&$E$&193415.33&193415.16&49.0&10.6&57.3&0.54\\
&4&-1&$E$&$\rightarrow$&3&-1&$E$&193441.61&193441.01&41.3&9.6&57.9&0.64\\
&4&0&$A^+$&$\rightarrow$&3&0&$A^+$&193454.37&193454.10&43.7&11.4&57.4&0.96\\
&4&1&$E$&$\rightarrow$&3&1&$E$&193506.57&193506.70&51.9&10.9&56.8&0.46\\
&1&1&$A^+$&$\rightarrow$&2&0&$A^+$&205791.23&205791.15&33.5&8.3&57.1&1.60\\
&18&3&$A^-$&$\rightarrow$&18&2&$A^+$&247611.04&247610.79&48.9&10.4&57.3&0.50\\
&23&1&$E$&$\rightarrow$&23&0&$E$&247967.93&247967.76&15.3&10.2&57.2&0.39\\
&16&3&$A^-$&$\rightarrow$&16&2&$A^+$&248885.48&248885.00&53.7&9.5&57.6&0.50\\
&16&-3&$E$&$\rightarrow$&15&-4&$E$&249192.86&249192.67&28.3&10.0&57.2&0.36\\
&15&3&$A^-$&$\rightarrow$&15&2&$A^+$&249419.90&249420.20&44.7&8.7&56.6&0.17\\
&7&4&$A^-$&$\rightarrow$&8&3&$A^-$&249443.34&249443.45&34.5&9.5&56.9&0.39\\
&7&4&$A^+$&$\rightarrow$&8&3&$A^+$&249451.89&249452.09&30.5&7.8&56.8&0.30\\
&13&3&$A^-$&$\rightarrow$&13&2&$A^+$&250291.13&250290.71&55.6&9.8&57.5&0.97\\
&11&0&$A^+$&$\rightarrow$&10&1&$A^+$&250507.02&250507.03&45.6&9.6&57.0&0.83\\
&2&1&$E$&$\rightarrow$&1&0&$E$&261805.74&261806.50&38.5&7.4&56.1&2.38\\
\hline 
&$J'$&$K_{\rm a}'$&$K_{\rm b}'$&$\rightarrow$&$J''$&$K_{\rm a}''$&$K_{\rm b}''$&&&&&& \\
\hline
CH$_2$NH&3&0&3&$\rightarrow$&2&0&2&191462.82&191461.44&8.5&9.5&59.2&0.31\\
&3&2&2&$\rightarrow$&2&2&1&191959.30&191957.43&5.3&7.5&59.9&0.05\\
&4&1&3&$\rightarrow$&4&0&4&192212.32&192210.28&13.6&10.4&60.2&0.26\\
&3&2&1&$\rightarrow$&2&2&0&192469.59&192467.57&4.5&6.0&60.1&0.07\\
&7&1&6&$\rightarrow$&7&0&7&250161.68&250159.51&16.4&9.4&59.6&0.44\\
\hline 
&$J'$&$K_{\rm a}'$&$\Gamma'$&$\rightarrow$&$J''$&$K_{\rm a}''$&$\Gamma''$&&&&&& \\
\hline
CH$_3$NH$_2$&16&2&$B_2$&$\rightarrow$&16&1&$B_1$&193795.82&193793.48&4.7&4.9&60.6&0.03\\
&12&2&$E_{1-1}$&$\rightarrow$&12&1&$E_{1+1}$&203050.98&203048.30&3.3&6.8&61.0&0.07\\
&12&1&$E_{1-1}$&$\rightarrow$&11&2&$E_{1+1}$&250398.51&250396.18&2.0&8.3&59.8&0.25\\
&11&1&$B_2$&$\rightarrow$&10&2&$B_1$&260963.40&260960.81&9.9&8.9&60.0&0.41\\
&4&1&$E_{2+1}$&$\rightarrow$&3&0&$E_{2+1}$&261024.31&261021.35&4.3&4.3&60.4&0.17\\
&4&1&$B_1$&$\rightarrow$&3&0&$B_2$&261219.28&261217.24&8.3&7.1&59.3&0.41\\
&6&2&$B_1$&$\rightarrow$&6&1&$B_2$&261252.83&261251.72&27.8&10.8&58.3&0.61\\
&8&0&$B_1$&$\rightarrow$&7&1&$B_2$&261562.18&261558.15&4.7&5.9&61.6&0.19\\
&4&1&$E_{1+1}$&$\rightarrow$&3&0&$E_{1+1}$&264172.18&264170.05&12.7&7.8&59.4&0.44\\
\enddata
\tablecomments{
Same as Table~\ref{tbl:NGC6334FMM2_line} but toward W51 e8.
}
\label{tbl:W51e8}
\end{deluxetable}
\clearpage

\startlongtable
\begin{deluxetable}{llllllllllrrrrrr}
\tabletypesize{\scriptsize}
\tablewidth{5pt}
\tablecaption{CH$_3$NH$_2$ Lines towards Orion KL Hot core from Archival Data}
\tablehead{
\colhead{Species}
& \colhead{}
& \colhead{}
& \colhead{}
& \colhead{$\rightarrow$}
& \colhead{}
& \colhead{}
& \colhead{}
& \colhead{\shortstack {Rest. \\ freq \\ (MHz)}}  
& \colhead{\shortstack {Obs \\ freq \\ (MHz)}}
& \colhead{ \shortstack {E$_{\rm u}$ \\ (K)}}
& \colhead{ \shortstack {S$\mu^2$ \\ (D$^2$)}}
& \colhead{\shortstack {T$_{\rm MB}$ \\ (K)} }
& \colhead{\shortstack {$\Delta$v \\ (km~s$^{-1}$)} }
& \colhead{\shortstack {V$_{\rm LSR}$ \\ (km~s$^{-1}$) }}
& \colhead{\shortstack {rms \\ (K) }}
}
\startdata
\hline 
&$J'$&$K_{\rm a}'$&$\Gamma'$&$\rightarrow$&$J''$&$K_{\rm a}''$&$\Gamma''$&&&&&& \\
\hline
CH$_3$NH$_2$& 12& 2& $B_{2}$& $\rightarrow$& 12& 1& $B_{1}$ &217758.43&217758.23&182.1&10.8&0.86&4.0&4.7&0.034\\
& 8& 2& $A_{2}$& $\rightarrow$& 8& 1& $A_{1}$ &229908.19&229908.12&92.7&6.8&0.65&3.2&4.9&0.064\\
& 8& 2& $B_{2}$ &$\rightarrow$& 8& 1& $B_{1}$ &235735.08&235734.97&92.8&6.8&1.70&5.6&4.9&0.081\\
& 6& 2& $B_{2}$& $\rightarrow$& 6& 1& $B_{1}$ &242262.12&242261.96&60.9&5.0&2.03&8.0&4.8&0.166\\
& 5& 2& $B_{1}$& $\rightarrow$& 5& 1& $B_{2}$ &244886.67&244886.86&48.1&4.1&1.29&4.2&5.2&0.043\\
& 12& 1& $B_{2}$& $\rightarrow$& 11& 2& $B_{1}$ &245202.56&245202.16&168.3&3.2&0.37&3.9&4.5&0.037\\
\enddata
\label{tbl:Ori}
\end{deluxetable}
\clearpage

%\begin{landscape}
\begin{deluxetable}{lrrrrrrr}
\tabletypesize{\scriptsize}
\tablecaption{The Derived Abundance}
\tablewidth{0pt}
\tablehead{
\colhead{Source}
& \colhead{ \shortstack {N[H$_2$] \\ (cm$^{-2}$)}}
& \colhead{ \shortstack {N[CH$_3$OH] \\ (cm$^{-2}$)}}
& \colhead{ \shortstack {Tex[CH$_3$OH] \\ (K)}}
& \colhead{ \shortstack {N[CH$_3$NH$_2$] \\ (cm$^{-2}$)}}
& \colhead{ \shortstack {Tex[CH$_3$NH$_2$] \\ (K)}}
& \colhead{ \shortstack {N[CH$_2$NH] \\ (cm$^{-2}$)}}
& \colhead{ \shortstack {Tex[CH$_2$NH] \\ (K)}}
}
\startdata
NGC6334I MM1&1.8 (25)&4.5 $\pm$ 1.8 (17)&120 $\pm$ 32&1.0 $\pm$ 0.1 (18)&120 $\pm$ 11&8.6 $\pm$ 4.8 (16)&[120]\\
NGC6334I MM2&1.2 (24)&6.5 $\pm$ 1.8 (17)&157 $\pm$ 36&6.8 $\pm$ 2.3 (16)&91 $\pm$ 16&8.7 $\pm$ 1.2 (15)&137 $\pm$ 46\\
NGC6334I MM3&1.8 (24)&6.2 $\pm$ 1.9 (17)&169 $\pm$ 47&$<$ 1.0 (16)&[170]&9.3 $\pm5.4$ (14)&[170]\\
G10.47+0.03&6.3 (24)&1.3 $\pm$ 0.4 (18)&235 $\pm$ 71&9.4 $\pm$ 2.7 (17)&166 $\pm$ 106&5.3 $\pm$ 1.2 (16)&90 $\pm$ 31\\
G31.41+0.3&7.5 (24)&8.3 $\pm$ 3.7 (17)&207 $\pm$ 55&3.2 $\pm$ 0.7 (17)&79 $\pm$ 18&7.4 $\pm$ 1.8 (16)&184 $\pm$ 141\\
W51~e2&9.1 (24)&1.3 $\pm$0.3 (17)&94 $\pm$ 19&1.9 $\pm$ 0.4 (17)&95 $\pm$ 14&1.9 $\pm$ 0.4 (16)&155 $\pm$ 98\\
W51~e8&4.1 (24)&6.5 $\pm$1.6 (17)&205 $\pm$ 40&2.5 $\pm$ 1.1 (17)&131 $\pm$ 51&2.7 $\pm$ 0.3 (16)&130 $\pm$ 28\\
\enddata
\tablecomments{
The column density of molecular hydrogen, and the column densities and the excitation temperatures for CH$_3$OH, CH$_3$NH$_2$, and CH$_2$NH derived by the rotaton diagram method are shown.
For CH$_2$NH in NGC6334I MM1 and MM3, excitation temperatures are assumed as they are shown in brackets.
The upper limit of CH$_3$NH$_2$ column density is shown for NGC6334I MM3.
a (b) means a$\times$10$^{b}$.
}
\label{tbl:observation_result}
\end{deluxetable}
%\end{landscape}
\clearpage

\begin{deluxetable}{lrrrr}
\tablecaption{Comparison with Chemical Model}
\tablewidth{0pt}
\tablehead{
\colhead{Source}
& \colhead{ \shortstack {CH$_3$NH$_2$/CH$_3$OH \\obs}}
%& \colhead{ \shortstack {CH$_3$NH$_2$/CH$_2$NH \\ obs}}
& \colhead{ \shortstack {X[CH$_3$OH] \\ obs}}
& \colhead{ \shortstack {X[CH$_3$NH$_2$] \\ obs}}
& \colhead{ \shortstack {X[CH$_2$NH] \\ obs}}
}
\startdata
NGC6334I MM1&2.2 $\pm$ 1.0&1.2 (-8)&2.8 (-8)&9.4 (-10)\\%&30 $\pm$ 14
NGC6334I MM2&0.11 $\pm$ 0.05&1.2 (-7)&2.8 (-8)&3.6 (-9)\\%7.8 $\pm$ 2.9&
NGC6334I MM3&$<$0.02&1.7 (-7)&$<$2.9 (-9)&2.6 (-10)\\%$<$11&
G10.47+0.03&0.72 $\pm$ 0.32&1.0 (-7)&7.5 (-8)&4.2 (-9)\\%18 $\pm$ 7&
G31.41+0.3&0.39 $\pm$ 0.19&5.3 (-8)&2.1 (-8)&4.9 (-9)\\%4.4 $\pm$ 1.4&
W51~e2&1.4 $\pm$ 0.5&7.2 (-9)&1.0 (-8)&1.0 (-9)\\%10 $\pm$ 3&
W51~e8&0.38 $\pm$ 0.19&8.0 (-8)&3.1 (-8)&3.3 (-9)\\%9.2 $\pm$ 4.1&
\hline
Model&CH$_3$NH$_2$/CH$_3$OH&&&\\
\hline
Our standard model&0.19-0.39&&&\\
B=0.7&0.49-1.0&&&\\
B=0.2&1.0-9.4&&&\\
B=0.1&1.0-75&&&\\
T=200~K&0.20-0.32&&&\\
n=1$\times$10$^{6}$~cm$^{-2}$&0.14-0.18&&&\\
n=1$\times$10$^{8}$~cm$^{-2}$&0.15-0.17&&&\\
long warm up&0.11-0.35&&&\\
short warm up&0.09-0.10&&&\\
Garrod (Fast)&0.007&&&\\
Garrod (Medium)&0.004&&&\\
Garrod (Slow)&0.001&&&\\
\hline
Previous Study&CH$_3$NH$_2$/CH$_3$OH&&&\\
\hline
Sgr B2(M)$^a$&8.0$\times$10$^{-3}$&&&\\
Sgr B2(N)$^a$&0.03&&&\\
Sgr B2(N)$^b$&0.10&&&\\
NGC6334I MM1$^c$&(2.5-5.9)$\times$10$^{-3}$&&&\\
NGC6334I MM2$^c$&(0.9-1.5)$\times$10$^{-3}$&&&\\
NGC6334I MM3$^c$&(4.8-5.4)$\times$10$^{4}$&&&\\
IRAS~16293-2422B$^d$&$<$5.3$\times$10$^{-5}$&&&\\
\enddata
\tablecomments{
(Top)The observed abundance ratios of CH$_3$NH$_2$/CH$_3$OH is shown.
X[CH$_3$OH], X[CH$_3$NH$_2$], and X[CH$_2$NH], respectively, are the fractional abundances of CH$_3$OH, CH$_3$NH$_2$, and CH$_2$NH, which are derived by deviding their column densities by that of molecular hydrogen column density.
(Middle)The predictions of these molecular ratios by chemical modeling are also summarized.
The age of the simulation is constrained so that the observed fractional abundances of CH$_3$OH and CH$_3$NH$_2$ are explained within a factor of 10.
For comparison, the peak abundance ratios under three warm-up models presented by \cite{Garrod13} are also shown.
(Bottom) The observed abundance ratios of ``CH$_3$NH$_2$/CH$_3$OH'' and ``CH$_3$NH$_2$/CH$_3$OH'' from the previous studies.
Citation: (a) \cite{Belloche13}, (b) \cite{Neill14}, (c) \cite{Bogelund19} (d)\cite{Ligterink18}
}
\label{tbl:ratios}
\end{deluxetable}
\clearpage

\clearpage


\clearpage
\clearpage
\clearpage
\begin{figure}
 \begin{tabular}{l}
\includegraphics[scale=0.7]{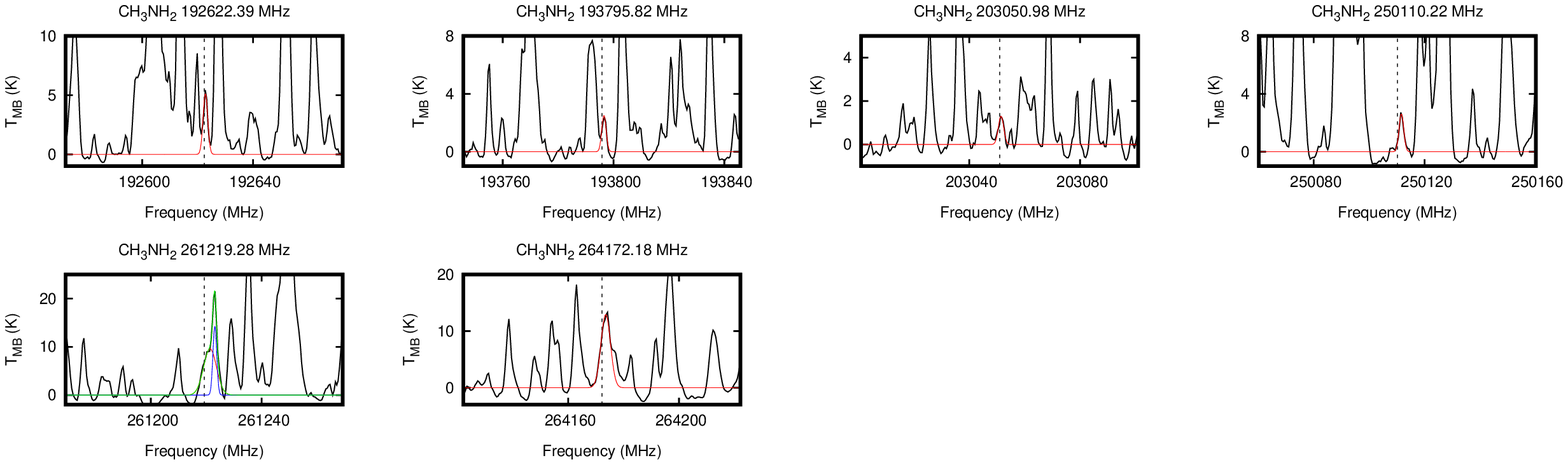}\\
\includegraphics[scale=0.7]{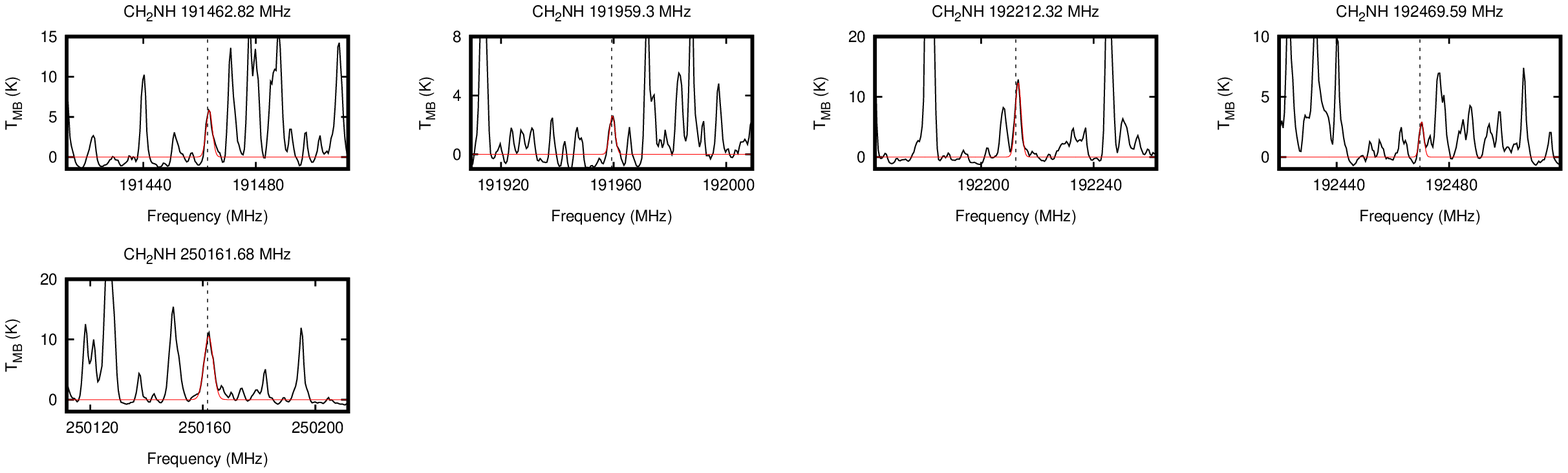}\\
 \end{tabular}
\vspace{15mm}
\caption{
The same as Figure~\ref{fig:spectle_N63MM1} but toward NGC6334I MM2.
\label{fig:spectle_N63MM2}
}
\end{figure}
\clearpage
\begin{figure}
 \begin{tabular}{l}
\includegraphics[scale=0.7]{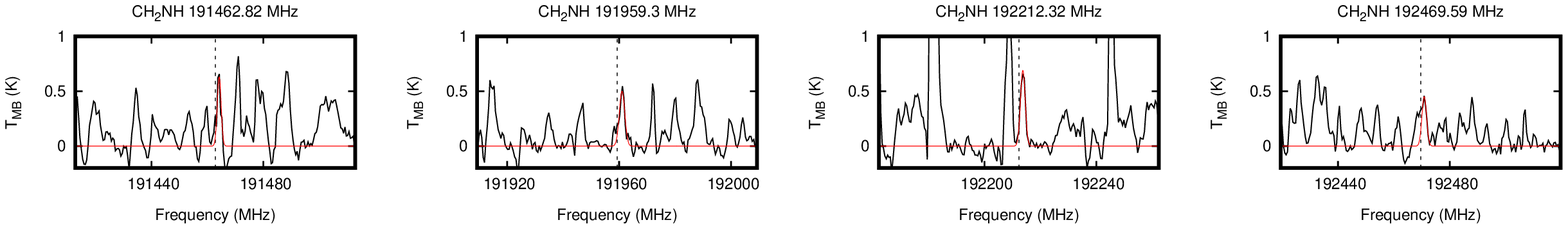}\\
 \end{tabular}
\vspace{15mm}
\caption{
The same as Figure~\ref{fig:spectle_N63MM1} but toward NGC6334I MM3.
\label{fig:spectle_N63MM3}
}
\end{figure}
\clearpage
\begin{figure}
 \begin{tabular}{l}
\includegraphics[scale=0.7]{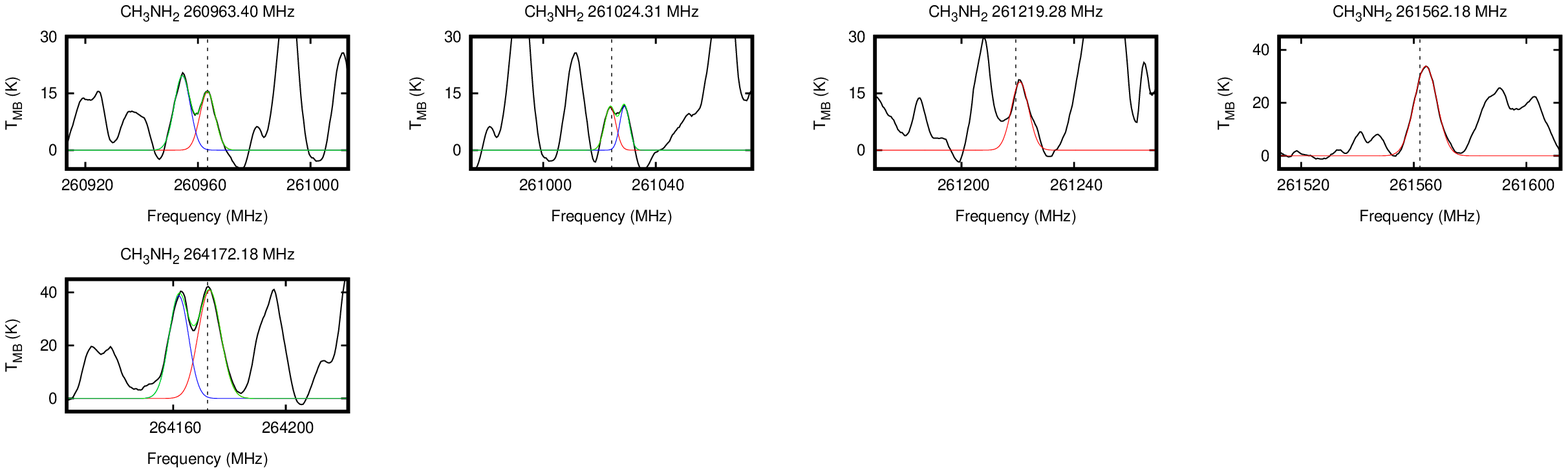}\\
\includegraphics[scale=0.7]{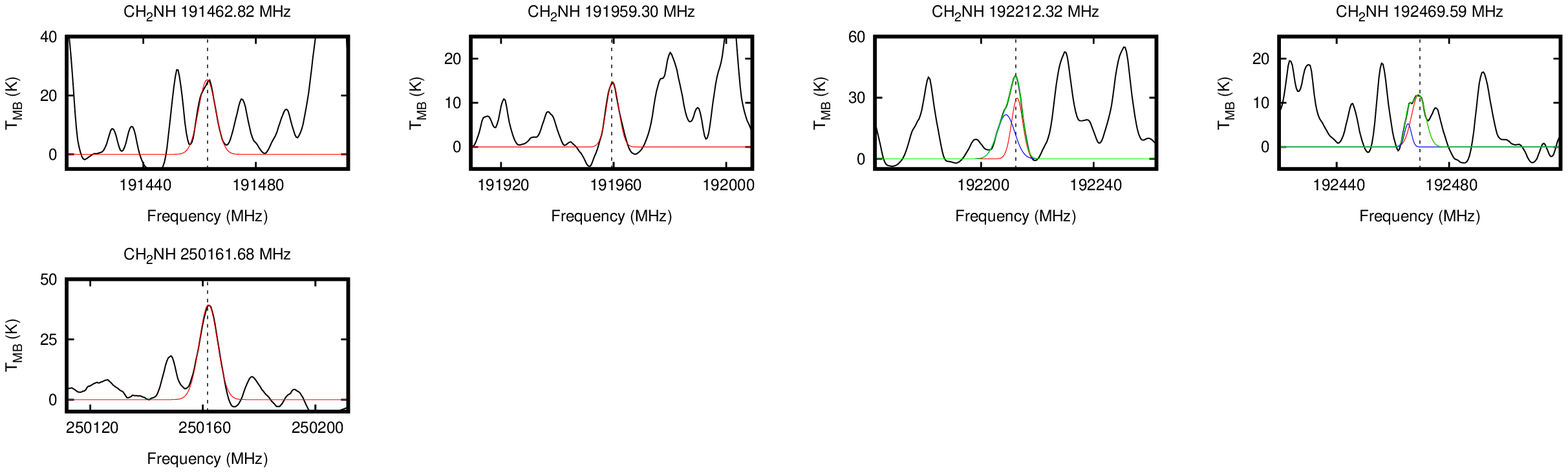}\\
 \end{tabular}
\vspace{15mm}
\caption{
The same as Figure~\ref{fig:spectle_N63MM1} but toward  G10.47+0.03. 
\label{fig:spectle_G10}
}
\end{figure}
\clearpage
\begin{figure}
 \begin{tabular}{l}
\includegraphics[scale=0.7]{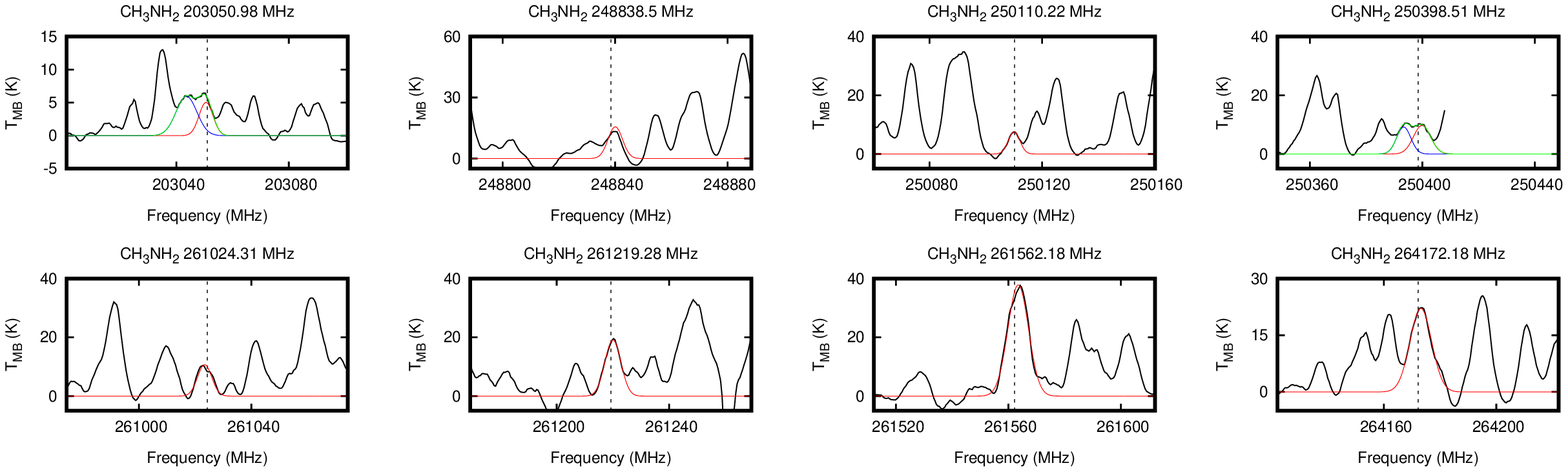}\\
\includegraphics[scale=0.7]{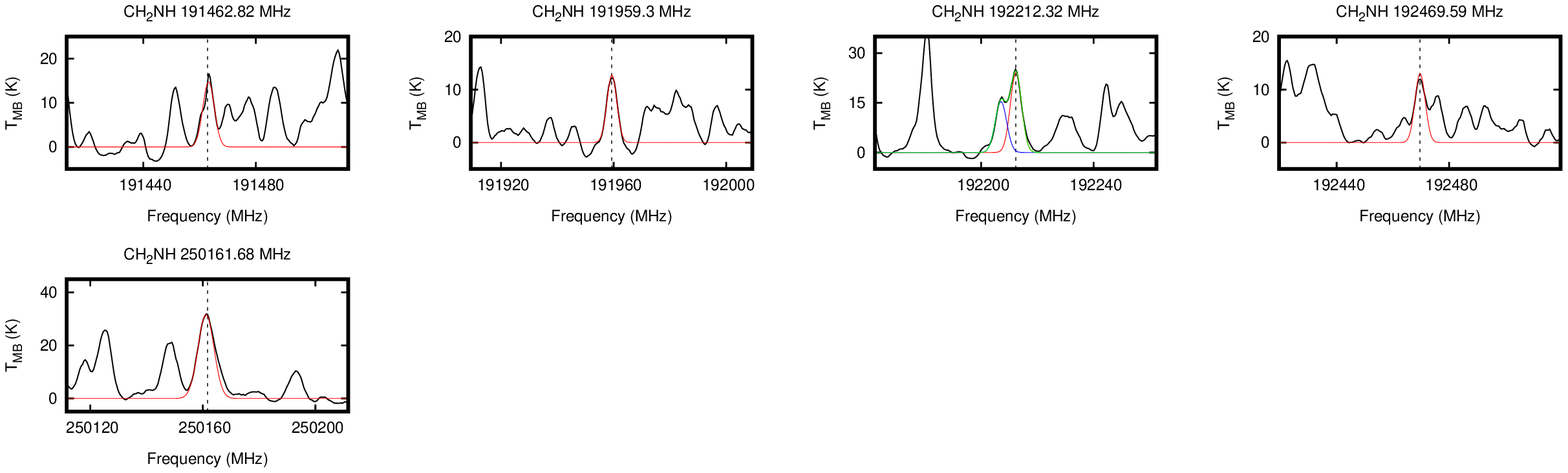}\\
 \end{tabular}
\vspace{15mm}
\caption{
The same as Figure~\ref{fig:spectle_N63MM1} but toward G31.41+0.3. 
\label{fig:spectle_G31}
}
\end{figure}
\clearpage
\begin{figure}
 \begin{tabular}{l}
\includegraphics[scale=0.7]{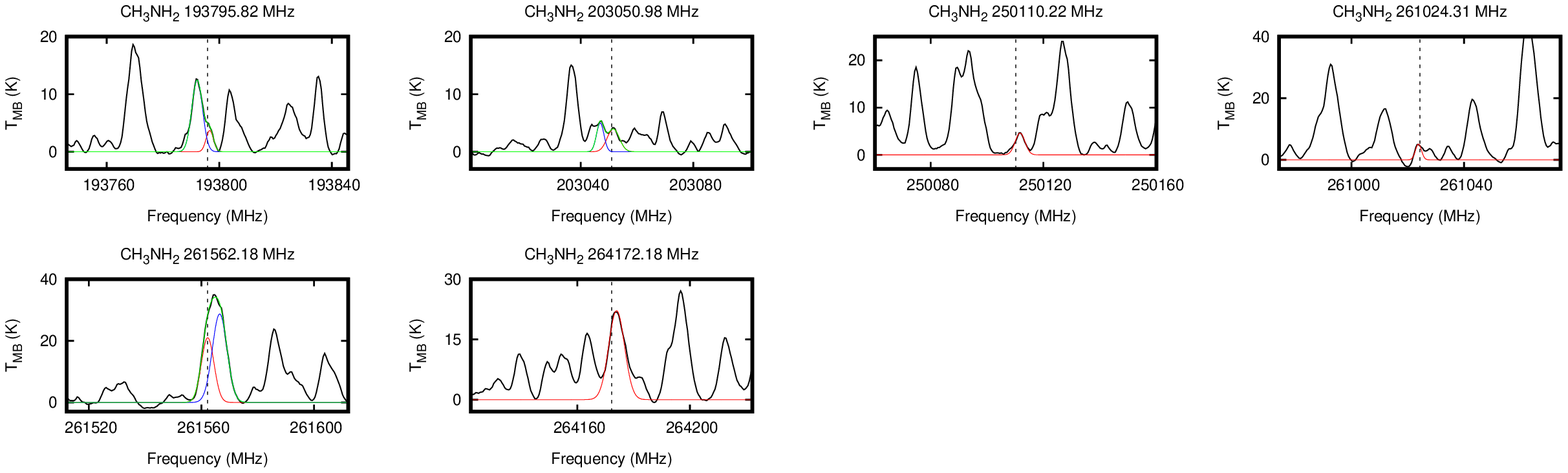}\\
\includegraphics[scale=0.7]{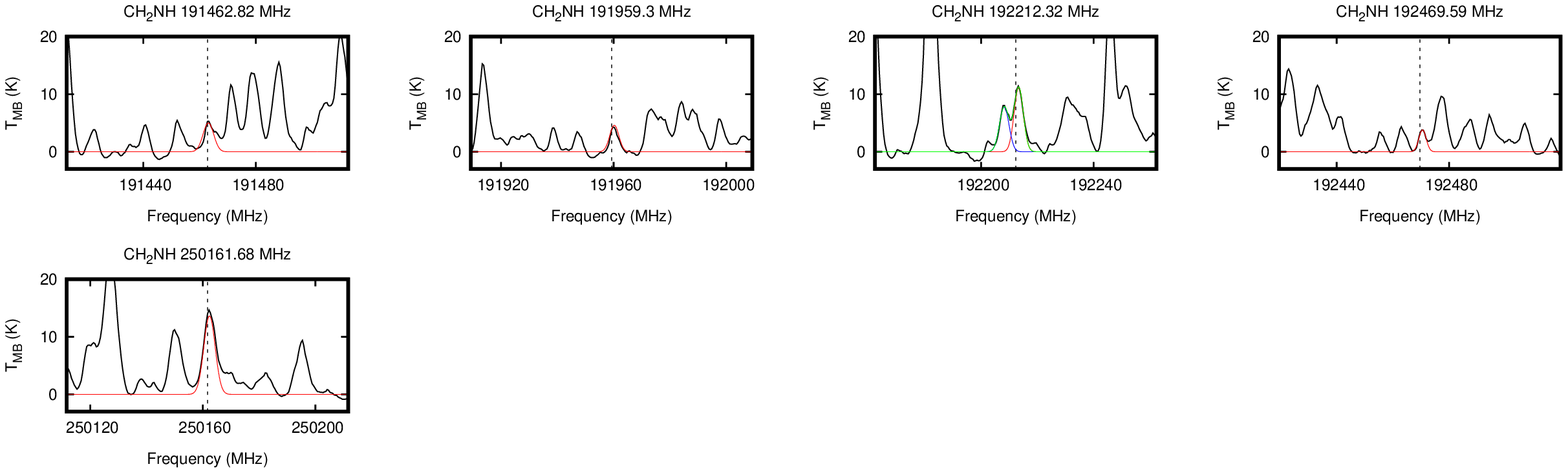}\\
 \end{tabular}
\vspace{15mm}
\caption{
The same as Figure~\ref{fig:spectle_N63MM1} but toward W51~e2. 
\label{fig:spectle_W51e2}
}
\end{figure}
\clearpage
\begin{figure}
 \begin{tabular}{l}
\includegraphics[scale=0.7]{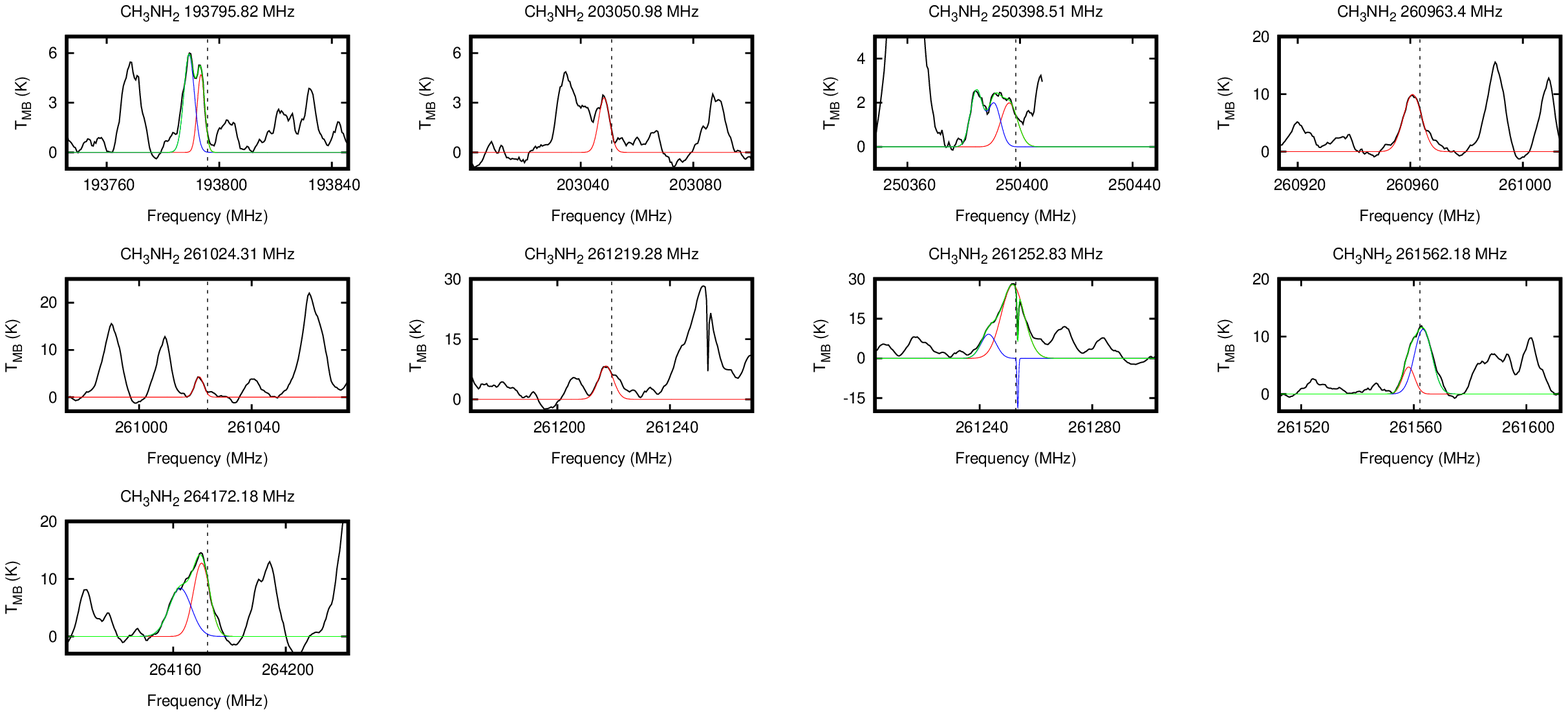}\\
\includegraphics[scale=0.7]{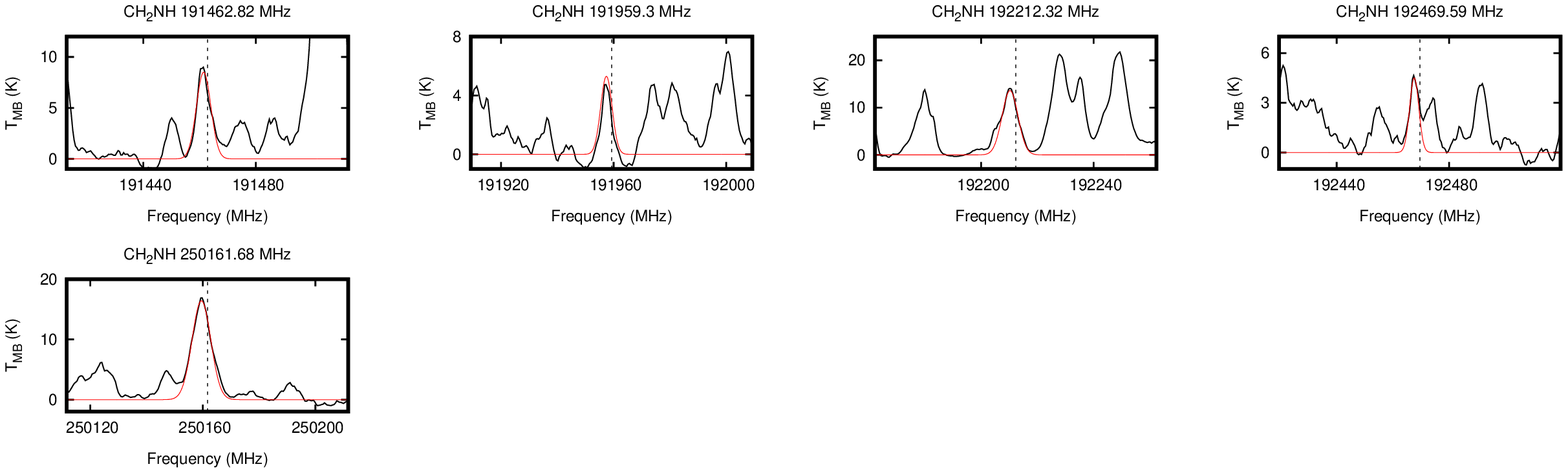}\\
 \end{tabular}
\vspace{15mm}
\caption{
The same as Figure~\ref{fig:spectle_N63MM1} but toward W51~e8. 
\label{fig:spectle_W51e2}
}
\end{figure}
\clearpage
\begin{figure}
 \begin{tabular}{ll}
\includegraphics[scale=0.6]{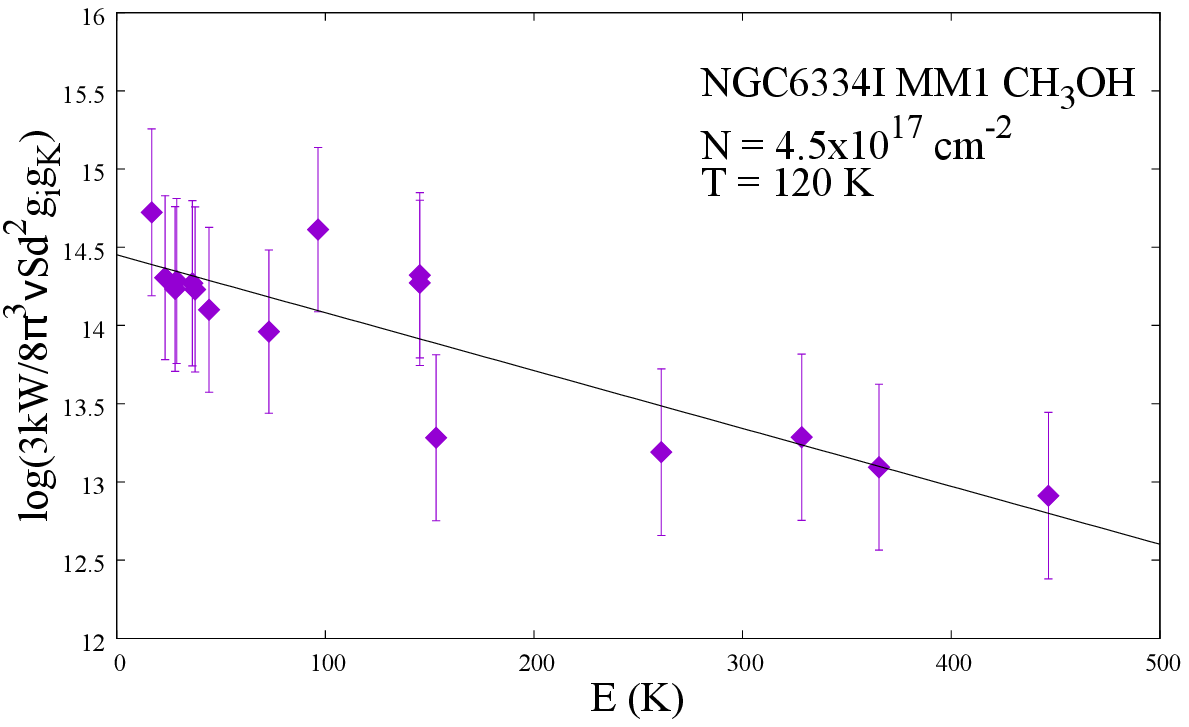}&
\includegraphics[scale=0.6]{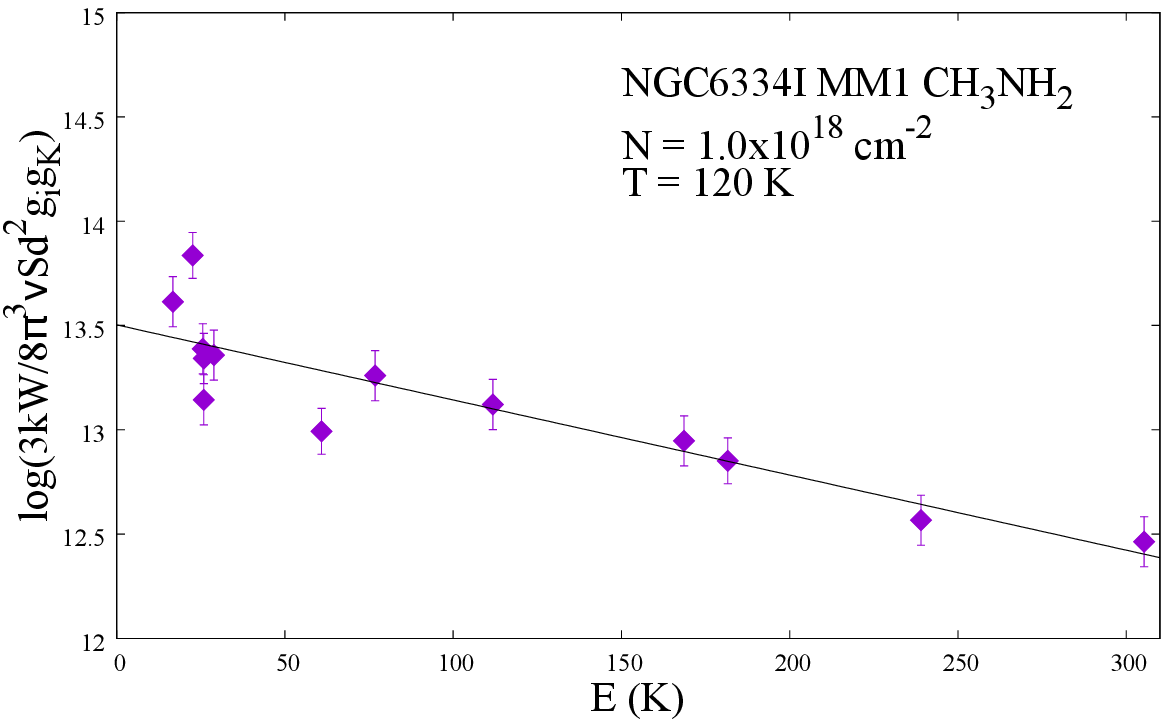}\\
\includegraphics[scale=0.6]{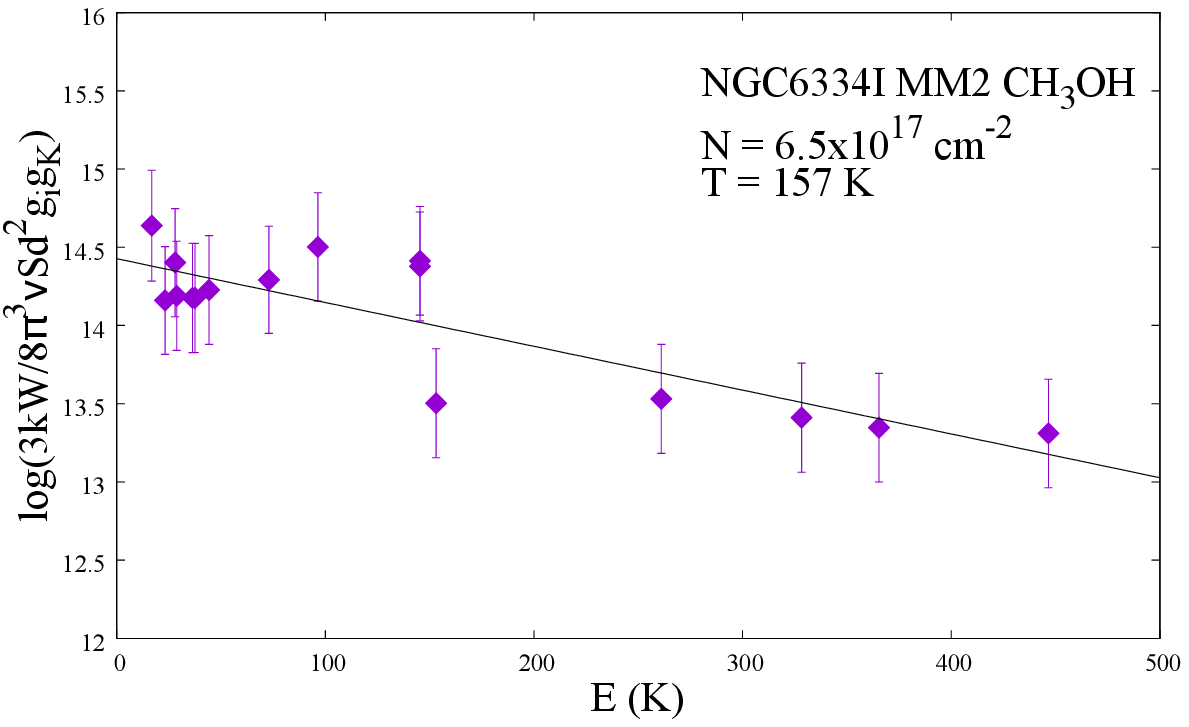}&
\includegraphics[scale=0.6]{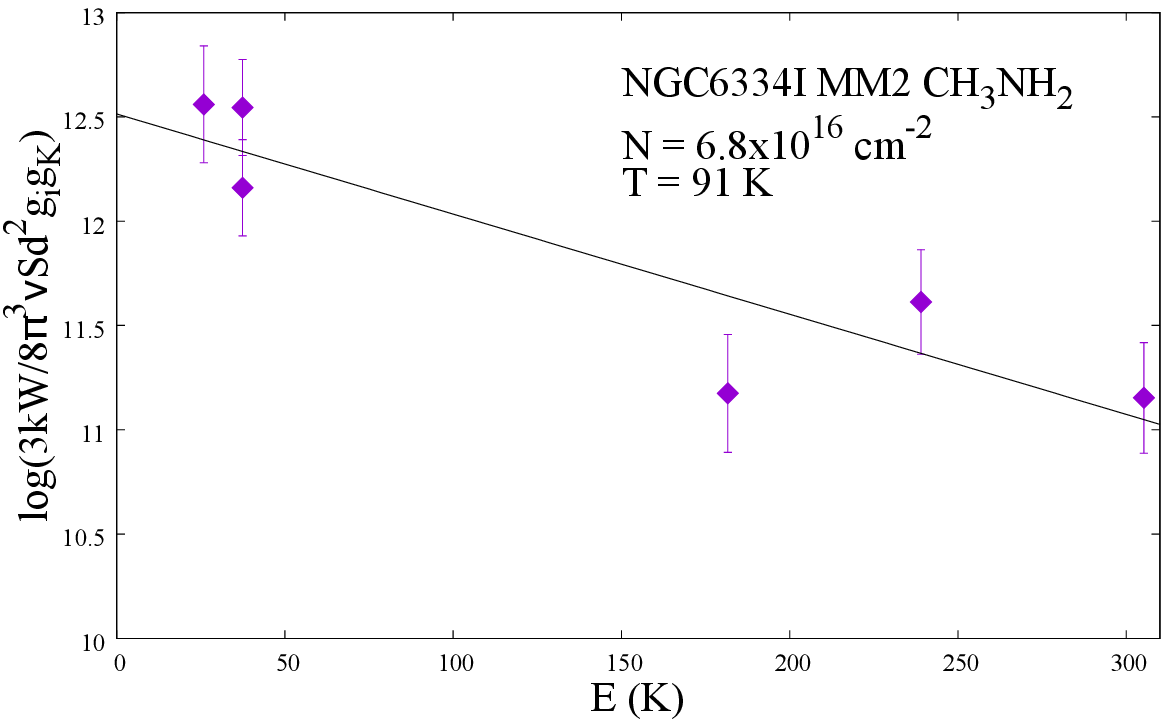}\\
\includegraphics[scale=0.6]{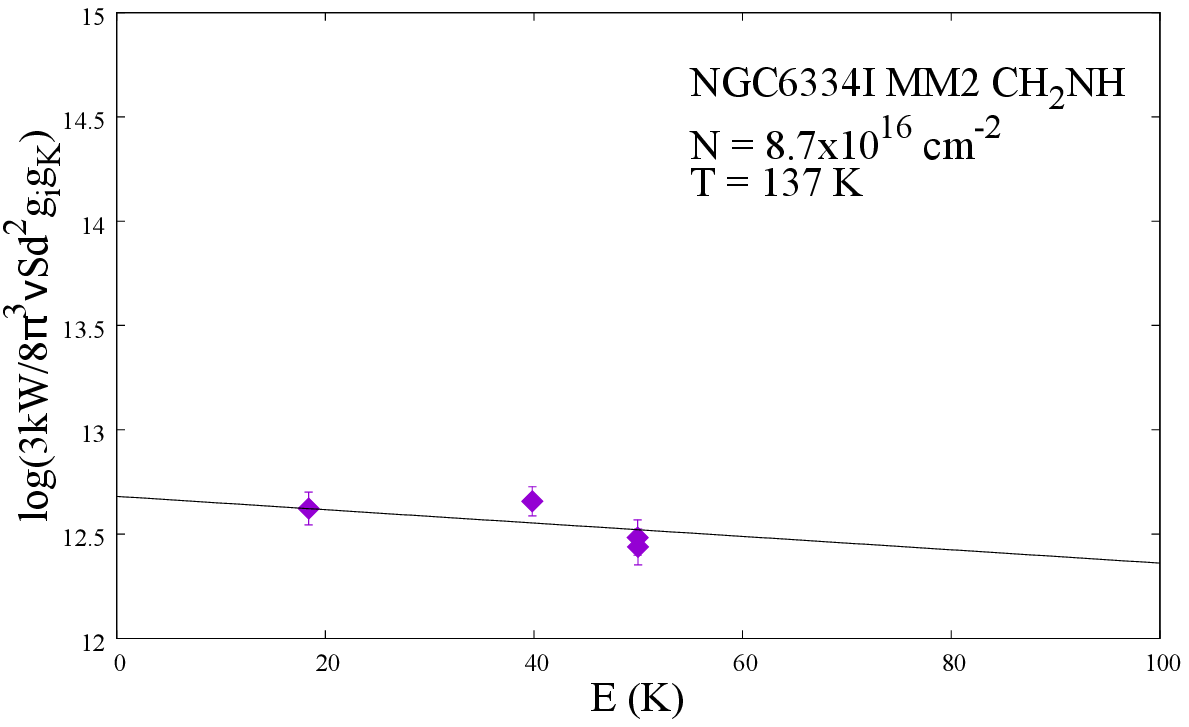}&
\includegraphics[scale=0.6]{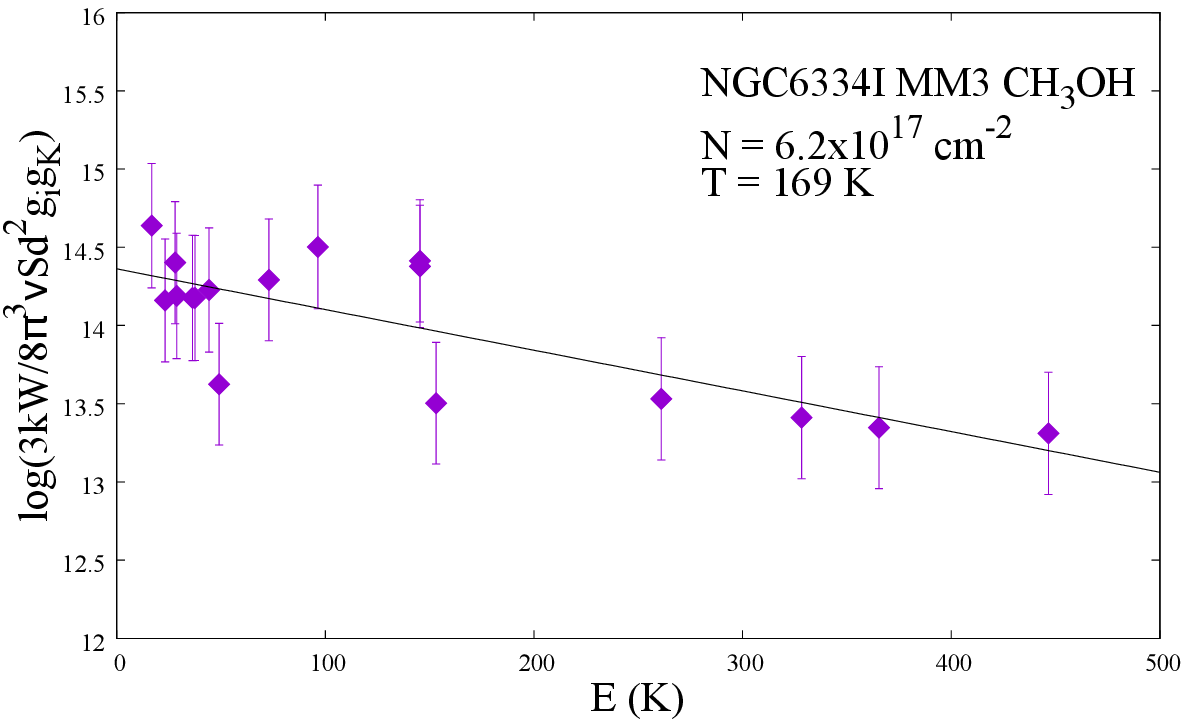}\\
 \end{tabular}
\caption{
The rotation diagrams of CH$_3$OH, CH$_3$NH$_2$, and CH$_2$NH for NGC6334I MM1, MM2, and MM3.
\label{fig:rotation}
}
\end{figure}

\clearpage
\addtocounter{figure}{-1}
\begin{figure}
 \begin{tabular}{ll}
\includegraphics[scale=0.6]{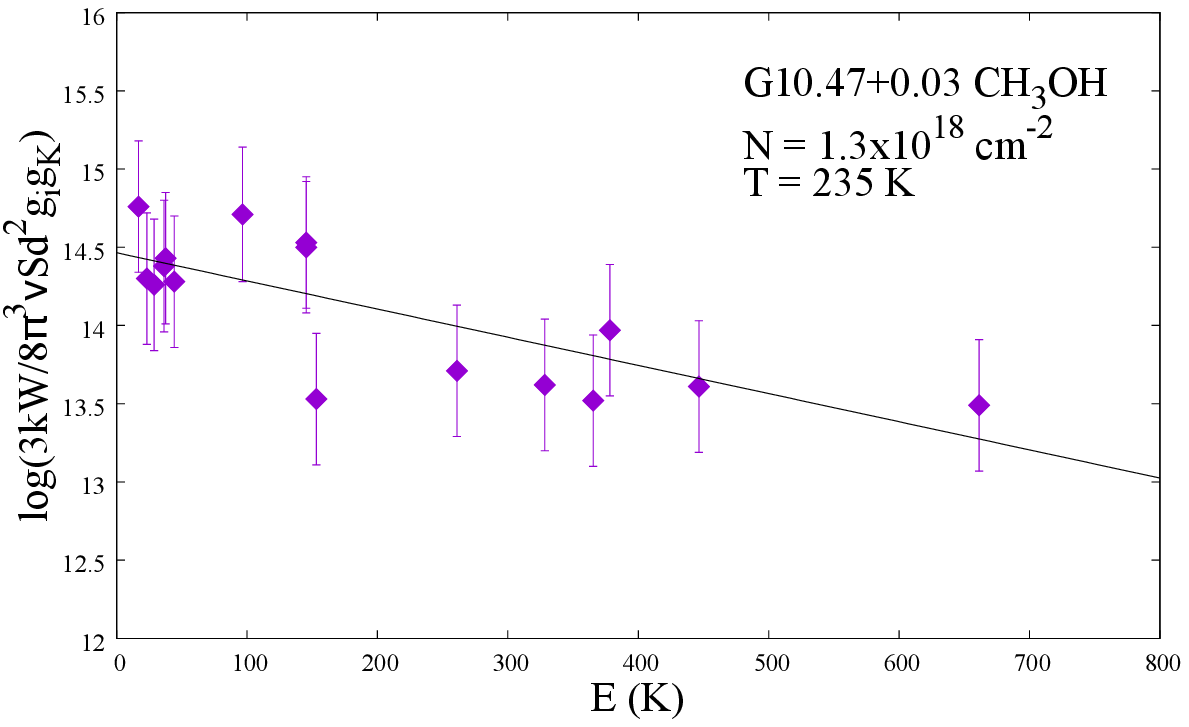}&
\includegraphics[scale=0.6]{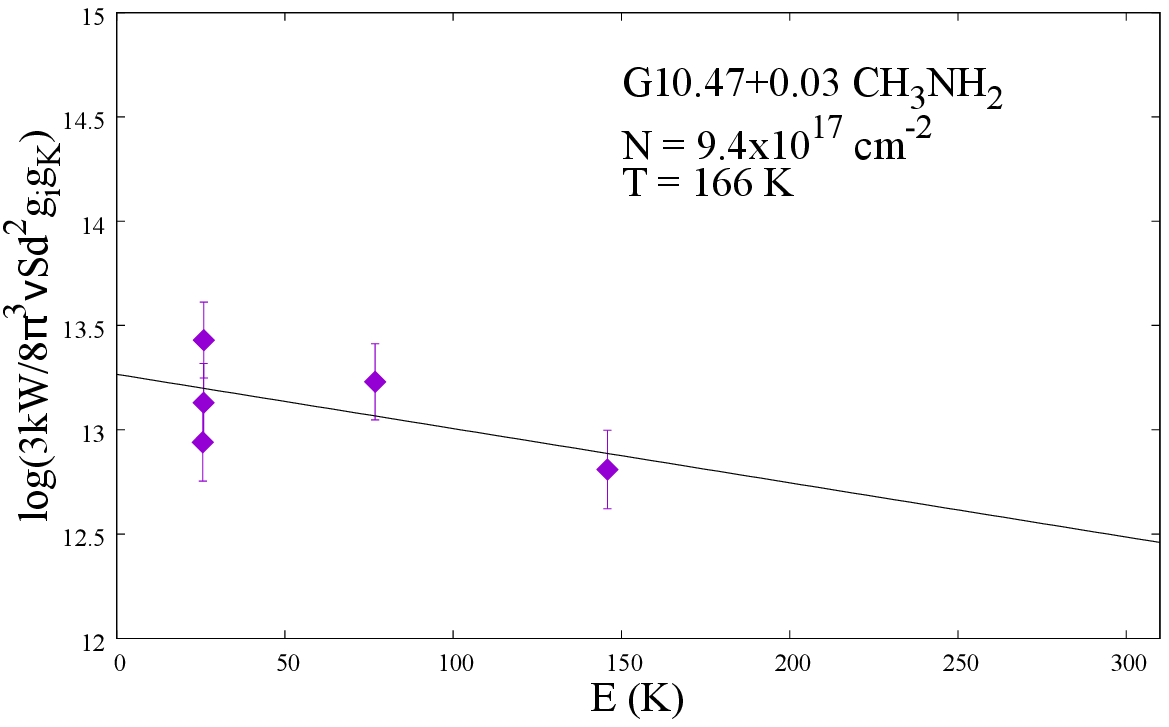}\\
\includegraphics[scale=0.6]{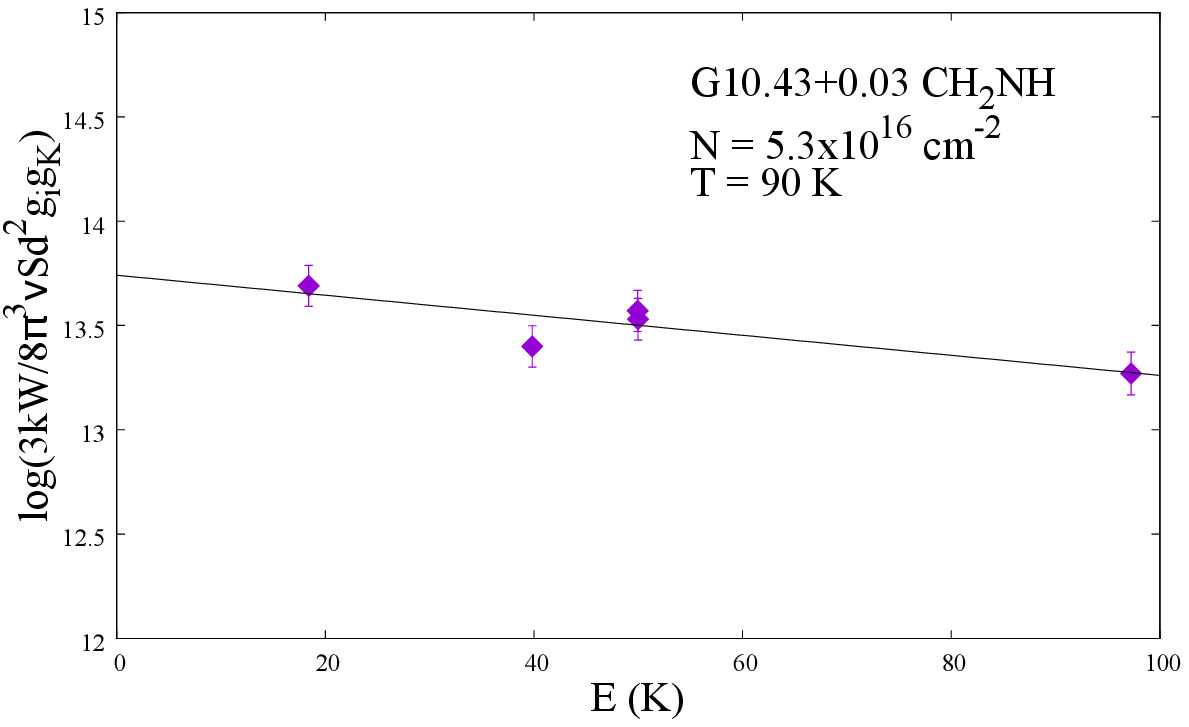}&
\includegraphics[scale=0.6]{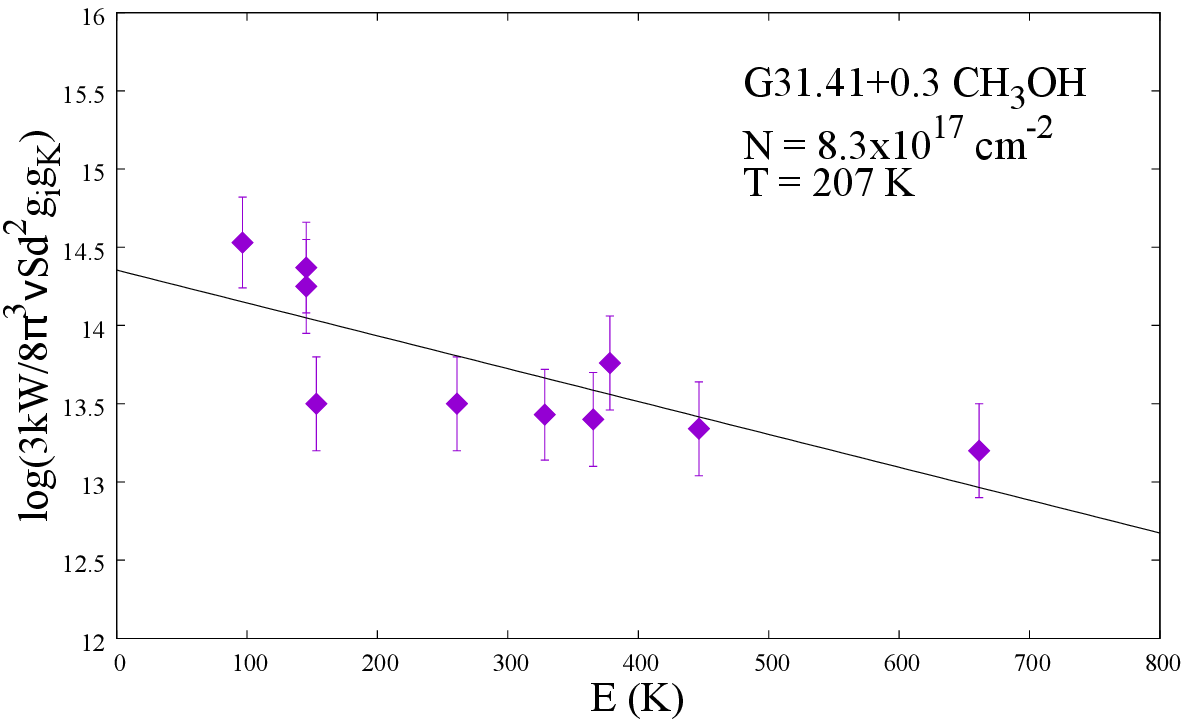}\\
\includegraphics[scale=0.6]{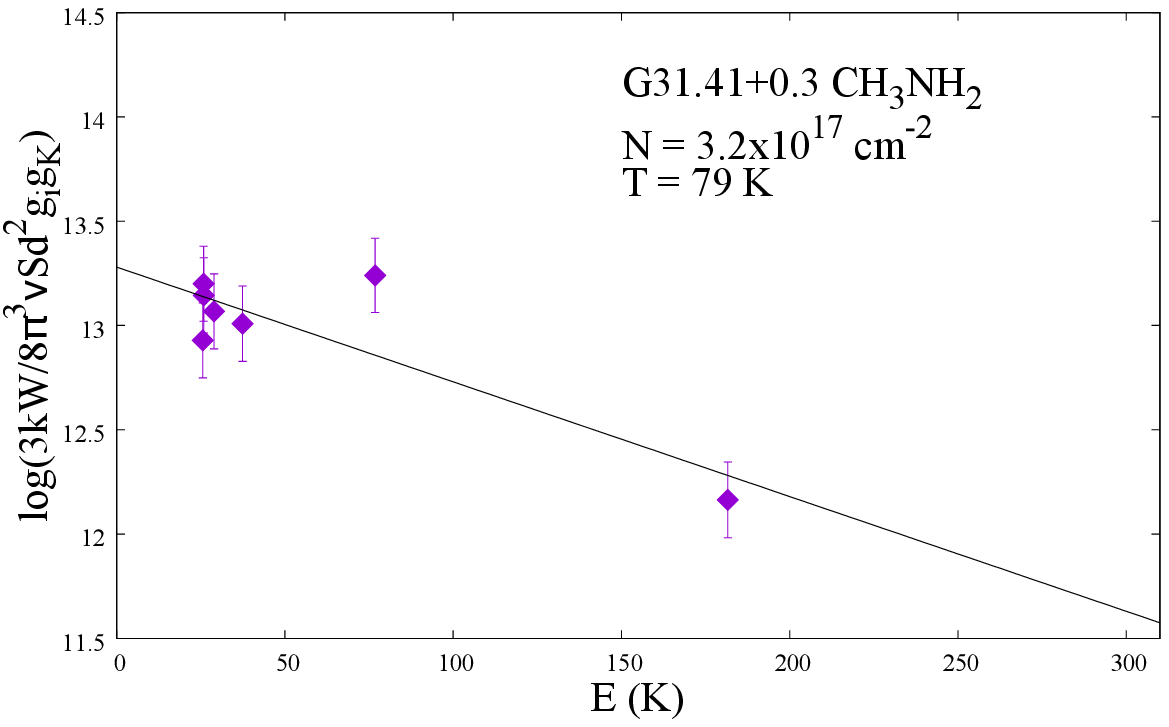}&
\includegraphics[scale=0.6]{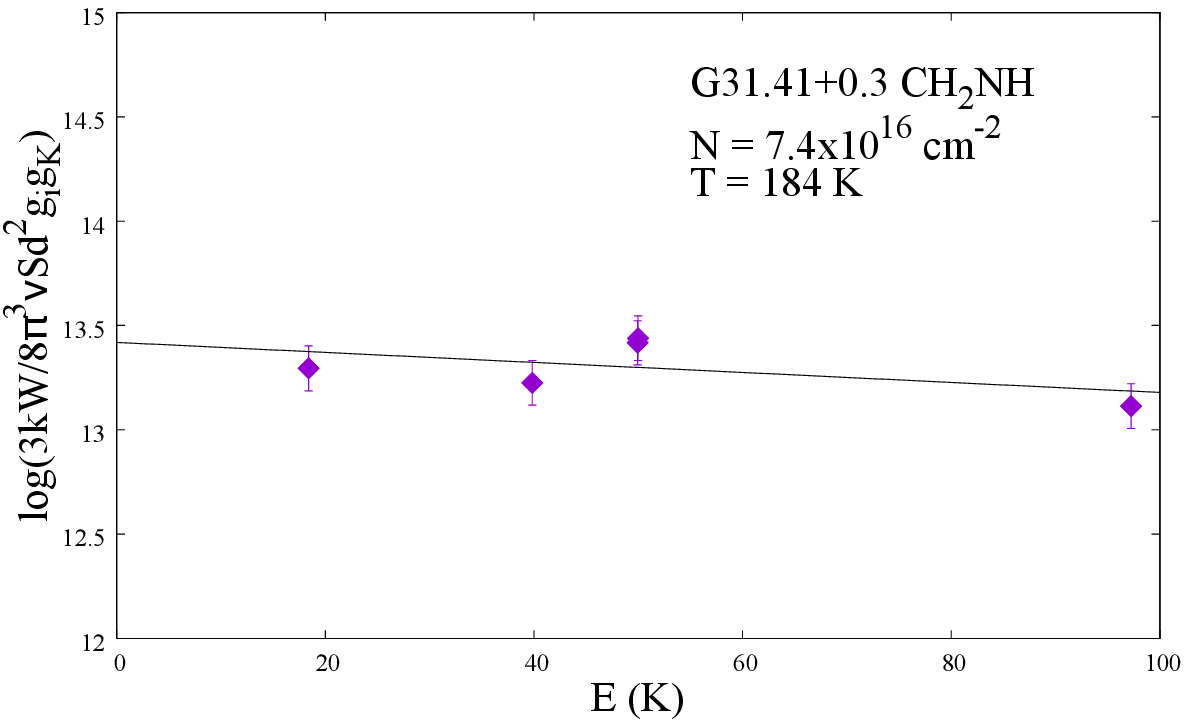}\\
 \end{tabular}
\caption{
(continued) The rotation diagrams of CH$_3$OH, CH$_3$NH$_2$, and CH$_2$NH for G10.47+0.03 and G31.41+0.3.
\label{fig:rotation2}
}
\end{figure}
\clearpage
\addtocounter{figure}{-1}
\begin{figure}
 \begin{tabular}{ll}
\includegraphics[scale=0.6]{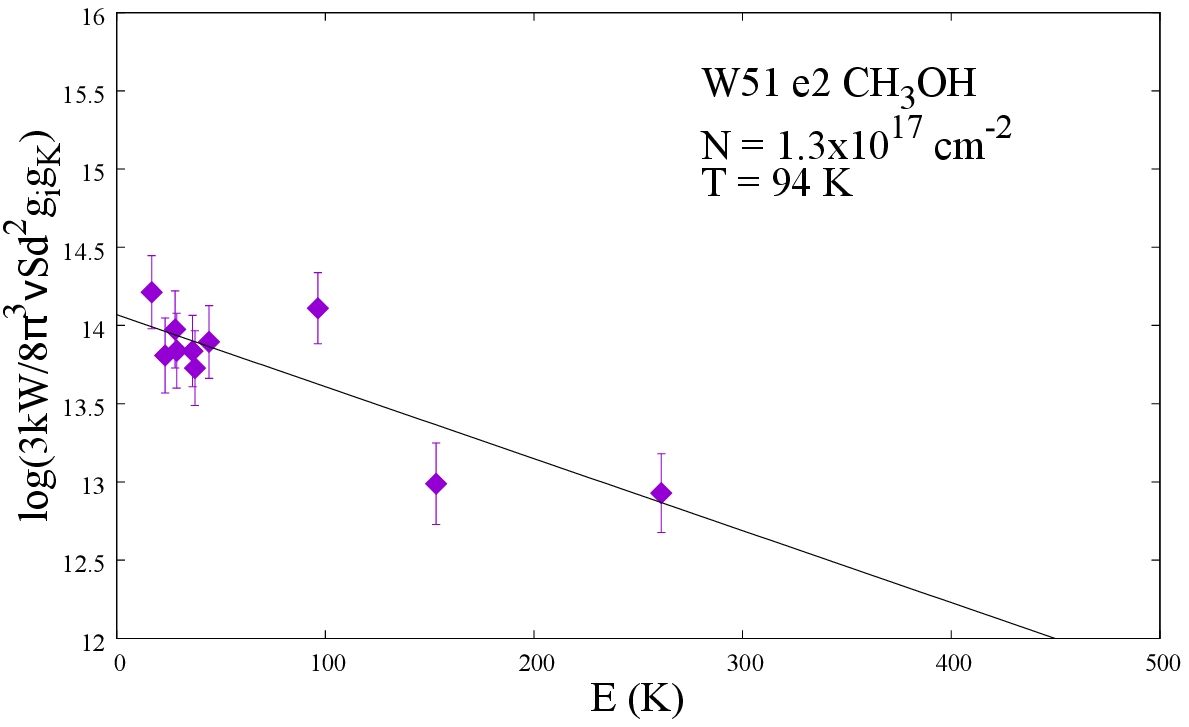}&
\includegraphics[scale=0.6]{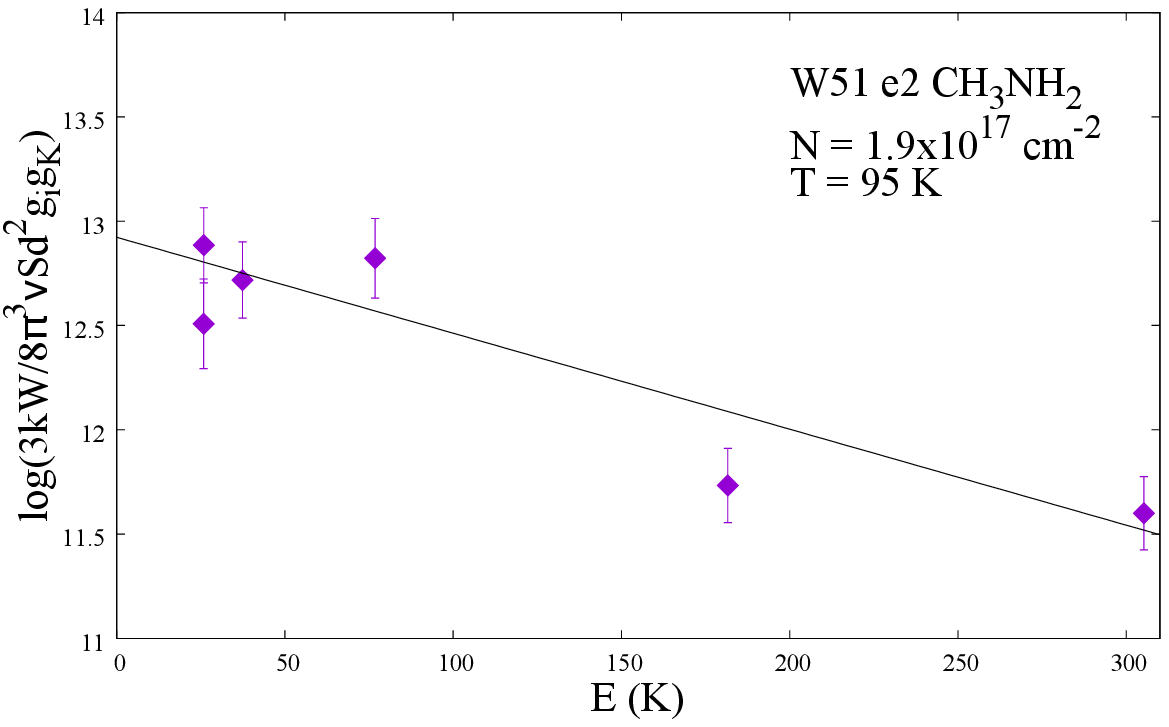}\\
\includegraphics[scale=0.6]{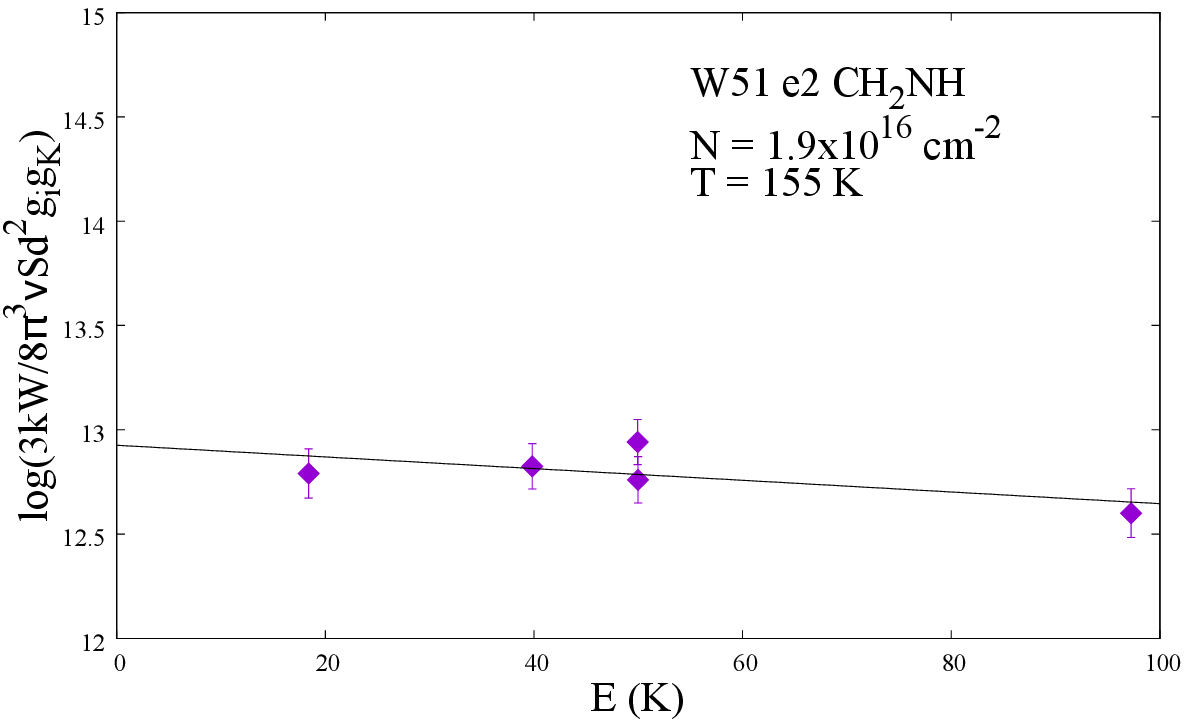}&
\includegraphics[scale=0.6]{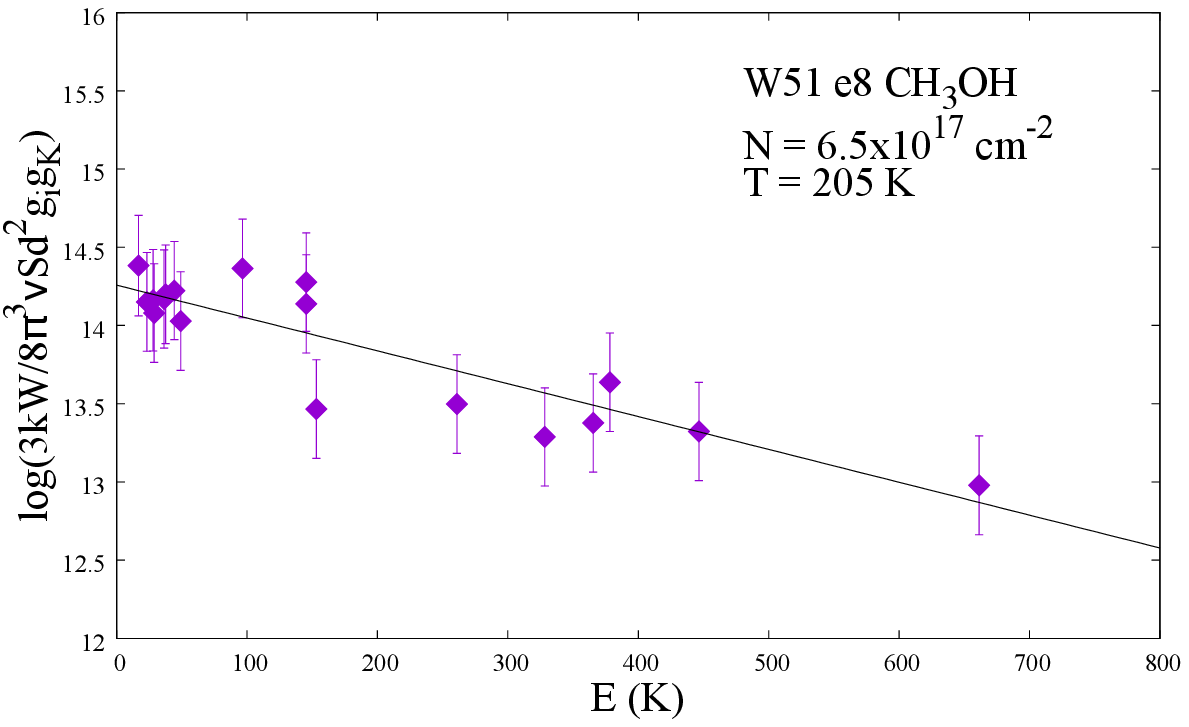}\\
\includegraphics[scale=0.6]{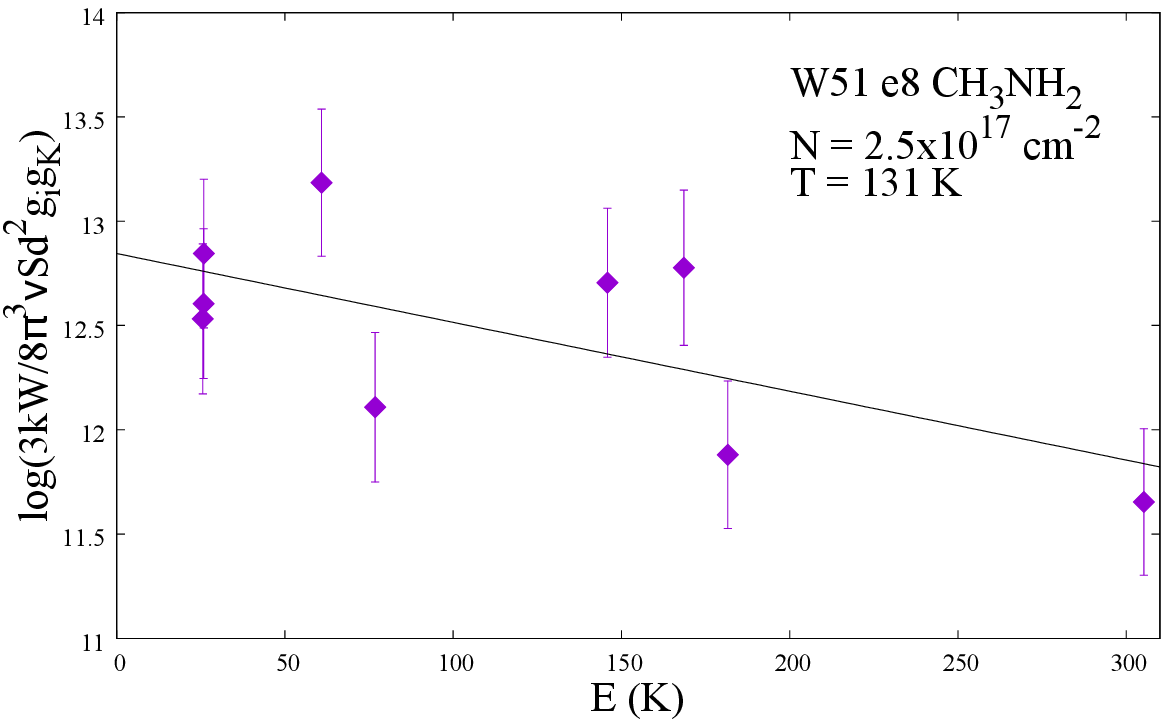}&
\includegraphics[scale=0.6]{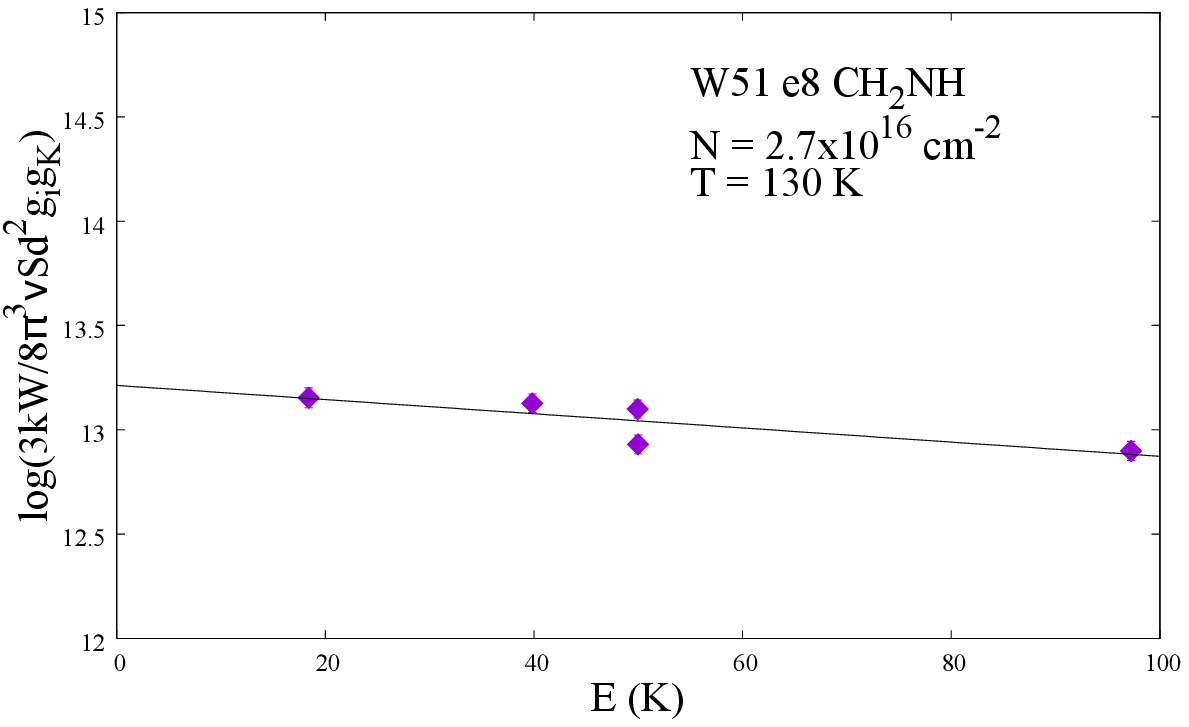}\\
 \end{tabular}
\caption{
(continued) The rotation diagrams of CH$_3$OH, CH$_3$NH$_2$, and CH$_2$NH for W51 e2 and e8.
\label{fig:rotation3}
}
\end{figure}
\clearpage

\begin{figure}
 \begin{tabular}{ll}
\includegraphics[scale=0.6]{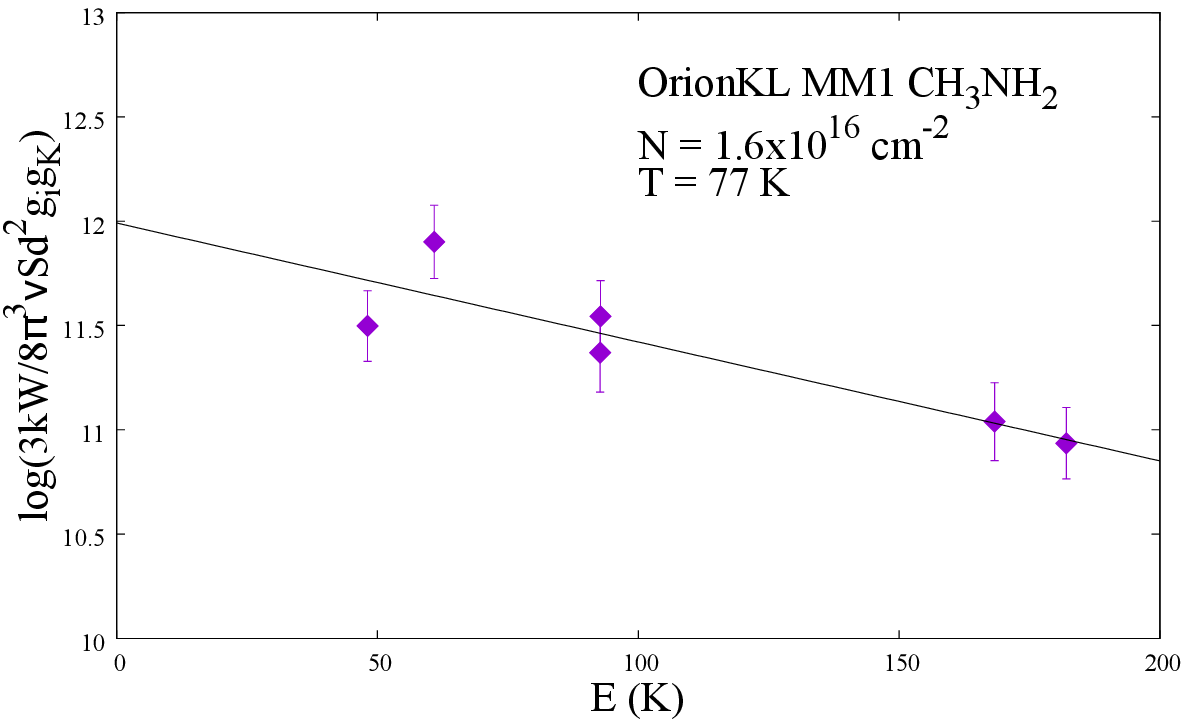}&\\
 \end{tabular}
\caption{
(continued) The rotation diagram of CH$_3$NH$_2$ for Orion KL Hot core analyzed by the archival data.
\label{fig:rotation}
}
\end{figure}
\clearpage
\begin{figure}
 \begin{tabular}{ll}
\includegraphics[scale=0.6]{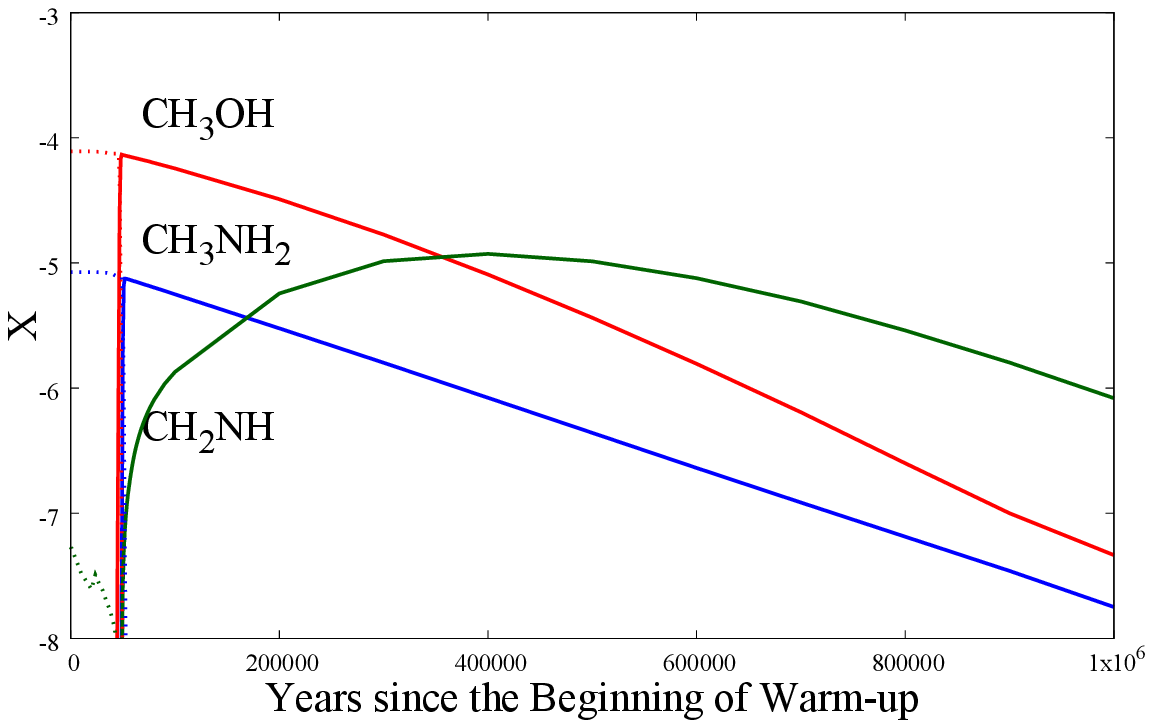}&
\includegraphics[scale=0.6]{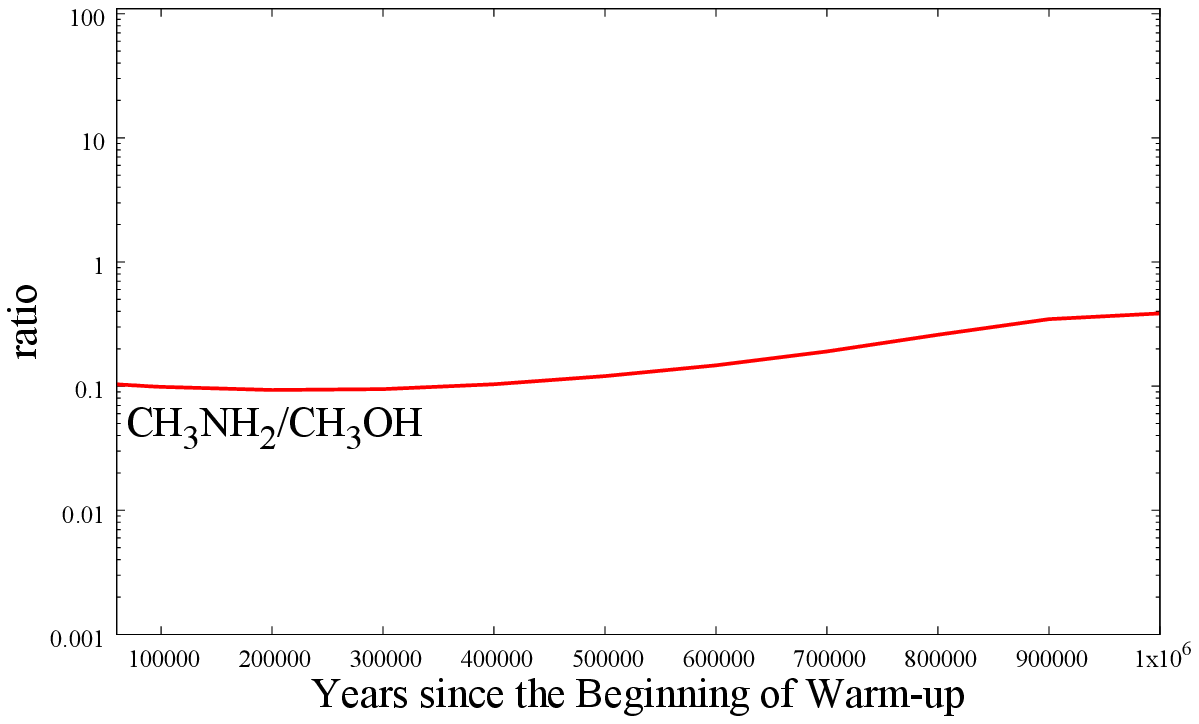}\\
 \end{tabular}
\caption{
(Left) The simulated fractionl abundance, X, for CH$_3$OH, CH$_3$NH$_2$, and CH$_2$NH in the gas phase are shown by red, blue, and green solid lines, respectively.
The dotted line represent the sum of the fractional abundances of those species in the grain mantle and on the grain surface.
(Right) The gas phase molecular abundance ratios of ``CH$_3$NH$_2$/CH$_3$OH'' and ``CH$_3$NH$_2$/CH$_2$NH'' are shown by red and green lines, respectively.
\label{fig:simulation}
}
\end{figure}

\end{document}